\documentclass[preprint,10pt]{elsarticle}
\usepackage[margin=3cm]{geometry}

\makeatletter
\def\ps@pprintTitle{%
  \let\@oddhead\@empty
  \let\@evenhead\@empty
  \let\@oddfoot\@empty 
  \let\@evenfoot\@oddfoot
}
\makeatother

\usepackage[T1]{fontenc}
\usepackage{subcaption}
\usepackage{amsmath}
\usepackage{amsfonts}
\usepackage{dsfont}
\usepackage{accents}
\usepackage[ruled,noend]{algorithm2e}
\usepackage{xcolor}
\usepackage[normalem]{ulem} 
\usepackage{hyperref}
\hypersetup{
    colorlinks=true,
    linkcolor=blue,
    }
\usepackage[title]{appendix}
\usepackage{booktabs}

\newcommand\thickbar[1]{\accentset{\rule{.4em}{.8pt}}{#1}}
\newcommand*{\vertbar}{\rule[-1ex]{0.5pt}{2.5ex}}

\usepackage[htt]{hyphenat}

\usepackage{xcolor}

\journal{ }

\begin{document}

\begin{frontmatter}



\title{Machine-Learning Enhanced Predictors for Accelerated Convergence of Partitioned Fluid-Structure Interaction Simulations}


\author[label1]{TIBA Azzeddine}
\author[label2,label5]{DAIRAY Thibault}
\author[label3]{DE VUYST Florian}
\author[label1]{MORTAZAVI Iraj}
\author[label4]{BERRO RAMIREZ Juan-Pedro}

\affiliation[label1]{organization={M2N, CNAM},
            addressline={2, Rue Conté}, 
            city={Paris},
            postcode={75003}, 
            country={France}}

\affiliation[label2]{organization={Manufacture Française des Pneumatiques Michelin},
            addressline={Place des Carmes-Dechaux}, 
            city={Clermont-Ferrand},
            postcode={63000}, 
            country={France}}

\affiliation[label5]{organization={Centre Borelli, CNRS, Université Paris Saclay, ENS Paris Saclay},
addressline={4, Avenue des Sciences}, 
city={Gif-sur-Yvette},
postcode={91190}, 
country={France}}

\affiliation[label3]{organization={Université de Technologie de Compiègne, CNRS, Laboratory of Biomechanics and Bioengineering},
            addressline={Rue du docteur Schweitzer}, 
            city={Compiègne},
            postcode={60203}, 
            country={France}}

\affiliation[label4]{organization={Altair Engineering France},
            addressline={5, Rue de la Renaissance}, 
            city={Antony},
            postcode={92160}, 
            country={France}}

\begin{abstract}
Stable partitioned techniques for simulating unsteady fluid-structure interaction (FSI) are known to be computationally expensive when high added-mass is involved. Multiple coupling strategies have been developed to accelerate these simulations, but often use predictors in the form of simple finite-difference extrapolations. In this work, we propose a non-intrusive data-driven predictor that couples reduced-order models of both the solid and fluid subproblems, providing an initial guess for the nonlinear problem of the next time step calculation. Each reduced order model is composed of a nonlinear encoder-regressor-decoder architecture and is equipped with an adaptive update strategy that adds robustness for extrapolation. In doing so, the proposed methodology leverages physics-based insights from high-fidelity solvers, thus establishing a physics-aware machine learning predictor. Using three strongly coupled FSI examples, this study demonstrates the improved convergence obtained with the new predictor and the overall computational speedup realized compared to classical approaches.

\end{abstract}

\begin{graphicalabstract}
\begin{center}
\includegraphics[width=1\textwidth]{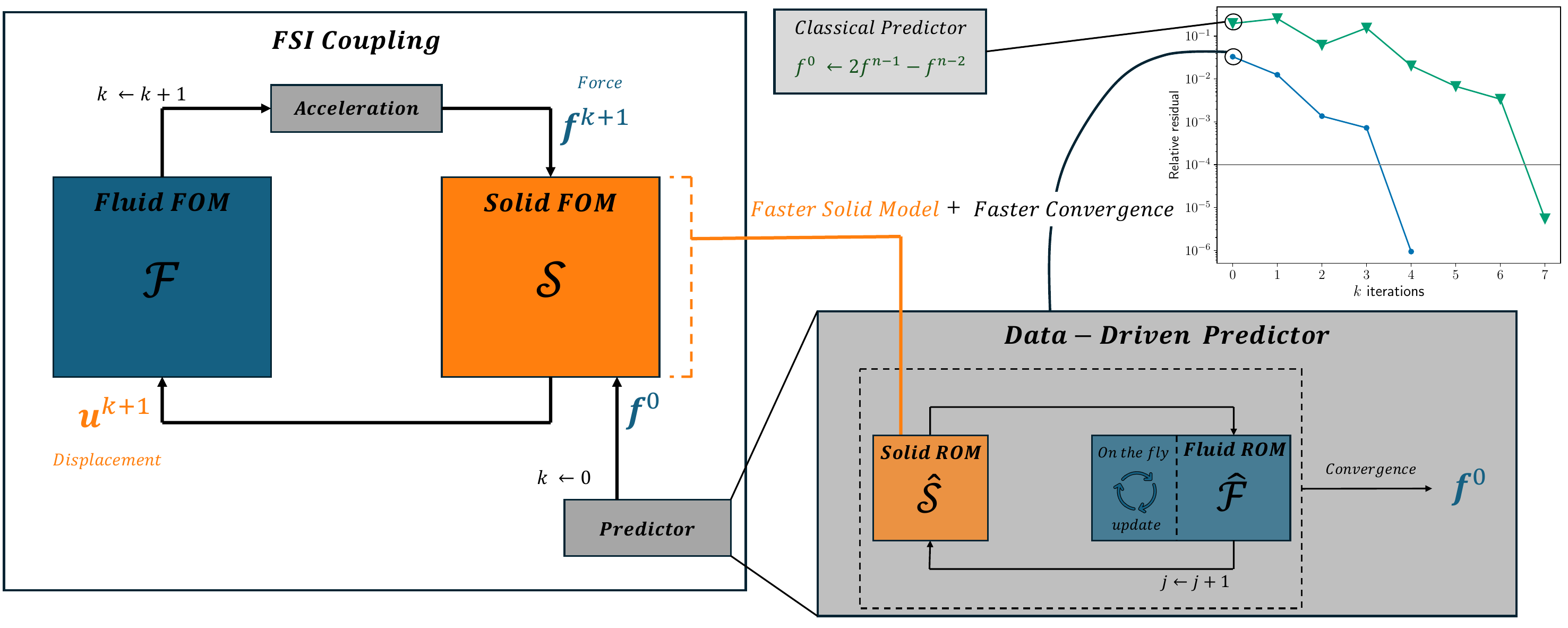}
\end{center}
\end{graphicalabstract}

\begin{highlights}
\item Novel predictor to accelerate convergence of fluid-structure interactions problems.
\item Coupled solution of solid and fluid reduced models is used as the next initial guess.
\item Reduced models take the form of encoder-regressor-decoder data-driven models.
\item Online adaptation of the reduced models is used for more robust extrapolation.
\item Faster convergence and speedups up to $3.2$ versus classical-predictor based coupling.
\end{highlights}


\end{frontmatter}


\section{Introduction}
\label{intro}

Engineering applications involving Fluid-Structure Interaction (FSI) phenomena are numerous, and are present in various fields. This includes aeroelasticity, biomechanics, microfluidics, hydrodynamics and many more. Modeling these FSI problems is a challenging task due to the usually high complexity of the coupling between the solid system and the fluid system.

Physics-based numerical simulations are considered as one of the leading options of modeling FSI problems, benefiting from years of advances in computational mechanics and from the increase of accessible computing power. These computations often have the goal of modeling the complicated coupling between the kinematics of a solid body and a fluid flow along with the mechanical loads associated with it. This results in very complicated problems, with nonlinearities present in both the fluid and the solid systems, in addition to the nonlinearities of the coupling itself. Furthermore, highly complex dynamics can be present due for example to the turbulent nature of the flow and/or to the interaction of the dynamical effects of the two systems involved.

Different discretization methods exist for handling FSI problems, especially from the fluid side, to deal with moving bodies: we mention for example immersed boundary methods~\cite{Peskin_2002}, smoothed particle hydrodynamics (SPH)~\cite{sph}, particle finite elements (PFEM)~\cite{pfem} and finite elements and finite volumes with the arbitrary Eulerian-Lagrangian formulation (ALE)~\cite{donea_arbitrary_1982} which is the method used in the present study.

The approaches to solve FSI problems numerically can be classified commonly into two main classes: monolithic and partitioned methods. In the former, the governing equations of the fluid and the solid behavior are solved simultaneously, with the coupling conditions (equality of the forces, displacements and velocities at the interface) respected exactly \cite{hron, heil_efficient_2004, FERNANDEZ20031337, Tezduyar}. However, computing the coupling terms (e.g the cross-derivatives in the Jacobian matrices) in the context of global Newton iterations can pose some serious computational problems. In addition to the expensive cost, it is very difficult to implement the monolithic schemes when different fluid and solid solvers are considered, especially if different discretization strategies are adopted for the two subproblems. This also becomes a significant disadvantage from a practical standpoint, in e.g an industrial context where there is a need for a non-intrusive combination of existing fluid and solid solvers. 

In the partitioned approach, each subproblem is solved separately, making it possible to use specialized solvers for the solid and the fluid, even when using different nonconforming grids on both sides and/or different discretization strategies on each domain (\emph{e.g. }finite elements on the solid side vs SPH~\cite{FOUREY201766} or PFEM~\cite{meduri} on the fluid side). This constitutes one of the main advantages of this approach, explaining its popularity. The partitioned coupling is achieved thanks to the communication of relevant quantities between the solvers at the interface, namely the displacement, velocity and stress fields at the interface, which are then used to enforce the adequate boundary conditions on each problem. This is the main idea behind the \textit{Dirichlet-Neumann} formulation for example, where the displacements and velocities stemming from the solid computations are enforced at the fluid boundary, and the forces resulting from the fluid stresses are imposed as a load on the solid domain. Unfortunately, an inherent instability appears when dealing with partitioned approaches: Due to the time delay that exists between the instants of solving the solid and the fluid problems, ensuring stability of the coupling conditions is not straightforward. In problems where the coupling is not very strong, i.e when the effect of one sub-problem on the other is much smaller than the effect of the other direction, coupling schemes can be made with one solver call per time step. This is commonly called \textit{staggered schemes} or \textit{loosely coupled schemes}. Extensive development of staggered schemes was done in early works~\cite{higherOrderLesoinne, pipernoPredictor, FARHAT_two_schemes, pipernoEnergy, felippa_partitioned_review, FARHATprovably2order}, where strategies were introduced with improved synchronization of the two subsystems\cite{FARHAT_two_schemes}, improved order of accuracy~\cite{higherOrderLesoinne, FARHATprovably2order}, and improved stability~\cite{pipernoPredictor, pipernoEnergy}, which generally involves clever choices of the time integrator in each solver, and a crucial choice of the good predictors of force or displacement at the beginning of each time step. Although the performance of such strategies was well demonstrated in aeroelastic applications with compressible flows, loosely coupled scheme perform much worse in problems with stronger coupling and incompressible flows. This is due to the added mass instability, a phenomenon related to the effect of the displaced (due to the solid displacement) fluid mass, on the solid itself. It was shown in~\cite{causin_added-mass_2005, BrummelenAdded, FORSTER20071278} that this effect is more significant when the ratio of the added fluid mass on the solid mass increases, and that it also depends on the geometry of the two subdomains. It was also shown that, when the added mass is large, the partitioned schemes - even with full subiterations - can, at worst, fail to converge quickly, and at best, be conditionally stable, irrespective of the time step. In fact, smaller time steps may even cause worse convergence~\cite{causin_added-mass_2005, BrummelenAdded, degroote_stability_2008}. As a consequence, FSI problems involving light structures and a high added mass become computationally expensive, due to the large number of sub iterations needed at each time step or the very small time steps needed for segregated schemes. To deal with this, important work was done using various strategies, some of which is intrusive, like semi-implicit coupling with a segregated fluid solver~\cite{fernandezSemiImplicit}, others are somewhat intrusive, like the \textit{Robin-Neumann} formulations~\cite{BADIA20087027}, adding artificial compressibility to the fluid system to ease convergence~\cite{viendereeelsDisatoleAICstatic, degrooteAIC}, or enforcing coupling conditions through domain decomposition methods \cite{liZhe, meduri2018, giber2022}. Other strategies can be implemented completely non-intrusively: One such approach is segregated \textit{Dirichlet-Neumann} schemes with specific force predictors to deal with relatively high added-mass and small time steps~\cite{dettmerSegr, kadapaSegr}. Another approach - suitable for even higher added mass ratios - is using iterative schemes with convergence acceleration techniques. This includes either Aitken acceleration~\cite{wallAitk}, or quasi-newton (QN) acceleration methods for the FSI problem formulated as a fixed point problem~\cite{ibqnls, degroote_performance_2009, bogaers_quasi-newton_2014, haeltermannFilter, uekermann_partitioned_2016} (see a review in~\cite{broyden}). The latter strategy is the one we focus on in this work thanks to its high flexibility for both problems with high added-mass and black-box coupling with existing solvers. We note that in all these partitioned schemes, there is often the notion of a predictor (for forces, displacements or velocities), the choice of which can be crucial for the acceleration of convergence.

To reduce the computational cost associated with strongly coupled FSI simulations, some non-intrusive reduced order models (ROMs) have been developed in the past years. These ROMs are generally data-driven models that combine various Machine Learning (ML) methods to predict solutions in the time-parameter space. We mention for example the works in~\cite{shinde_galerkin-free_2019, xiao_non-intrusive_2016, Dupont_DVuyst, fluidsFresca} where Proper Orthogonal Decomposition (POD) was combined with machine learning to predict parametric dynamical FSI solutions, or using spectral submanifolds for non-linearizable cases \cite{haller2, haller}. In a recent work~\cite{tiba2023nonintrusive}, we constructed data-driven ROMs for the solid subproblem intended to be coupled with high fidelity (HF) fluid full order models (FOMs), making thus a strong ROM-FOM coupling, capable of finding the FSI problem solutions with a much lower cost than the FOM-FOM while maintaining good accuracy.

In the present work, we combine ideas of partitioned coupling acceleration and data-driven ROM-FOMs to construct more effective predictors for a faster convergence of iterative partitioned FSI schemes. Moreover, we propose a strategy to adaptively update these predictors using HF data generated from the fluid FOM that is still activated during online computations. In doing so, the proposed methodology leverages physics-based insights from the fluid solver, thus establishing a physics-aware machine learning predictor. This method was inspired by our previous work in~\cite{tiba2023nonintrusive} where the replacement of the solid FOM with a ROM, albeit reduces the computational cost, can contribute to slower convergence due to the ROM inaccuracies. Hence, equipping this ROM-FOM coupling with stronger predictors can ensure even greater speedups. Importantly, this enhanced predictor can be used with classical FOM-FOM FSI coupling as well.

Other recent works have been focused on accelerating convergence of solving nonlinear problems using ML; interesting examples include~\cite{pinL, kadeethum_enhancing_2022, zandbergen2023improving, silva_machine_2021} where ML techniques (PCA dimensionality reduction, neural networks and random forests) are used to obtain correct parameters of nonlinear numerical solutions (initial guess, pseudo time steps for pseudo-transient continuation and Aitken relaxation parameter). In a similar work~\cite{DELAISSE2022106720}, although not using data-driven models, surrogate models (using e.g simplified faster models) have been utilized to obtain faster convergence of the Interface Quasi-Newton Inverse Least-Squares (IQN-ILS) method~\cite{degroote_performance_2009}. The novelty of the work presented here however is the use of data-driven techniques to accelerate convergence of partitioned FSI coupling.

\begin{figure}
    \centering
    \includegraphics[width=1\textwidth]{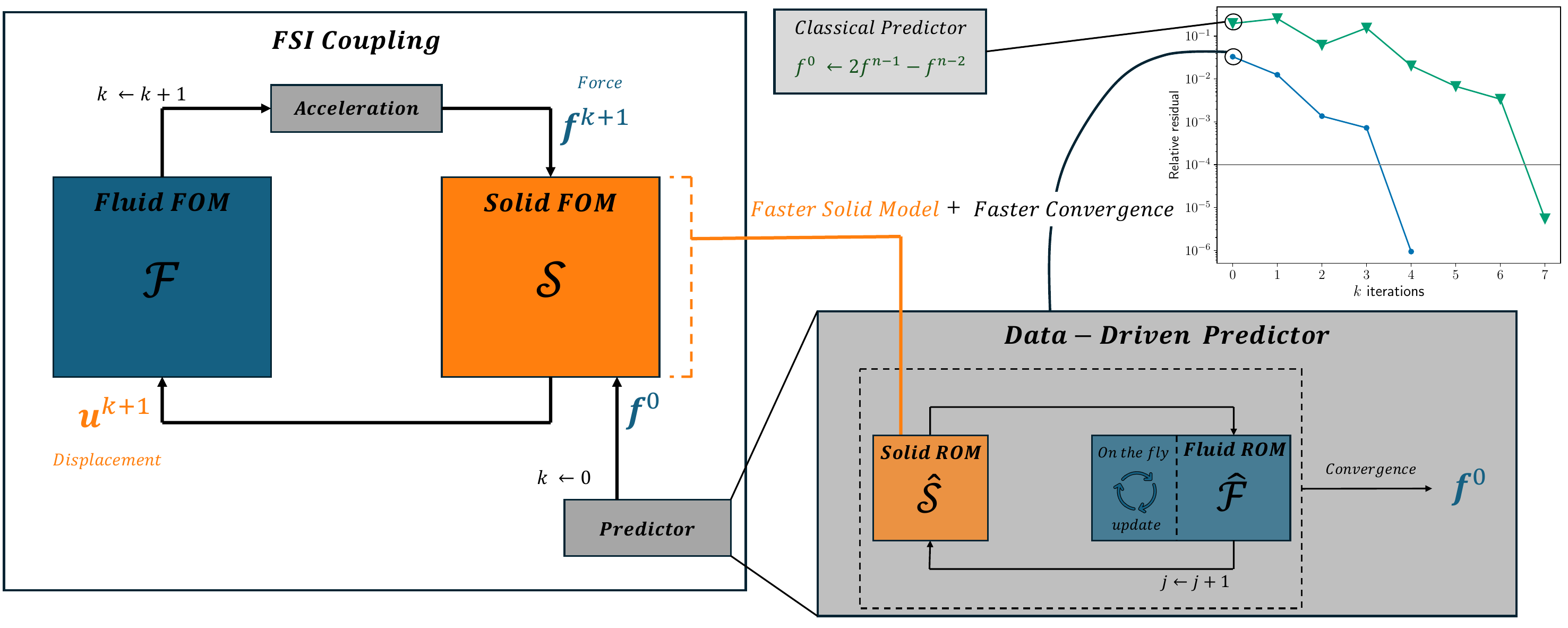}
    \caption{Brief illustrative representation of the presented approach.}
    \label{fig:graphAbstract}
\end{figure}

Particularly, we use POD dimensionality reduction, followed by two regression models approximating the load to displacement (solid solver) and the displacement to load (fluid solver) operators to construct fast ROMs that can be coupled in few iterations to produce an initial guess for the next time step problem. In addition, the fluid ROM can be updated using the HF data obtained online and thus keeps a satisfying accuracy. The predictor thus takes advantage of the available HF data in a smarter way than a simple extrapolator does. A brief illustrative explanation of the complete methodology is presented in Figure~\ref{fig:graphAbstract}. In this study, we focus on solid models with neglected inertia, but a natural extension can be used for dynamical models, which we intend to pursue in future work.

The remainder of this paper is structured as follows: In Sect. \ref{sect:fomfom}, the governing equations of FSI and the partitioned formulation are presented. Then, in Sect. \ref{sect:romfom}, the proposed data-driven predictor approach is detailed. The results of the evaluation of this in terms of convergence acceleration are presented in Sect. \ref{sect:results}: 3 test cases are used to demonstrate the efficiency of this approach. Finally, a conclusion is given in Sect. \ref{sect:concl}.

\section{FOM-FOM fluid-structure interaction black-box coupling}\label{sect:fomfom}

In FSI problems, the nonlinear global problem is expressed by both the dynamic and kinematic coupling conditions at the interface ${\Gamma_{fsi}}$: 
\begin{equation}\label{fsiCondition1}
    \begin{cases}
	  \sigma_f \cdot \pmb{n}  =  - \sigma_s \cdot \pmb{n} \, \, \, \, at \, \, \, \, {\Gamma_{fsi}} \\
	  \pmb{v} = \dot{\pmb{u}} \, \, \, \, at \, \, \, \, {\Gamma_{fsi}}\\
   \end{cases}
\end{equation}
where $\sigma_f$ and $\sigma_s$ are the Cauchy stresses applied by the fluid and the solid respectively, $\pmb{v}$ is the fluid velocity and $\dot{\pmb{u}}$ is the solid velocity. In mesh-based methods for the fluid problem, the superposition of the two \textit{wet} interfaces translates into an additionnal coupling condition
\begin{equation}\label{fsiCondition2}
      \pmb{x} = {\pmb{u}} \, \, \, \, at \, \, \, \,{\Gamma_{fsi}}\\
\end{equation}
where $\pmb{x}$ is the fluid mesh displacement field and $\pmb{u}$ is the solid displacement.

The \textit{Dirichlet-Neumann} coupling formulation represents the two solvers as distinct operators that exchange their input and output at each iteration. We represent the fluid solver operator as $\mathcal{F}$: 
\begin{equation}
  \mathcal{F}: \mathbb{R}^{2N} \rightarrow \mathbb{R}^N\;;\; (\pmb{u}_{|\Gamma_{fsi}}, \dot{\pmb{u}}_{|\Gamma_{fsi}}) \rightarrow \pmb{f}_{|\Gamma_{fsi}} 
\end{equation}
where $\pmb{u}_{|\Gamma_{fsi}}$ is the displacement field $N$ is the number of interface degrees of freedom (\textit{e.g} the number of interface grid points times the number of components in mesh-based methods) and $\pmb{f}_{|\Gamma_{fsi}}$ represents the fluid viscous and pressure forces at $\Gamma_{fsi}$:
\begin{equation}
  \pmb{f}_{|\Gamma_{fsi}} = \sigma_f \cdot \pmb{n}_{f|\Gamma_{fsi}}.
\end{equation}
%
Similarly, the solid operator $\mathcal{S}$ is defined as:
\begin{equation}\label{solid_operator}
  \mathcal{S}: \mathbb{R}^N \rightarrow \mathbb{R}^{2N}\;;\; \pmb{f}_{|\Gamma_{fsi}} \rightarrow (\pmb{u}_{|\Gamma_{fsi}}, \dot{\pmb{u}}_{|\Gamma_{fsi}})  .
\end{equation}
In strongly coupled schemes, the coupling conditions are enforced at the iterations convergence. This nonlinear problem can be written as the fixed-point:
\begin{equation}\label{dirich_neum}
    (\mathcal{F} \circ \mathcal{S})(\pmb{f}_{|\Gamma_{fsi}}) = \pmb{f}_{|\Gamma_{fsi}}.
\end{equation}

In this work, we consider problems where the solid inertia is negligible compared to the fluid one, i.e the solids studied here have negligible mass and are solved under quasi-static load. These assumptions have been relevant in modeling many FSI problems, like collapsible channels \cite{mpRastChannel, HEIL_1997, LUO_PEDLEY_1998, heil_efficient_2004}. Together with the flow incompressiblity, this means that the FSI coupling in theses cases is very strong and that the added mass is extremely large, as shown in \cite{causin_added-mass_2005, BrummelenAdded}. 
Standard Gauss-Seidel will not converge in the majority of such cases. Otherwise, QN-accelerated iterative schemes  will take many iterations for each time step, making the potential acceleration provided by the new predictor even more advantageous. Moreover, given that only elastic (albeit nonlinear) material laws are used, the solid problem considered in this work is hence path-independent.

We note that the quasi-static load setting means that no velocity field is computed from the solid solver, and the mesh velocity can be computed using a proper time differentiation scheme \cite{lesoinne_geometric_1996} in the ALE mesh motion solver (usually part of the fluid solver). Consequently, the Dirichlet-Neumann formulation (\ref{dirich_neum}) becomes:

\begin{equation}\label{fluid_static_operator}
  \mathcal{F}: \mathbb{R}^{N} \rightarrow \mathbb{R}^N\;;\; \pmb{u}_{|\Gamma_{fsi}} \rightarrow \pmb{f}_{|\Gamma_{fsi}} 
\end{equation}

\begin{equation}\label{solid_static_operator}
  \mathcal{S}: \mathbb{R}^N \rightarrow \mathbb{R}^{N}\;;\; \pmb{f}_{|\Gamma_{fsi}} \rightarrow \pmb{u}_{|\Gamma_{fsi}}  .
\end{equation}

In the following, for clarity, we will remove the interface subscript from $\pmb{f}_{|\Gamma_{fsi}}$ and $\pmb{u}_{|\Gamma_{fsi}}$. In order to pass the field of interest at the interface, a mapping should be done to interpolate between the two grids. Energy-conserving mapping methods should be performed in the case of non-matching grids between the two systems. In the paper, we represent the mapping operators as $\mathcal{M}_{\mathcal{F} \rightarrow \mathcal{S}}$ (from fluid grid to solid grid) and $\mathcal{M}_{\mathcal{S} \rightarrow \mathcal{F}}$ (from solid grid to fluid grid). In the presented cases here, all the grids are matching on the interface, and the nearest-neighbor method is used. The reader is referred to \cite{Gatzhammer2015EfficientAF, mapping_FARHAT199895, boer_comparison_2008} for additional details on mapping algorithms. In the rest of the paper, each field will be represented on the grid defined for the solver from which it is computed, \textit{i.e} $\pmb{u}$ on the solid mesh and $\pmb{f}$ on the fluid mesh, otherwise, proper subscripts will indicate if they are defined elsewhere, using the notations: $\pmb{f}_{\mathcal{F}}$, $\pmb{f}_{\mathcal{S}}$, $\pmb{u}_{\mathcal{F}}$ and $\pmb{u}_{\mathcal{S}}$.

\subsection{\textbf{Coupling scheme}}\label{CouplingSchemeParagraph}
To solve (\ref{dirich_neum}), an iterative solution can be found by calling the 2 solvers sequentially at each iteration $k$:
\begin{equation}
    \begin{cases}
	  \pmb{f}^k  = \mathcal{F}(\pmb{u}^k) \\
	  \pmb{u}^{k+1} = \mathcal{S}(\pmb{f}^k) \\
 \end{cases}.
\end{equation}
These are sometimes called Picard iterations, and the scheme is referred to as the Gauss-Seidel scheme.
Note that, depending on which subproblem is computed first, the unknown may change: if the fluid solver is called first, (\ref{dirich_neum}) is replaced with the -generally- equivalent formulation:
\begin{equation}\label{dirich_neum2}
    (\mathcal{S} \circ \mathcal{F})(\pmb{u}) = \pmb{u}.
\end{equation}

In this work, we use (\ref{dirich_neum}) because:
\begin{enumerate}
    \item Force predictors (and (\ref{dirich_neum})) were reported to be more efficient and result in fewer iterations than displacement predictors (and (\ref{dirich_neum2})) \cite{Sachs_Streitenberger_Sternel_Schafer_2011, dettmerSegr}.
    \item Although out of the scope of this work, when dynamical solutions are solved in the solid problem, enforcing the kinematic coupling conditions at the interface means that the velocity field needs to be passed from the solid solver to the fluid solver, especially in black-box coupling when there is no knowledge about the time integration scheme in the solid solver. In this context, it is easier to accelerate the fluid forces field than it is for the displacement, where careful considerations should be done and/or additional cost to update the velocity field is spent.
    \item The accuracy of the regression models of our data-driven ROMs will be higher when these models are trained on updated/relaxed force fields compared to when they are trained on non-relaxed fields. This is due to the low variance of the data points when they correspond to relaxed fields. This point will be highlighted with a numerical experiment in the results Section \ref{casTest2}.
\end{enumerate}

Equation (\ref{dirich_neum}) can also be rewritten as a block fixed-point system
\begin{equation}\label{block}
  \left(\begin{array}{cc}
    \mathcal{S} \circ \mathcal{F} & \pmb{0} \\
    \pmb{0} & \mathcal{F} \circ \mathcal{S}
  \end{array}\right) \left(\begin{array}{c}
     \pmb{u} \\
    \pmb{f}
  \end{array}\right) = \left(\begin{array}{c}
     \pmb{u} \\
    \pmb{f}
  \end{array}\right).
\end{equation}
Alternatively, a parallel formulation of the fixed-point can be sought if we write it as a Jacobi system: 
\begin{equation}
  \left(\begin{array}{cc}
    \pmb{0} & \mathcal{S}  \\
    \mathcal{F} & \pmb{0}
  \end{array}\right) \left(\begin{array}{c}
     \pmb{u} \\
    \pmb{f}
  \end{array}\right) = \left(\begin{array}{c}
     \pmb{u} \\
    \pmb{f}
  \end{array}\right)
\end{equation}
which allows the simultaneous solution of the two solvers in parallel. While the ROM strategy we present in this work can be equivalently used in all these different formulations, we will focus here on the Gauss-Seidel system (\ref{dirich_neum}). The FSI fixed-point iterations to find the solution of (\ref{dirich_neum}) are called henceforth the \textit{global iterations}.

\subsection{\textbf{Convergence acceleration}}
The simple use of Picard iterations, where the resultant forces field ($\Tilde{\pmb{f}}$) is passed directly as an input to the solid solver ($\pmb{f} = \Tilde{\pmb{f}}$), shows poor convergence or may diverge in strongly coupled problems due to the added-mass instabilities~\cite{causin_added-mass_2005}. To remedy this, convergence acceleration 
is done at the end of each iteration. A simple method is to use Aitken acceleration~\cite{aitkenOrg}, where at the end of iteration $k$, $\Tilde{\pmb{f}}_{|\Gamma_{fsi}}$ is modified as:
\begin{equation}
    {\pmb{f}^k} = w^k \Tilde{{\pmb{f}}}^k + (1-w^k){{\pmb{f}}}^{k-1}
\end{equation}
where 
\begin{equation}\label{aitkenR}
     w^k  = - w^{k-1} \frac{\pmb{r}^{k-1\;T}(\pmb{r}^{k}-\pmb{r}^{k-1})}{||\pmb{r}^{k}-\pmb{r}^{k-1}||_2^2}
\end{equation}
and $\pmb{r}^{k}$ is the residual at iteration $k$ : 
$\pmb{r}^k = \Tilde{{\pmb{f}}}^{k} - {\pmb{f}}^{k-1}$

Another approach is to formulate the problem (\ref{dirich_neum}) as a nonlinear problem to be solved using Quasi-Newton (QN) iterations:
\begin{equation}
    \pmb{\mathcal{R}}(\pmb{x}) = \pmb{0}
\end{equation}
where $ \pmb{\mathcal{R}}(\cdot)$ represents $ \pmb{\mathcal{R}}(\pmb{f}) = (\mathcal{F} \circ \mathcal{S})(\pmb{f}) - \pmb{f}$ in the case of (\ref{dirich_neum}) and $ \pmb{\mathcal{R}}(\pmb{u}) = (\mathcal{S} \circ \mathcal{F})(\pmb{u}) - \pmb{u}$ in the case of (\ref{dirich_neum2}).
Then, at iteration $k$, instead of passing the solver output $\Tilde{\pmb{x}}^k$ directly, a QN relaxation is computed as 
\begin{equation}
    \pmb{x}^{k} = \pmb{x}^{k-1} - \pmb{J}^{-1} \pmb{r}^k
\end{equation}
where
\begin{equation}
    \pmb{r}^k = \Tilde{\pmb{x}}^k - \pmb{x}^k .
\end{equation}
In this class of acceleration methods, the Jacobian (or inverse Jacobian) needed for the QN algorithm is approximated non-intrusively using snapshots of the past iterations. This generally gives faster convergence rate than Aitken relaxation~\cite{degroote_performance_2009, uekermann_partitioned_2016, broyden}. Several acceleration methods have been developed, each differing in how the Jacobian is approximated, which unknown to consider (For example, $\pmb{x} = \pmb{u}$, $\pmb{x} = \pmb{f}$ or $\pmb{x} = \left(\begin{array}{c}
     \pmb{u} \\
    \pmb{f}
  \end{array}\right)$ in the case of block formulation (\ref{block})) or how the past iteration information is used. In the test cases we show, we used the Quasi-Newton Inverse Least-Squares (IQN-ILS) method, introduced first in \cite{degroote_performance_2009}. The original IQN-ILS algorithm \ref{alg:iqnils} is recalled in the appendix.

\subsection{\textbf{Predictors}}
Another component of iterative FSI schemes is the predictor used to "kick-start" the next time step. Because of the time delay between the two solvers, there is no available "updated" solution for the solid solver to use in the first iteration. Usually, the last converged solution of the previous time step is used as a first guess
\begin{equation}\label{Const-Extrap}
    \pmb{f}^{0, n} = \pmb{f}^{n-1}
\end{equation}
where the first superscript refers to the iteration number, and the second to the time step index for the current time step, and one superscript for the previous time steps indicates the converged solution. This notation will be used in the rest of the paper.

A linear extrapolation from the previous time steps such as 
\begin{equation}\label{Lin-Extrap}
    \pmb{f}^{0, n} = 2\pmb{f}^{n-1} - \pmb{f}^{n-2}
\end{equation}
can also be used, or, alternatively, a quadratic variation:
\begin{equation}\label{Quad-Extrap}
    \pmb{f}^{0, n} = 3\pmb{f}^{n-1} - 3\pmb{f}^{n-2} + \pmb{f}^{n-3}
\end{equation}
 but whether this gives better convergence highly depends on each FSI problem and can result in bad performance if the time step is not sufficiently small. In this work, we propose an alternative approach to construct this predictor, as explained in the next section.

\section{Non-intrusive ROM-FOM fluid-structure interaction acceleration}\label{sect:romfom}

In this work, the strategy of data-driven acceleration of FSI problems consists in a combination of two approaches:

\setlength\parindent{24pt}
$1.$ A data-driven ROM of the solid subproblem similar to the one proposed in \cite{tiba2023nonintrusive}. 

\setlength\parindent{24pt}
$2.$ An adaptive data-driven force predictor that acts as an efficient generator of a good initial guess of the fixed point FSI problem at each time step, enabling an easier convergence and thus faster overall computations.

Note that these two components, although sharing the same methods internally, are independent and can be used separately from one another. In particular, the force predictor can be used in more general cases, and can achieve good computational economies with practically no loss of accuracy, since the final solution depends on the state at the convergence, and rarely on the initial guess. On the other hand, a ROM for the solid part can achieve a speedup up of orders of magnitudes on the solid subproblem, with a small loss of accuracy, and this can be particularly useful if the computation time is predominantly due to the solid solver. In addition, using the two components enable the accumulation of the speedups as well as preventing the slowdown of the convergence rate due to the addition of the solid ROM.

\subsection{\textbf{Solid reduced-order model}}\label{solidROM}
A data-driven reduced order model for the solid behavior is constructed following the approach presented in \cite{tiba2023nonintrusive}. The ROM approximates the force to displacement function 
\begin{equation}\label{solid_ROMoperator}
  \Hat{\mathcal{S}}: \mathbb{R}^N \rightarrow \mathbb{R}^{N_S}\;;\; \pmb{f}_{|\Gamma_{fsi}} \rightarrow {\pmb{d}}
\end{equation}
where $\pmb{d}$ is the approximated displacement field of the solid domain and $N_S$ is the dimension of the discretized solid domain. The approximation of the solid operator output is then :
\begin{equation}
\pmb u = {\mathcal{S}}(\pmb{f}) \approx \pmb d_{|\Gamma_{fsi}} = \Hat{\mathcal{S}}(\pmb{f})_{|\Gamma_{fsi}}. 
\end{equation} Hence, the solid ROM can predict the displacement solution in online computations with a much reduced cost and in a non intrusive manner. The ROM has three main components: the \textit{encoder}, the \textit{regressor} and the \textit{decoder}.
\subsubsection{Forces encoder $\mathcal{E_F(\cdot)}$}
At the input of the ROM, an encoder reduces the dimensionality of the forces field at the interface. This is done using the popular POD method \cite{sirovich_turbulence_1987}, hereby looking for a low dimensional linear subspace where the force field is projected: 
\begin{equation}
     \pmb{f}(\pmb{X}, t, \pmb{\mu}) = \sum_i^{r_f} \pmb{\Phi}_{fi}(\pmb{X}) {\pmb{f}}_{r,i}(t, \pmb{\mu})
\end{equation}
where $\pmb{\mu}$ is a parameter of the FSI problem simulated, which, in this work, is associated to the fluid subproblem alone. The rank $r_f$ is the dimension of the POD subspace, ideally very small $r_f << N$, and $\pmb{\Phi}_{fi}$ are the POD modes. Written in the discretized form, the force field $\pmb{f} \in \mathbb{R}^{N}$ can be represented as:
 \begin{equation}
    \pmb{f} = \pmb{\Phi}_f {\pmb{f}}_r
 \end{equation}
 where $\pmb{\Phi}_{f} \in \mathbb{R}^{N \times r_f}$ the matrix of POD modes and ${\pmb{f}}_r \in \mathbb{R}^{r_f}$ is the coordinates vector of the force in the reduced POD subspace.
 
 In the offline phase, the POD modes are computed using Singular Value Decomposition (SVD) of snapshot matrix of HF results data $ \pmb{F}  \in \mathbb R ^{N \times m}$ where m is the total number of available snapshots. 
  \begin{equation}\label{svdF}
     \pmb{F} - \Bar{\pmb{F}}= \pmb{\Phi}_{f} \pmb \Sigma_f \pmb{V}^*_f
 \end{equation}
 where $\Bar{\pmb{F}}$ is the temporal mean of the fluid forces, $\Sigma_f$ is the singular values diagonal matrix, and $\pmb{V}^*_f$ is the conjugate transpose of the POD time coefficients.
 In fact, in this work, simulation results are collected from different values of $\pmb{\mu} \in \mathcal{P}$ of cardinal $n_\mu$ and from all the simulated $n_t$ timesteps corresponding to $t \in [0., T]$, including the non converged global iterations $n_{k, i} \; \forall i \in \{1, \dots n_t n_\mu\}$., giving 
 \begin{equation}\label{num_snaps}
    m = \sum_i^{ n_t n_\mu} n_{k, i}.
 \end{equation}

Accordingly, and since the POD modes are orthogonal, the encoder acts as the dimensionality reducer of the force field quantity as :
 \begin{equation}
     \mathcal{E}_F(\pmb{f}): \mathbb{R}^{N} \rightarrow \mathbb{R}^{r_f}\;;\; {\pmb{f}}_r =  \pmb{\Phi}_f^T  (\pmb{f} -  \Bar{\pmb{F}}).
 \end{equation}

\subsubsection{Regressor $\mathcal{I_S(\cdot)}$} The regressor approximates the relationship between the two low dimensional representations of the forces and the displacements 
\begin{equation}
    \mathcal{I}_S: \mathbb{R}^{r_f} \rightarrow \mathbb{R}^{r_u}\;;\; {\pmb{f}}_r  \rightarrow  {\pmb{u}}_r.
\end{equation}
Different existing methods can accomplish this task.
In our experiments, the regression methods that provided the best accuracy are Reduced Basis Function (RBF) interpolation \cite{wahba_spline_1990, audouzeRBF} and second-order polynomial sparse approximation. In the former, the function is modeled as 
\begin{equation}\label{regr_rbf}
    \mathcal{I}_S({\pmb{f}}_r) = \sum_i^m \pmb{w_i}\, \phi(||{\pmb{f}_r} - {\pmb{f}}_{r,i}||) + \pmb{P}({\pmb{f}}_r)
\end{equation}
where $\phi( \cdot )$ is a kernel function, $\pmb{P}$ is the $1^{st}$ order polynomial and ${\pmb{f}}_{r,i}$ are the RBF centers, chosen as the training points of the reduced forces, resulting eventually in a linear system to be solved for the RBF weights $w_i$. Alternatively, a polynomial regression of order 2 is used. The Force-Displacement relationship is thus modeled as a second order polynomial, here written in a discretized form:
\begin{equation}\label{regr_poly}
    \mathcal{I}_S({\pmb{f}_r}) = \pmb{W} [{\pmb{f}_r} {\otimes}  {\pmb{f}_r}]
\end{equation}
where $\otimes$ is the Kronecker product. The polynomial coefficients are arranged in $\pmb{W} \in \mathbb{R}^{r_u \times \hat{r}_f }$, where $\hat{r}_f = (r_f+1)(r_f+2)/2$.
The number of polynomial coefficients can be very large if $r_f = \mathcal{O}(10)$, and it is highly unlikely that all the polynomial terms are important for modeling $\mathcal{I(\cdot)}$. We thus propose using the Lasso regularization \cite{LassoRef} in order to obtain a sparse model with as fewest terms as possible, the minimization is written as:
\begin{equation}\label{lasso}
    \pmb{W}_i = arg \min_{{W}_i} ||{\pmb{u}}_{r,i} - \sum_{j=1}^{\hat{r}_f} {W}_{ij} [{\pmb{f}_r} \otimes  {\pmb{f}_r}]_j ||_2^2  + \lambda \sum_{j=1}^{\hat{r}_f} |{W}_{ij}| \qquad
\forall \ i \in \{1 \cdots r_u\}.
\end{equation} \medskip
The parameter $\lambda$ promotes the sparsity of the solution $\pmb{W}$ and usually requires fine-tuning. In this work, we find the solution of (\ref{lasso}) using the LARS algorithm presented in \cite{LASSO} and implemented in \texttt{scikit-learn} \cite{sktlrn}. In order to avoid additional bias from the training data points distribution, a standardization should be done to the input before the inference.
As a general rule of thumb, we recommend using (\ref{regr_poly}) when a large amount of data is available (compared to the number of $2^{nd}$ order polynomial terms) and (\ref{regr_rbf}) otherwise.

\subsubsection{Displacement decoder $\mathcal{D_S(\cdot)}$} 
We use quadratic manifolds as nonlinear decoders for the reconstruction of the displacement field from obtained points in the latent space. Quadratic manifolds model the POD reconstruction error using the $2^{nd}$ order polynomial terms of the reduced coordinates in the POD subspace, and associating them with a mapping operator $\thickbar{\pmb{\Phi}}_{U}$ \cite{geelen_operator_2023} :
\begin{equation}
     \mathcal{D}_S({\pmb{u}_r}): \mathbb{R}^{r_u} \rightarrow \mathbb{R}^{N_S}\;;\; \pmb{d} =  \Bar{\pmb{U}} + \pmb{\Phi}_U  {\pmb{u}_r}  + \sum_j^{r_u (r_u+1)/2} \thickbar{\pmb{\Phi}}_{U} ({\pmb{u}_r}  \otimes {\pmb{u}_r})_j.
 \end{equation}
 Obtaining the POD modes $\pmb{\Phi}_{U}\in \mathbb R ^{N_S \times r_u}$ in the offline phase is similar to (\ref{svdF}) :
  \begin{equation}\label{svdU}
     \pmb{U} - \Bar{\pmb{U}} = \pmb{\Phi}_{U} \pmb \Sigma_U \pmb{V}^*_U
 \end{equation}
and a least-squares problem is solved for the quadratic operator $\thickbar{\pmb{\Phi}}_{U}$ :
\begin{equation}\label{vbar}
    \thickbar{\pmb{\Phi}}_{U} =  arg \min_{\thickbar{\pmb{\Phi}}  \in \mathbb{R}^{N_S \times r_u (r_u+1)/2} } \frac{1}{2} ||(\pmb{I} - \pmb{\Phi}_U \pmb{\Phi}_U^T) (\pmb{U} - \Bar{\pmb{U}})  -  \thickbar{\pmb{\Phi}} [{\pmb{u}_r}  \otimes {\pmb{u}_r}]   ||^2_F.
\end{equation}

This generally enables a greater reconstruction accuracy for a practically negligible added cost (see \cite{geelen_operator_2023, tiba2023nonintrusive} for more details).
 Regarding the choice of the latent spaces dimensions, $r_u$ and $r_f$: for $r_u$, we use the energy criterion to select the number of modes retaining $\varepsilon = 99.99\%$ of the energy as:
\begin{equation}\label{energy_sing}
\begin{aligned}
arg \min_{r_u\in[1, min(N_S, m)]} \quad & S = \frac{\sum_i^{r_u} \sigma_i^2}{\sum_i^{min(N_S, m)} \sigma_i^2}\\
\textrm{s.t.} \quad &  S \leq \varepsilon .\\
\end{aligned}
\end{equation}
As for $r_f$, we use a cross-validation strategy on a small percentage of the available data as test data, until a plateau of the validation error is attained (see \cite{tiba2023nonintrusive} for more details about this cross-validation strategy).

\subsubsection{In a nutshell:} 
The solid ROM can finally be symbolically represented as : 
\begin{equation}
\Hat{\mathcal{S}}(\cdot) = \mathcal{D}_S\circ \mathcal{I}_S \circ\mathcal{E}_F(\cdot)
\end{equation}
predicting the displacement field for a given applied force field on the interface:
\begin{equation}
\pmb{d}(\pmb{f}) = \Hat{\mathcal{S}}(\pmb{f}) = \mathcal{D}_S(\mathcal{I}_S(\mathcal{E}_F(\pmb{f}))).
\end{equation}
Note that, for predicting the displacement at the interface only, which is the only necessary information at each global FSI iteration, a simple row selection of the POD modes matrix is used in the decoding phase, giving a new decoding operator $\mathcal{D}_{S, \Gamma}$
  \begin{equation}
     \mathcal{D}_{S, \Gamma}({\pmb{u}_r}): \mathbb{R}^{r_u} \rightarrow \mathbb{R}^{N}\;;\; \pmb{u} = \pmb{d}_{\Gamma_{FSI}} = \Bar{\pmb{u}} + \pmb{\Phi}_u  {\pmb{u}_r}
 \end{equation}
where $\pmb{\Phi}_u$ is the matrix of POD modes after removing the rows corresponding to the degrees of freedom (dofs) outside of the interface $\Gamma_{FSI}$ from $\pmb{\Phi}_U$. We accordingly define 
\begin{equation}\label{solid_local_operator}
  \Hat{\mathcal{S}}_\Gamma: \mathbb{R}^N \rightarrow \mathbb{R}^{N}\;;\; \pmb{f} \rightarrow {\pmb{u}} = \mathcal{D}_{S, \Gamma}(\mathcal{I}_S(\mathcal{E}_F(\pmb{f}))).
\end{equation}

Some important remarks are to be made here :
\begin{itemize}
    \item Since only the displacement at the interface is needed at each iteration, only the local version of the ROM $\Hat{\mathcal{S}}_{\Gamma}$ is used at each iteration to pass the displacement to the fluid solver. The reduced coordinates ${\pmb u}_r$ of the converged iteration are stored at each time step (in a small $\pmb{U}^n_r \in \mathbb R^{r_u \times n_t}$ matrix), so that a reconstruction of the full displacement field in all the time steps can be made in parallel when needed, enabling then a computation of the stress and strain fields for example.
    \item If the solid ROM is constructed to take as input the fluid forces on the fluid grid directly, the ROM prediction can be made without the need of mapping the forces, simplifying further the coupling procedure and making a slight additional computational gain.
    \item In the context of mixed formulations of the solid problem, where other degrees of freedom are present (for example, pressure dofs in incompressible solid problems, or rotation dofs in shell elements), additional ROMs should be constructed for these unknowns. These fields are however not needed to pass the interface displacement to $\mathcal{F}$ and the reduced force coordinates can be stored at each converged iteration to compute the full solution when needed (for example at the end of the simulation). An example of this will be demonstrated in test case n°3.
\end{itemize}

\subsection{\textbf{Adaptive data-driven predictors with fluid and solid ROMs}}
In this work, we propose to construct a data-driven predictor, based on the use of information from past data, for example past iterations/time steps, or historical simulation results obtained at different parameters values, thus providing a better initial guess than a finite-differences-based extrapolation. In addition to the solid ROM described in \ref{solidROM}, another surrogate is constructed to approximate the inverse of $\Hat{\mathcal{S}}$ at the interface, \textit{i.e} approximating the $\pmb{u}$ to $\pmb{f}$ relationship. Note that, contrary to the solid ROM presented above, this fluid surrogate should take into account the dynamical effects 
\begin{equation}
    \hat{\mathcal{F}}(\pmb{u}, t) = \pmb{f} .
\end{equation}
This is done using a discrete approach, where the input of the surrogate contains not only the current displacement but also the forces at the previous time step
\begin{equation}
    \hat{\mathcal{F}}(\pmb{u}^{k, n+1}, \pmb{f}^{k, n}) = \pmb{f}^{k, n+1} .
\end{equation}
This is in principle similar to the Dynamic Mode Decomposition with Control (DMDc) \cite{proctorDMDc} concept, the only difference here being the consideration of the system nonlinearity through the nonlinear regressor and/or the update strategy adopted.

\subsubsection{Prediction of a better initial guess: Inner coupling}
A "reduced FSI coupling" is launched at the start of each time step, where each subsystem $\mathcal{F}$ and $\mathcal{S}$ is replaced by its reduced equivalent $\hat{\mathcal{F}}$ and $\hat{\mathcal{S}}$, and the solution
\begin{equation}
    \hat{\mathcal{F}}(\hat{\mathcal{S}}(\pmb{f}^0)) = \pmb{f}^0
\end{equation} 
is sought at a fraction of the computational time needed for the FOM-FOM problem solver calls. Those iterations will be called henceforth \textit{local iterations}. In \cite{DELAISSE2022106720}, reduced physics-based models were used as surrogates to compute the local iterations, specifically to enhance the inverse Jacobian approximation in IQN-ILS, while the new initial guess did not particularly produce less overall computational time.

We note that using our proposed data-driven models, and constructing them so that they consider inputs and outputs on the same grid (the fluid grid in our case), allows to bypass the need for grid mapping. In addition, we propose an adaptation strategy to update the prediction capability of the fluid surrogate, especially since new HF data from the fluid FOM solver is available at the online computations. We note that the tolerance required for the convergence of these local iterations $\delta_r$ should not be very small, since the goal here is merely to obtain a closer initial guess than the previous time step solution, and to avoid a large number of iterations that could slow down computations. In the same direction, we propose that these local iterations use a simple Aitken relaxation (\ref{aitkenR}), since using QN-acceleration would add a computational cost to find and store an additional inverse Jacobian approximate, while the tolerance required is already not very small. 
Furthermore, if the convergence is not reached, the predictor should be deactivated, and the previous time step solution should be used instead, since poor convergence could suggest that the dynamical nonlinearity is not captured by the two ROMs involved, and thus their predictions may be too inaccurate and lead to poor global convergence. In Algorithm \ref{local}, we detail the local iterations' procedure of the reduced models. 

\definecolor{mynicegreen}{RGB}{48,168,48}
\begin{algorithm}
\caption{Local ROM-ROM iterations for finding an initial guess - $Pred(\cdot)$}\label{local}
\SetKwInput{KwData}{Input}
\KwData{
Fluid ROM $\Hat{\mathcal{F}}$

Solid ROM $\Hat{\mathcal{S}}$

Initial relaxation $w_0$

Convergence tolerance $\delta_r = \mathcal{O}(10^{-2})$

Maximum number of iterations $M$

Previous time step fluid forces $\pmb{f}^{n-1}$,  $\pmb{f}^{n-2}$, $\pmb{f}^{n-3}$
}
\KwResult{Next initial guess $\pmb{f}^{0, n}$}
\vline

\nl $\pmb{f}^{0, n} = 2\pmb{f}^{n-1} - \pmb{f}^{n-2}$ or $\pmb{f}^{0, n} = 3\pmb{f}^{n-1} - 3\pmb{f}^{n-2} + \pmb{f}^{n-3}$\label{firstPred_local}
    
\nl $j = 1$

\nl $e = 1$

  \While{$e > \delta$ \textit{and} $k<M$}{
    \nl $\pmb u^j_r = \Hat{\mathcal{S}}(\pmb{f}^{j-1})$

    \nl $\Tilde{\pmb f}^j = \Hat{\mathcal{F}}(\pmb{u}^j, \pmb f^{n-1})$

    \nl $\pmb r^j = \Tilde{\pmb f}^j - \pmb f^{j-1}$

    \nl $e = ||\pmb r^j||_2/||\Tilde{\pmb f^j}||_2$

    \uIf{$e > \delta_r$}{
    \nl $w^j = - w^{j-1} \frac{\pmb{r}^{j-1\;T}(\pmb{r}^{j}-\pmb{r}^{j-1})}{||\pmb{r}^{j}-\pmb{r}^{j-1}||_2^2}$

    \nl ${\pmb f^j} = w^j \Tilde{\pmb f}^j + (1-w^j) \pmb f^{j-1}$\label{relaxed_local}
  }\Else{
  \nl $\pmb{f}^{0, n} = \Tilde{\pmb f^j}$

  \nl \textbf{End algorithm}
  }
  \nl $j = j+1$
    }
\nl $\pmb{f}^{0, n} = \pmb f^{n-1}$ \textcolor{mynicegreen}{// \textit{Failure of convergence}}
\end{algorithm}

In the next part, the components of $\hat{\mathcal{F}}$ will be detailed.
\subsubsection{Fluid ROM components:} 
\begin{figure}
   \makebox[\textwidth][c]{\includegraphics[width=.65\textwidth]{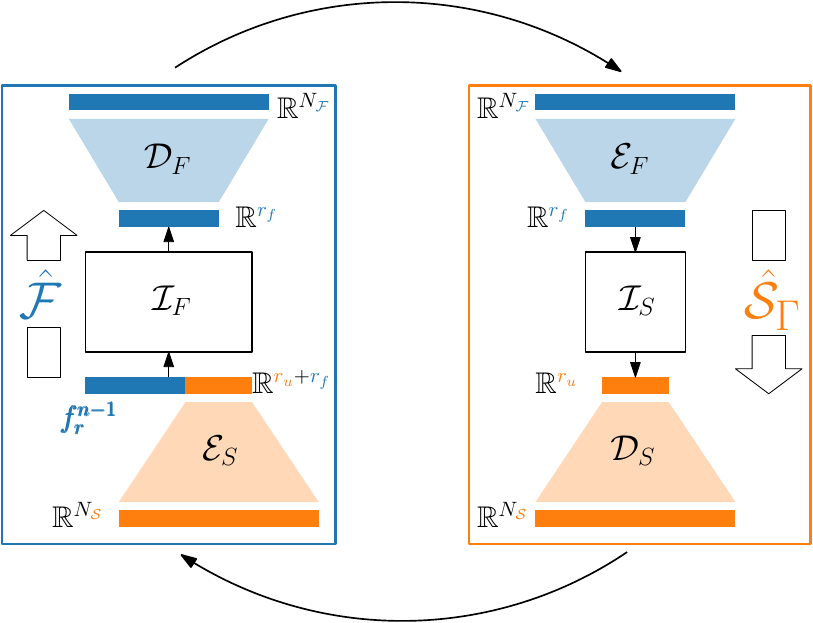}}
     \caption{Fluid and solid ROM components}
     \label{fig:ROM-guts}
 \end{figure}

\begin{figure}
   \makebox[\textwidth][c]{\includegraphics[width=.65\textwidth]{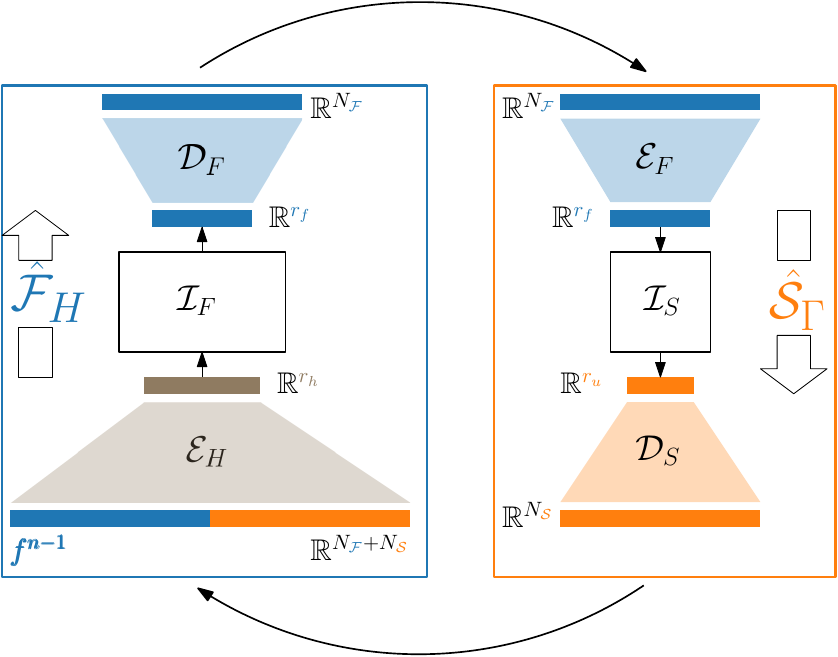}}
     \caption{Fluid and solid ROM components - Hybrid variant}
     \label{fig:ROM-guts-2}
 \end{figure}

The fluid ROM $\Hat{\mathcal{F}}$ is constructed with the same philosophy as its solid counterpart $\Hat{\mathcal{S}}_{\Gamma}$, in the sense of finding a relationship in a latent space. This can be done using two approaches: using separate encoders for the displacement and forces and the reduced representation is composed of the two encoded variables:
\begin{equation}\label{fluid_operator}
  \Hat{\mathcal{F}}: \mathbb{R}^{2 N} \rightarrow \mathbb{R}^{N}\;;\; (\pmb{u}, \pmb{f}^{n-1}) \rightarrow {\pmb{f}} = \mathcal{D}_F(\mathcal{I}_F([\mathcal{E}_S(\pmb{u}), \mathcal{E}_F(\pmb{f}^{n-1})]^T)) .
\end{equation}
The second approach, denoted as $\Hat{\mathcal{F}}_H$ consists of constructing a new unknown from the concatenation of the 2 variables in the high-dimensional space, with a new \textit{hybrid} encoder $\mathcal{E}_H(\cdot)$:
\begin{equation}\label{fluid_operator_hybrid}
  \Hat{\mathcal{F}}_H: \mathbb{R}^{2 N} \rightarrow \mathbb{R}^{N}\;;\; (\pmb{u}, \pmb{f}^{n-1}) \rightarrow {\pmb{f}} = \mathcal{D}_F(\mathcal{I}_F(\mathcal{E}_H(\pmb{u}, \pmb{f}^{n-1}))) .
\end{equation}
The two ROMs' components are summarized in the illustrative Figures \ref{fig:ROM-guts} and \ref{fig:ROM-guts-2}. 
 
The fluid ROM $\Hat{\mathcal{F}}$ differs from $\Hat{\mathcal{S}}_{\Gamma}$ in that: 
\begin{itemize}
    \item The decoder $\mathcal{D}_F(\cdot)$ and encoder  $\mathcal{E}_S(\cdot)$ are the equivalent of the inverses of their encoder and decoder counterparts $\mathcal{E}_F(\cdot)$ and $\mathcal{D}_{S, \Gamma}(\cdot)$, respectively. They are then defined as : 
  \begin{equation}
     \mathcal{D}_F({\pmb{f}_r}): \mathbb{R}^{r_f} \rightarrow \mathbb{R}^{N}\;;\; \pmb{f} =  \Bar{\pmb{F}} + \pmb{\Phi}_f  {\pmb{f}_r}
 \end{equation}
  \begin{equation}
     \mathcal{E}_S(\pmb{u}): \mathbb{R}^{N} \rightarrow \mathbb{R}^{r_u}\;;\; {\pmb{u}_r} = \pmb{\Phi}_u^T ( {\pmb{u}} - \Bar{\pmb{U}}) .
 \end{equation}

This means that in the case of $\Hat{\mathcal{F}}$, no additional training cost is spent for obtaining $\mathcal{D}_F(\cdot)$ and $\mathcal{E}_S(\cdot)$, since the modes are already computed during the training of $\Hat{\mathcal{S}}_{\Gamma}$. As for $\mathcal{E}_H(\cdot)$, this is also done using the POD but on the new variable $(\pmb{u}, \pmb{f}^{n-1})$
 \begin{equation}
     \mathcal{E}_H(\pmb{u}, \pmb{f}^{n-1}): \mathbb{R}^{2N} \rightarrow \mathbb{R}^{r_h}\;;\; {\pmb{h}}_r =  \pmb{\Phi}_H^T  ([\pmb{u}, \pmb{f}^{n-1}]^T -  [\Bar{\pmb{U}}, \Bar{\pmb{F}}]^T) .
 \end{equation}
 
    \item A new regressor $\mathcal{I}_F$ is used that takes into account the dynamical effect using the previous timestep force $\pmb{f}^{n-1}$ in the augmented regression input
    \begin{equation}\label{fluidRegressor}
        \mathcal{I}_F: \mathbb{R}^{r_f+r_u} \rightarrow \mathbb{R}^{r_f}\;;\; [{\pmb{u}}_r^{k} ,{\pmb{f}_r}^{n-1}]^T  \rightarrow  {\pmb{f}_r} .
    \end{equation}
While different regression forms can be used, we use here linear regression or the RBF regressor (\ref{regr_rbf}), which in our view, represent a good trade-off option between fitting efficiency and training efficiency, which is crucial for our method since an online update strategy is adopted (more details will be given in Section \ref{updateParagraph}).
\item Note that in order to obtain accurate evaluations of $\Hat{\mathcal{F}}$ and thus reliable predictors, the direct output of the HF fluid solver ${\mathcal{F}}$ should be "\textit{seen}" during training. We recall that in the available HF data, we distinguish between the forces computed from the fluid solver $\pmb{f}^k$, and the QN-updated forces $\Tilde{\pmb{f}^k}$, arranged in a new snapshot matrix that we call $\pmb{\Tilde{F}}$. Although, for obvious reasons, the new regressor $\mathcal{I}_F$ should necessarily use $\Tilde{\pmb{F}}$ (or rather, its reduced coordinates $\pmb{\Phi_f}^T \Tilde{\pmb{F}}$). Only a combination of $\pmb{F}$ and $\Tilde{\pmb{F}}$ suffices to compute the POD modes $\pmb{\Phi_f}$. In our experiments, we used the concatenation of all the snapshots $[\pmb{F}, \Tilde{\pmb{F}}]$, but a more efficient choice could be for example to only include the first iteration results in $\Tilde{\pmb{F}}$ since the other iterations' resultant forces --closer to FSI convergence-- would be very close to the updated forces, and would only add little information for computing the POD subspace.
\end{itemize}

In our numerical experiments, the two approaches $\Hat{\mathcal{F}}$ and $\Hat{\mathcal{F}}_H$ yield nearly identical results, we then proceed with detailing only the first approach $\Hat{\mathcal{F}}$ since its associated offline step is computationally more efficient (dropping the need of learning $\mathcal{E}_H$).

\textbf{Remark:} We should note here that, with our choices of the dimensionality reduction methods, the evaluation of $\mathcal{D}_S(\cdot)$ and $\mathcal{E}_S(\cdot)$ can be dropped at the local iterations (Algorithm \ref{local}) during the inference of the two ROMs $\Hat{\mathcal{F}}$ and $\Hat{\mathcal{S}}$, since, in our case, using POD as our encoder-decoder implies that $\mathcal{E}_S(\mathcal{D}_S(\pmb{u}_r)) = \pmb{u}_r$. However, the full force field must be recovered at each iteration (of the reduced fixed-point problem) because the relaxation used in line \ref{relaxed_local}, together with an initial guess from the fluid solver (line \ref{firstPred_local}) means that $\pmb{f}^k$ contains a part the falls outside the POD subspace $\pmb{\Phi}_f$. Thus, the convergence should be verified at the high-dimensional space rather than at the latent space. Furthermore, as stated earlier, and as shown in \cite{tiba2023nonintrusive}, the force field is not easily compressed using POD, and the ignored modes may be necessary as in the contribution to the force values used when checking the convergence.

The offline strategy of the solid and fluid ROMs are outlined in Algorithm \ref{alg:offline}.

\subsubsection{Fluid ROM online update:}\label{updateParagraph}
At the end of the offline phase, the full snapshot matrices are no longer of use, since their reduced representation is now learned, and thus can be freed from memory. However, we store a chunk of size $p$ from the reduced coordinates of the training input and output data of $\Hat{\mathcal{F}}$: ${\pmb{U}_r} \in \mathbb R^{r_u \times p}$ and ${\Tilde{\pmb{F}_r}} \in \mathbb R^{r_f \times p}$ as:
\begin{equation}
    \pmb{U}_r =   \left[ \begin{array}{cccc}
        \vertbar & \vertbar &        & \vertbar \\
         {\pmb{u}}_{r}^1    & {\pmb{u}}_{r}^2   & \ldots & {\pmb{u}}_{r}^p    \\
        \vertbar & \vertbar &        & \vertbar 
      \end{array}     \right]
\, \, \, and \, \, \, 
    \Tilde{\pmb{F}}_r =   \left[ \begin{array}{cccc}
        \vertbar & \vertbar &        & \vertbar \\
         \Tilde{\pmb{f}}_{r}^1    & \Tilde{\pmb{f}}_{r}^2   & \ldots & \Tilde{\pmb{f}}_{r}^p    \\
        \vertbar & \vertbar &        & \vertbar 
      \end{array}     \right] .
\end{equation}
Theses reduced representations of data points will be updated in the online computations by replacing the data points with the least relevant information by new snapshot couples of $[{\pmb{u}}_r^k \; | \; \Tilde{\pmb{f}}_r^k]$ available at each iteration :
\begin{equation}\label{updating}
    \pmb{U}_r \leftarrow{}   \left[ \begin{array}{ccccc}
        \vertbar & \vertbar &        & \vertbar & \vertbar \\
         {\pmb{u}}_{r}^2    & {\pmb{u}}_{r}^3   & \ldots & {\pmb{u}}_{r}^p & {\pmb{u}}_{r}^k    \\
        \vertbar & \vertbar &        & \vertbar & \vertbar
      \end{array}     \right]
\, \, \, and \, \, \, 
    \Tilde{\pmb{F}}_r \leftarrow   \left[ \begin{array}{ccccc}
        \vertbar & \vertbar &        & \vertbar & \vertbar \\
         \Tilde{\pmb{f}}_{r}^2    & \Tilde{\pmb{f}}_{r}^3   & \ldots & \Tilde{\pmb{f}}_{r}^p & \Tilde{\pmb{f}}_{r}^k    \\
        \vertbar & \vertbar &        & \vertbar & \vertbar
      \end{array}     \right] .
\end{equation}
These new (reduced) snapshots represent the high fidelity information stemming from the HF fluid solver operations. It can thus be used to enrich the fluid ROM. Specifically, after a certain advancement along the transient simulation, defined for example by a number $Z$ (chosen by the user) of global FSI iterations, these now updated matrices ${\pmb{U}_r}$ and ${\Tilde{\pmb{F}}_r}$ can be used to retrain the fluid regressor (\ref{fluidRegressor}). By doing so, we ensure that the fluid ROM $\hat{\mathcal{F}}$ maintains a high enough fidelity so that the provided initial guess does indeed help the FSI converge faster, especially since the newest information, from the latest time steps and iterations will be used.

Note that the matrices ${\pmb{U}_r}$ and ${\Tilde{\pmb{F}}_r}$ are of small sizes since $r_u << N$ and $r_f << N$. In addition, the "width" of these matrices can be kept at its maximum $p$, also defined by the user, keeping the memory cost small and constant. As for the CPU cost, retraining the regressor (\ref{regr_rbf}) or linear regression, i.e solving a linear system of size $(p+r_u+r_f) \times (p+r_u+r_f)$ or a $p \times (1+r_u+r_f)$ least-square system, and only after (a large) number of iterations $Z$, represents only a small fraction of the FSI timestep solution.

It should be emphasized that this online update strategy is limited to the update of the regressor for simplicity. An extension of this approach could be done by updating the encoder-decoder as well. For example, a recursive method like the one presented in \cite{taleb} could be used to update the POD bases using data obtained on the fly.

\subsection{\textbf{Summary}}
\begin{figure}
   \makebox[\textwidth][c]{\includegraphics[width=1.2\textwidth]{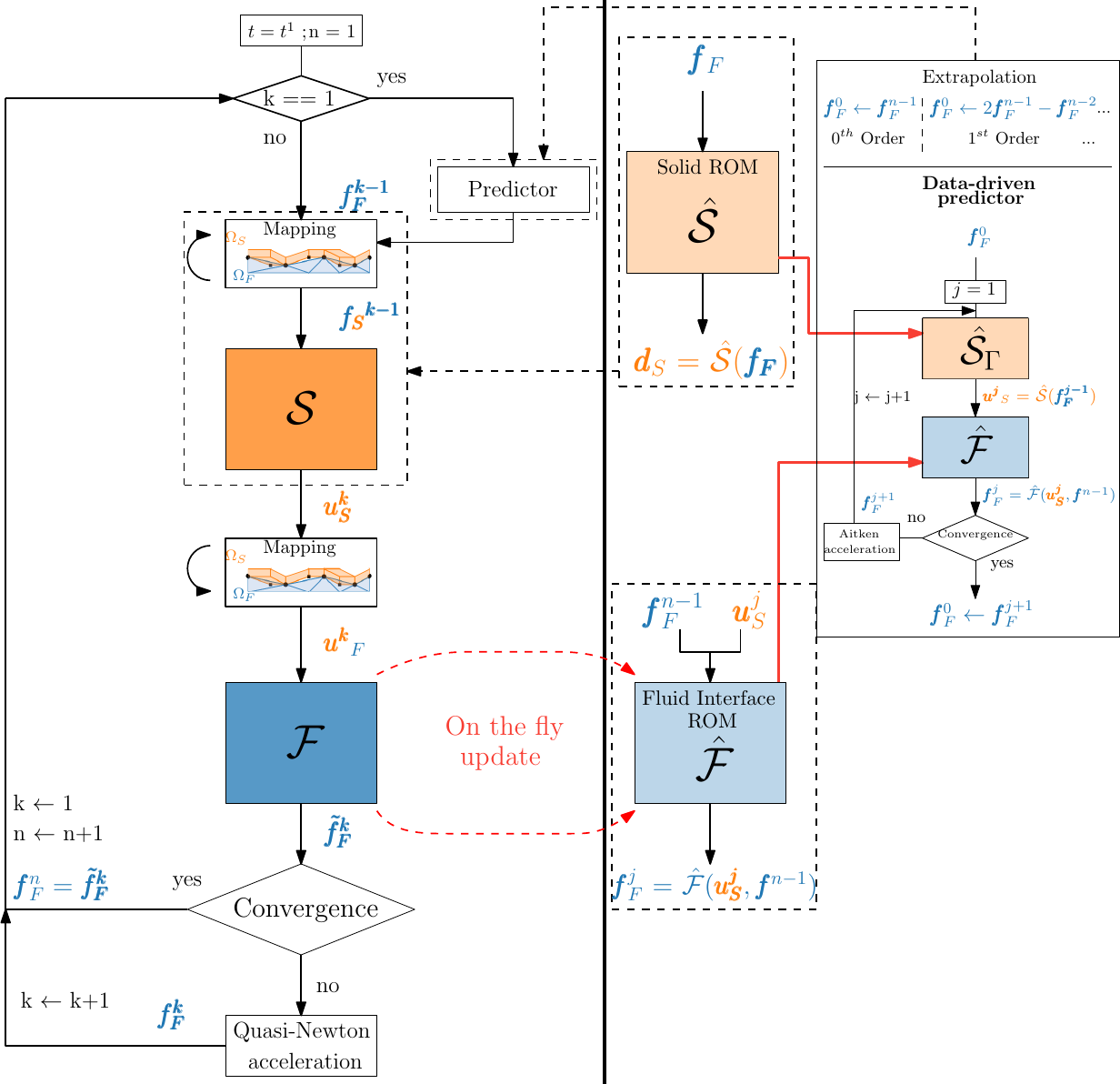}}
     \caption{Global FSI coupling scheme with the ROM-FOM coupling and data-driven adaptive predictors.}
     \label{fig:global-alg}
 \end{figure}

To summarize, the proposed acceleration method can be easily implemented in a non-intrusive manner. Using HF data obtained with varying fluid parameter simulations, two separate ROMs can be trained efficiently and then used in a new simulation with unseen values of the parameters. This added block in the FSI scheme provides an initial force field that will "jump-start" the coupling at each time step resulting in faster convergence and faster overall computation. During the new unseen simulation, the fluid ROM can be adapted using the online results coming from the FOM fluid solver, constituting a phsyics-aware predictor that can be used effectively even in extrapolative regions. The global FSI iterations with the new prediction approach is outlined in Algorithm \ref{globalAlg}. We stress once again the nonintrusivity of this approach, since calling external solvers as black-boxes as done in lines \ref{callingFluid} and \ref{callingSolid} is completely possible. Moreover, additional calculations can be done in line \ref{callingPostProcess} using the imposed (already computed) displacements to compute for example the stress and strain fields. A sketch of the global methodology is shown in Figure \ref{fig:global-alg} for illustration.

\begin{algorithm}
\caption{ROM-FOM - Global FSI scheme}\label{globalAlg}

\nl $n = 1$

\nl $z = 1$

\While{$n < N_t$}{

\nl $k = 1$

\nl $e = 1$

\nl Predictor : $\pmb{f}^{0, n} = Pred(\pmb f^{n-1}, \pmb f^{n-2})$ using Algorithm \ref{local}

  \While{$e > \delta$ \textit{and} $k<N_k$}{

    \nl Solid solver : $\begin{cases}
	  \pmb{f}^{k-1}_{|\mathcal{S}}  = \mathcal{M}_{\mathcal{F} \rightarrow \mathcal{S}}(\pmb{f}^{k-1}_{|\mathcal{F}}) \\
	  \Tilde {\pmb u}^k = \mathcal{S}(\pmb{f}_{|\mathcal{S}}^{k-1}) \\
     \end{cases}$ or $\Tilde{\pmb u}^k = \Hat{\mathcal{S}}_{\Gamma}(\pmb{f}_{|\mathcal{F}}^{k-1})$\label{callingSolid}

    \uIf{Block Formulation (\ref{block}) is used}{
    \nl $\Delta \pmb{u}^k = Acc(\Tilde{\pmb u}^{k-1}, \pmb r^k)$ using block QN acceleration, e.g \cite{bogaers_quasi-newton_2014}

    \nl ${\pmb u^k} = {\pmb u}^k + \Delta \pmb{u}^k$
    }\Else{\nl $\pmb u^k = \Tilde{\pmb{u}}^k$}

    \nl Displacement mapping : $\pmb{u}^k_{|\mathcal{F}}  = \mathcal{M}_{\mathcal{S} \rightarrow \mathcal{F}}(\pmb{u}^{k}_{|\mathcal{S}})$

    \nl Fluid solver : $\Tilde{\pmb f}^k = \Hat{\mathcal{F}}(\pmb{u}^k_{|\mathcal{F}}, t)$\;\label{callingFluid}

    \nl Update matrices ${\pmb{U}_r}$ and $\Tilde{\pmb{F}}_r$ with $\pmb u^k$ and $\Tilde{\pmb f}^k$ with (\ref{updating})

    \nl $z = z + 1$

    \uIf{$z > Z$}{\nl Refit $\mathcal{I}_F(\cdot)$ using the updated ${\pmb{U}_r}$ and ${\Tilde{\pmb{F}}_r}$
    
    \nl $z = 0$}

    \nl $\pmb r^k = \Tilde{\pmb f}^k - \pmb f^{k-1}$

    \nl $e = ||\pmb r^k||_2/||\Tilde{\pmb f^k}||_2$\label{alg:tolerance}

    \uIf{$e > \delta$}{
    \nl $\Delta \pmb{f}^k = Acc(\Tilde{\pmb f}^k, \pmb r^k)$ using e.g Algorithm \ref{alg:iqnils}

    \nl ${\pmb f^k} = {\pmb f}^{k-1} + \Delta \pmb{f}^k$
  }\Else{
  \nl $\pmb{f}^{n} = \Tilde{\pmb f^k}$  \textcolor{mynicegreen}{// \textit{Convergence}}

  \nl Store converged reduced coordinate $\pmb{u}_r \in \mathbb R ^{r_u}$ in $\pmb{U}^n_r \in \mathbb R ^{r_u \times n}$
  }
  \nl $k = k+1$
    }
    \nl $n = n + 1$}

\nl Retrieve the full displacement field using $\pmb{U}^n = \pmb{\Phi}_U \pmb{U}_r^n$\label{callingPostProcess}
\end{algorithm}


\SetKwComment{Comment}{/* }{ */}
\SetKwInput{KwData}{Input}

\begin{algorithm}
\caption{ROM-FOM - Offline stage}\label{alg:offline}
\KwData{Force snapshots matrix $\pmb{F}$

Direct Force snapshots matrix $\Tilde{\pmb{F}}$

Displacement snapshots matrix $\pmb{U}$,

Number of selected force modes $r_f$,

Number of selected displacement modes $r_u$,

Update batch size $p$}
\KwResult{$\{\pmb{\Phi}_f$, $\Bar{\pmb{F}}$, $\pmb{\Phi}_u$, $\Bar{\pmb{\Phi}}_u$, $\Bar{\pmb{U}}$, $\mathcal{I}_{F}(\cdot)$, $\mathcal{I}_{S}(\cdot)$, ${\pmb{U}}_r$, $\Tilde{\pmb{F}}_r\}$}
\vline

\nl Arrange a combination of $\pmb{F}$ and $\Tilde{\pmb{F}}$: $\Hat{\pmb{F}}$

\nl Compute mean of $\Hat{\pmb{F}}$ : $\Bar{\pmb{F}}$

\nl SVD : $\Hat{\pmb{F}} - \Bar{\pmb{F}}= \pmb{\Phi}_{f} \pmb \Sigma_f \pmb{V}^*_f$

\nl Truncation of $r_f$ modes:  $\pmb{\Phi}_{f}$ $\leftarrow$ $\left[ \begin{array}{cccc}
        \vertbar & \vertbar &        & \vertbar \\
         {\pmb{\Phi}}_{1, f}    & {\pmb{\Phi}}_{2, f}   & \ldots & {\pmb{\Phi}}_{r_f, f}   \\
        \vertbar & \vertbar &        & \vertbar
      \end{array}     \right]$

\nl Compute mean of ${\pmb{U}}$ : $\Bar{\pmb{U}}$

\nl SVD : $\pmb{U} - \Bar{\pmb{U}}= \pmb{\Phi}_{U} \pmb \Sigma_U \pmb{V}^*_U$

\nl Solve least-squares problem $\thickbar{\pmb{\Phi}}_u [{\pmb{u}_r}  \otimes {\pmb{u}_r}] = (\pmb{I} - \pmb{\Phi}_U \pmb{\Phi}_U^T) (\pmb{U} - \Bar{\pmb{U}})$ for $\Bar{\pmb{\Phi}}_u$

\nl Truncation of $r_u$ modes:  $\pmb{\Phi}_{U}$ $\leftarrow$ $\left[ \begin{array}{cccc}
        \vertbar & \vertbar &        & \vertbar \\
         {\pmb{\Phi}}_{1, U}    & {\pmb{\Phi}}_{2, U}   & \ldots & {\pmb{\Phi}}_{r_u, U}   \\
        \vertbar & \vertbar &        & \vertbar
      \end{array}     \right]$

\nl Retrieve the reduced coordinates $\pmb{F}_r$ and $\Tilde{\pmb{F}}_r$ $\equiv$ $\pmb \Sigma_f \pmb{V}^*_f$ 

\nl Retrieve the reduced coordinates $\pmb{U}_r$ $\equiv$ $\pmb \Sigma_U \pmb{V}^*_U$ 

\nl Fit $\mathcal{I}_{S}(\cdot)$ : $\mathcal{I}_{S}(\pmb{F}_r) \approx \pmb{U}_r$

\nl Retrieve the reduced coordinates of converged iterations $\pmb{F}_r^n$ by column-extraction from  $\Tilde{\pmb{F}}_r$

\nl Fit $\mathcal{I}_{F}(\cdot)$ : $\mathcal{I}_{F}(\pmb{U}_r, \pmb{F}_r^n) \approx \Tilde{\pmb{F}}_r$

\nl Store $p$-size batches from ${\pmb{U}}_r$ and $\Tilde{\pmb{F}}_r$

\end{algorithm}

\section{Results}\label{sect:results}

In the following, we intend to evaluate the performance of the ROM strategy proposed in terms of both the gained of CPU time (speedup) and number of fixed-point iterations. We demonstrate this on three transient FSI test cases with a very strong coupling, and high added mass in $1D$, $2D$ and $3D$, respectively, with low $Re$ (laminar) flows.
In each case, we apply the proposed ROM-based predictors on the problem in 2 different configurations: FOM-FOM and ROM-FOM, with the latter using a solid ROM in the actual solid computations. Since the computational time is almost always dominated by the solvers' internal iterations, and the addition of the predictors only add a fraction of that time, the comparisons will be mainly done using the number of coupling subiterations. We assess how the addition of the new data-driven predictors affects the number of needed iterations. For the sake of completeness, we also compare in terms of the total CPU time spent to ensure that speedups are realized, especially in the last 2 (more realistic) cases. We also check that, in the case of replacing the solid solver with a solid ROM, the errors of the computed fields are sufficiently low due to the predictive ability of the solid ROM. The coupling operations are done using the coupling component in the multiphysics simulation code \texttt{KratosMutiphysics} \cite{dadvand_object-oriented_2010}. In what follows, we will denote $\Hat{\mathcal{S}}_\Gamma$ and $\Hat{\mathcal{F}}$ as \textit{SROM} and \textit{FROM} respectively. The standard predictors in (\ref{Const-Extrap}), (\ref{Lin-Extrap}) and (\ref{Quad-Extrap}) will be denoted by \textit{Constant Extrapolator}, \textit{Linear Extrapolator} and \textit{Quadratic Extrapolator} respectively, while the new proposed predictor will be called \textit{Data-Driven Predictor}.

\subsection{Example 1: 1D flexible tube model}

\begin{figure}
\begin{center}
\includegraphics[width=1\textwidth]{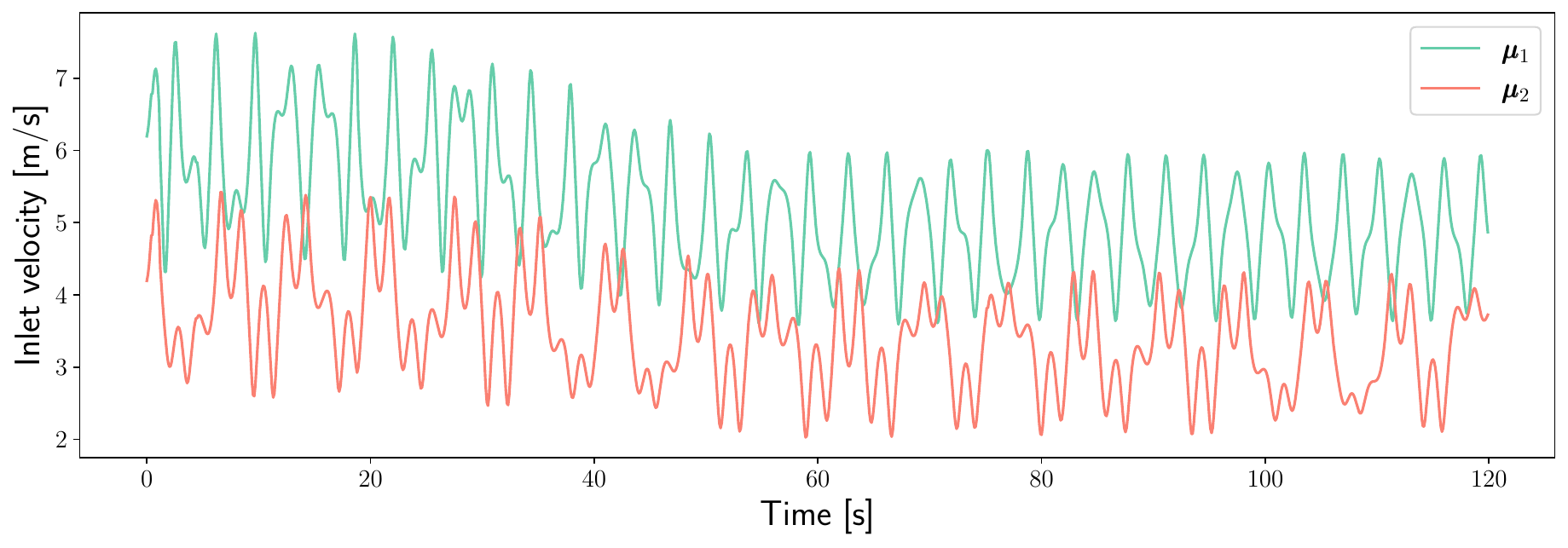}
\end{center}
\caption{Inlet velocity signals corresponding to $\pmb{\mu}_1$ and ${\pmb{\mu}}_2$ parameters values.}
\label{inlet_v}
\end{figure}

The model of flexible tube and related HF partitioned solvers proposed by \citet{degroote_stability_2008} are used here. The flow is assumed to be incompressible with constant density $\rho$. Both fluid mass and momentum conservation equations (neglecting viscosity) read
\begin{equation}\label{fluid_flow}
    \begin{cases}
	 \partial_t a + \partial_x(a v) = 0, \\
        \partial_t(av)+ \partial_x(a v^2) + \dfrac{a}{\rho}\,\partial_x p = 0,
	\quad t>0, \ x\in [0,L]
 \end{cases}
\end{equation}
where $v$ is the velocity, $a$ is the time-dependent tube cross section and
$p$ is the pressure. From the fluid side, the unknowns are both
velocity and pressure.
For the thin flexible tube with a thickness $h_s$, a quasi-static model 
\[
a = a(p)
\] 
is used (retaining only the vessel stress in the circumferential direction). The following nonlinear elastic stress-strain law $\sigma_{\varphi \varphi} ( \epsilon_{\varphi \varphi})$ is used \cite{tiba2023nonintrusive}:
\begin{equation}\label{strs_strain_law}
    \begin{cases}
	 \sigma_{\varphi \varphi} = E\, \epsilon_{\varphi \varphi}\quad \text{if} \ |\epsilon_{\varphi \varphi}|< \epsilon_0 \\
   \sigma_{\varphi \varphi} = E/5\, \epsilon_{\varphi \varphi} + 20\quad \text{if} \ \epsilon_{\varphi \varphi}\geq \epsilon_0 \\
   \sigma_{\varphi \varphi} = E/5\, \epsilon_{\varphi \varphi} - 20\quad \text{if} \ \epsilon_{\varphi \varphi}\leq -\epsilon_0 \\
 \end{cases}
\end{equation}
with $\epsilon_0=2\,\, 10^{-3}$ and $E=12500~Pa$.
\begin{figure}[t]
\begin{center}
\includegraphics[width=.6\textwidth]{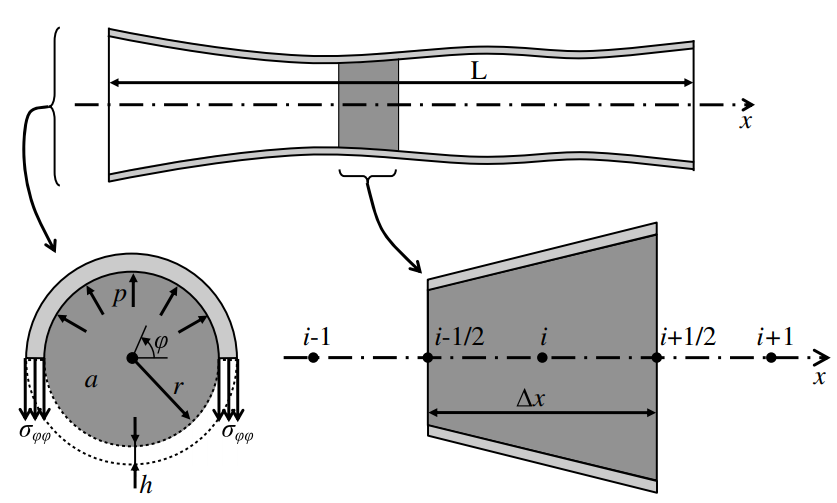}
\end{center}
\caption{Example 1 - 1D flexible tube test case schematic explanation (from \cite{degroote_stability_2008}). $\sigma_{\varphi \varphi}$ is the vessel hoop stress, $h$ is the thickness and $\Delta x $ is the length of the finite volume cell.}
\label{testcase1}
\end{figure}
Figure \ref{testcase1} shows a schematic description of this problem.
A non-reflective boundary condition is used on the $x=L$ boundary as $\frac{du}{dt} = \frac{1}{c} \frac{dp}{t}$ with $c$ the fluid wave speed $c=\sqrt{\frac{a}{\frac{da}{dp}}}$. The prescribed inlet ($x=0$) velocity $v_0$ is computed using the solution of a nonlinear Duffing equation in order to evaluate the ROM performances in problems with complex dynamics, and the capacity to benefit from the HF fluid solver output to handle the dynamics:
\begin{equation}
    \begin{cases}
	 \ddot u(t) = a\, u(t) + b\, u(t)^2 + c\, u(t)^3 + d + p\,cos(f t) + e\, \dot u(t)\,\,\,\, \forall t \in [0, 120] \\
   u(0) = 10\,\,\,;\,\,\,\dot u (0) = 0. \\
   v_{0}(t) = (g u(t) + h) r(t) \\
   r(t) = 1\,\,\,\,\,\,\,\,\,\,\,\,\,\,\,\,\,\,\,\,\,\,\,\,\,\,\,\,\,\,\,\,\,\,\,\,\,\,\,\,\,\,\,\,\,\,\,\,\,\,\,\,\,\,\,\, \forall t \in [0, 20] \\
   r(t) = 0.9+0.1 sin(t \pi / 40)\,\,\,\,\,\,\,\,\, \forall t \in [20, 60] ,\,\,\,\,\,\, {\#Negative \,\, ramp\,\, function}\\
   r(t) = 0.8\,\,\,\,\,\,\,\,\,\,\,\,\,\,\,\,\,\,\,\,\,\,\,\,\,\,\,\,\,\,\,\,\,\,\,\,\,\,\,\,\,\,\,\,\,\,\,\,\,\,\, \forall t \in [60, 120]  . \\
 \end{cases}
\end{equation}
We fix $(a,\ b\ ,c\ ,d\ ,e\ ,g\ , p) = (-1,0 ,-0.002 ,-1 ,-0.02 ,1/60,360)$ and we parameterize this signal with the parameter vector $\pmb{\mu} =(f\,\,\, ,h )^T$ allowing the generation of different frequencies and amplitudes. For this study, the two values $\pmb{\mu}_1 = [2,\, 6]^T$ and $\pmb{\mu}_2 = [0.9,\, 4]^T$ have been selected, leading to the signals shown in Figure \ref{inlet_v}.

 The fluid flow equations (\ref{fluid_flow}) are solved using a second order finite volume scheme with $100$ cells and the solid section $\pmb{a}(\pmb{p})$ is computed at each iteration as the solution of the scalar minimization problem 
 \begin{equation}
     \pmb{p} \sqrt{\frac{\pmb{a}}{\pi}} =  \sigma_{\varphi \varphi}(\frac{\sqrt{\frac{\pmb{a}}{\pi}} - r_0}{r_0}) h_s .
 \end{equation}
The solid subdomain in this case is the interface itself, and the nodes from the solid and fluid sides match each other. The nondimensional time step $\tau = \frac{u_0 \Delta t}{L}$ and nondimensional stiffness $\kappa = \sqrt{\frac{{E h_s}}{{2\rho r_0 v_0(0)^2}}}$ are chosen as $\tau = 0.05$ and $\kappa=21$. Note that the authors in \cite{degroote_stability_2008} showed that the coupling with values of $\tau$ and $\kappa$ of this order is strong and standard Gauss-Seidel iterations quickly fail to converge. FSI subiterations tolerance used here is $\delta = 10^{-4}$ and the reuse number of IQN-ILS used is $q=2$.

In order to train the ROM model, a FOM-FOM computation is done on a single inlet velocity case corresponding to $\pmb{\mu}_1$ until $T = 35$~s. The offline computations of the ROM models are then performed on the available results, giving data snapshots of size $m = 1759$ (see (\ref{num_snaps})).  

The latent dimension of the displacement POD subspace is $r_u=4$ and $r_f=10$ for the pressure field. Regarding the regressor, a thin plate spline kernel RBF \cite{wood_thin_2003} is used with 
\begin{equation}
    \phi( \pmb{x} ) = \pmb{x}^2 \log(\pmb{x})
\end{equation}
for both $\mathcal{I}_S(\cdot)$ and $\mathcal{I}_F(\cdot)$.
For the local iterations the convergence criterion is  $\delta_r = 0.02$, the iteration-frequency of the model update is $Z = 200$ and $p = 1640$ is chosen as the batch size.

\begin{figure}
    \centering
    \includegraphics[width = 0.9\textwidth]{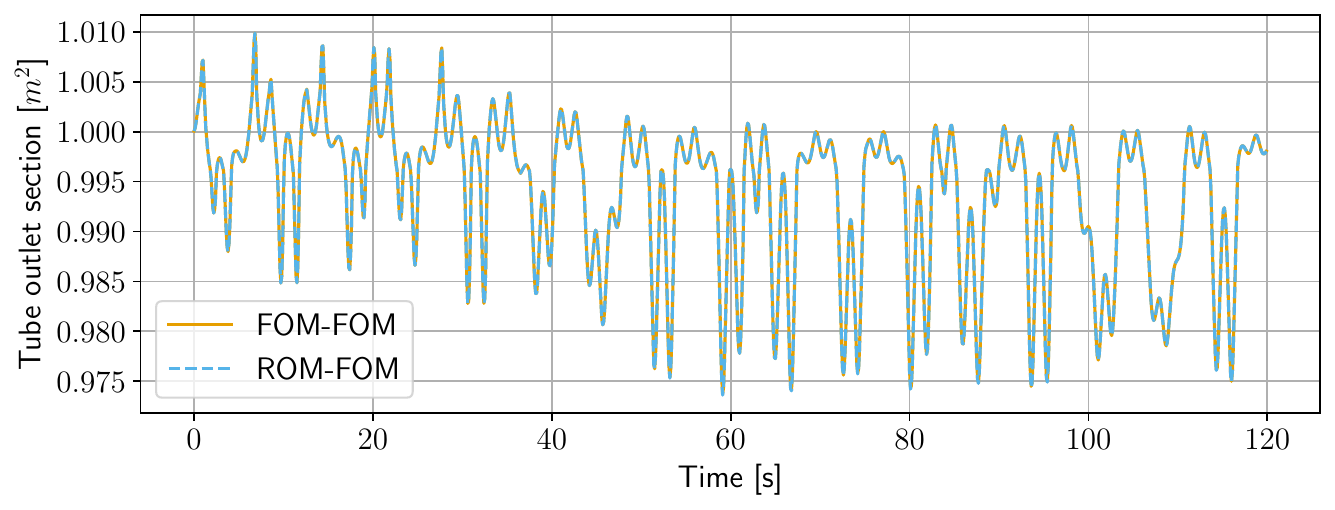}
    \caption{Example 1 - Time-evolution of the outlet tube section, comparing the ROM-FOM and FOM-FOM solutions at $\pmb{\mu} =  \pmb{\mu}_2$. The time-series are quasi-superposed 'at the eye norm'}
    \label{fig:case1-prediction}
\end{figure}

\begin{figure}
    \centering
    \includegraphics[width = 0.6\textwidth]{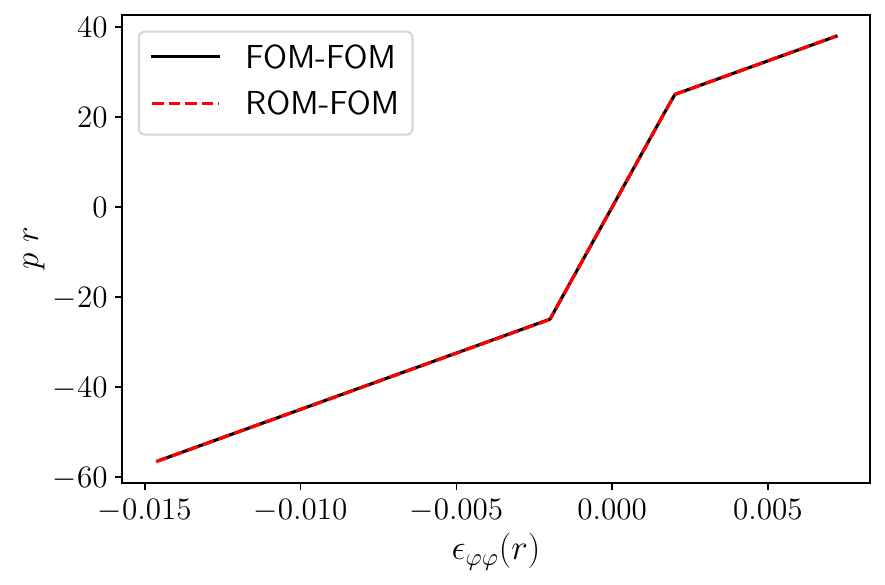}
    \caption{From the results of the ROM-FOM problem, the pressure $p$ and tube radius $r$ are retrieved from the results and $\epsilon_{\varphi \varphi}$ is computed as $\frac{\sqrt{\frac{\pmb{a}}{\pi}} - r_0}{r_0}$. We can see the nonlinear law used in (\ref{strs_strain_law}) is well reproduced. Case of $\pmb{\mu} =  \pmb{\mu}_2$}
    \label{fig:case1-nonlin}
\end{figure}

The performance of the ROMs is evaluated on the future time prediction of the simulation with $\pmb{\mu} = \pmb{\mu}_1$, i.e for $t \in [35 s~,~120 s]$. In addition, we also test the ROMs and predictors on the unseen parameter value $\pmb{\mu}_2$. 
First, we check the accuracy of the solid ROM prediction in the ROM-FOM coupling: we show in Figure \ref{fig:case1-prediction} the outlet section evolution in time comparing the FOM-FOM and the ROM-FOM solutions. We also demonstrate the solid nonlinearity well predicted by the ROM in Figure \ref{fig:case1-nonlin}, which is expected since the region of deformations reached during the prediction was well present in the training data.


\begin{figure}
    \centering
    \includegraphics[width = 1\textwidth]{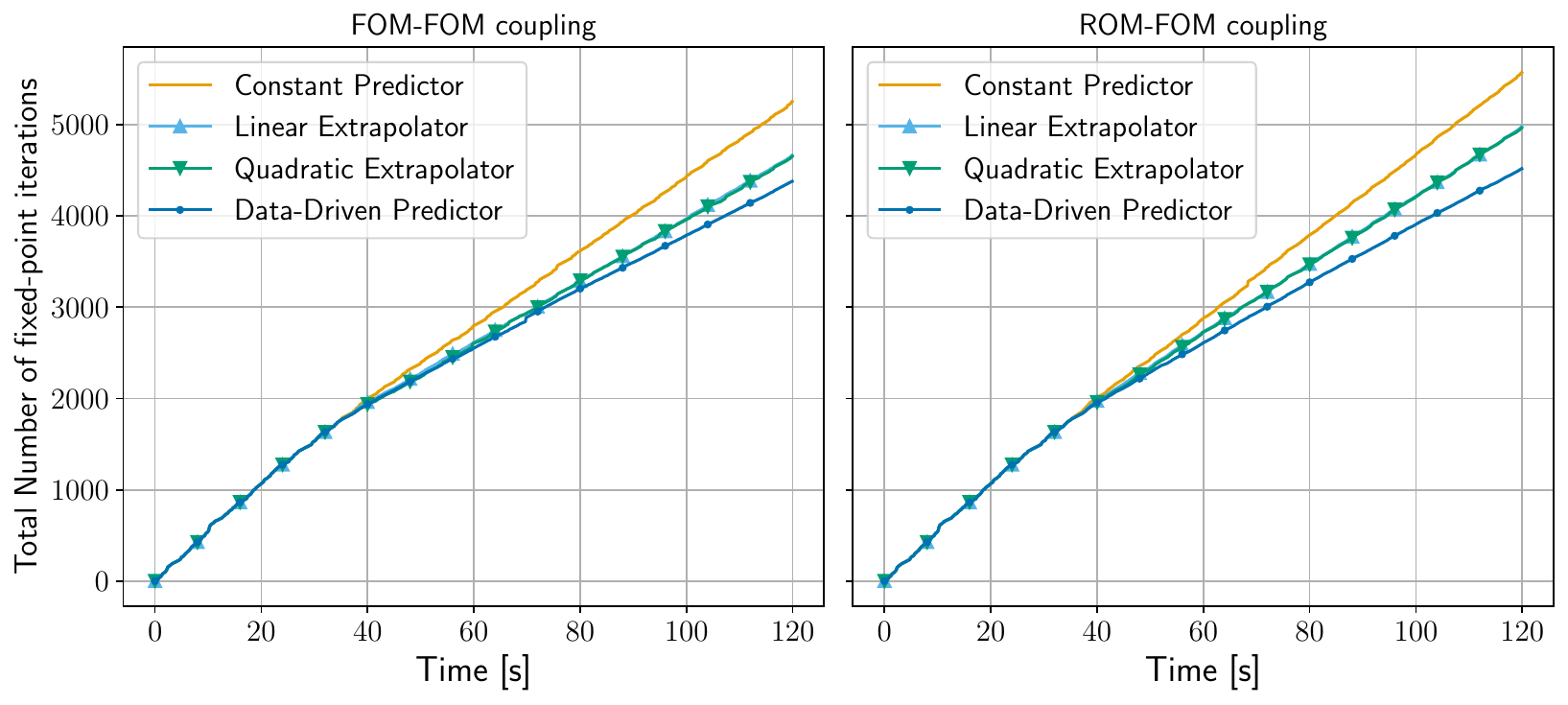}
    \caption{Example 1 - Comparison of total iterations needed for the simulation in a time-prediction regime using $\pmb{\mu} = \pmb{\mu}_1 = [2, 6]^T$}
    \label{fig:case1-iterations-mu1}
\end{figure}

\begin{figure}
    \centering
    \includegraphics[width = 1\textwidth]{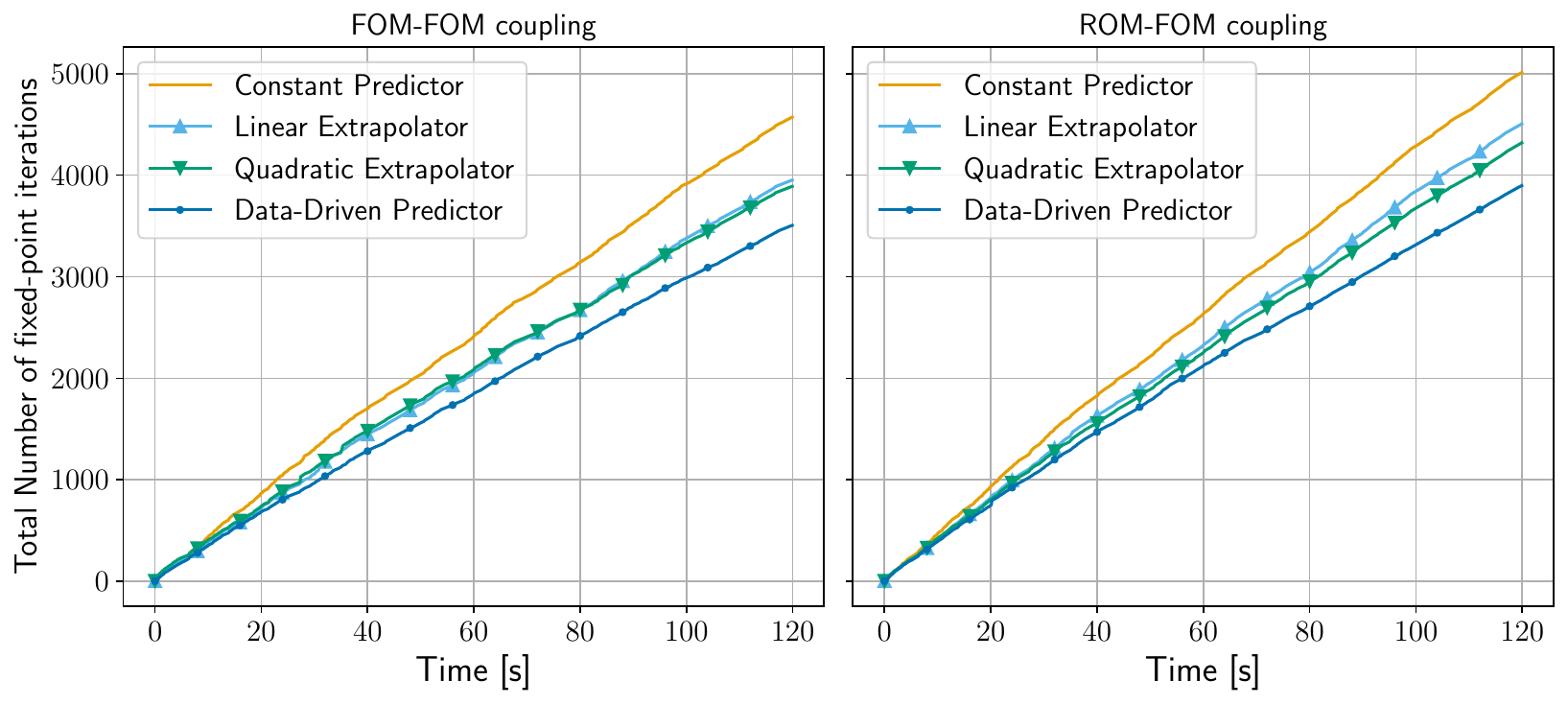}
    \caption{Example 1 - Comparison of total fixed-point iterations needed for the simulation using the unseen parameter $\pmb{\mu} = \pmb{\mu}_2 = [0.9, 4]^T$ (Time-parameter prediction)}
    \label{fig:case1-iterations-mu2}
\end{figure}

We show in figure \ref{fig:case1-iterations-mu1} the total number of iterations performed using the different predictors for $\pmb{\mu}_1$, and figure \ref{fig:case1-iterations-mu2} for $\pmb{\mu}_2$.


\begin{figure}
    \centering
    \includegraphics[width=0.57\textwidth]{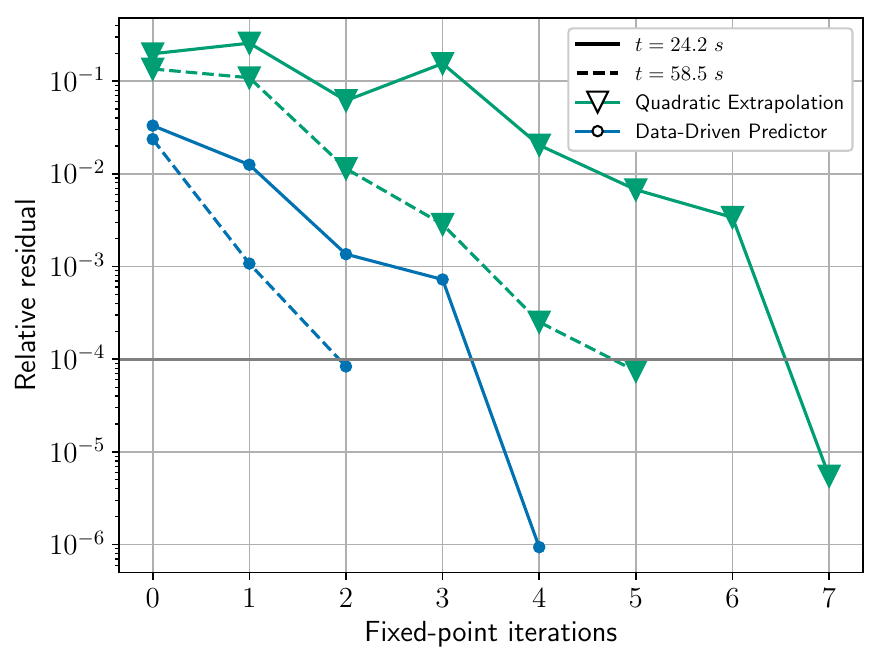}
    \caption{Fixed-point convergence rate using a data-driven predictor and a quadratic extrapolation at two time steps, $t = 24.2~s$ and $t = 58.5~s$, example 1. Convergence relative tolerance $\delta = 10^{-4}$.}
    \label{fig:someIterationsCase1}
\end{figure}

The results clearly show that the new data-driven predictor provides the best efficiency in terms of number of iterations. We can see that using the ROM-FOM with constant predictors (right figures) leads to additional iterations due to the inaccuracy of the solid ROM. This effect is no longer observed when using the data-driven predictor. Moreover, while the use of linear and quadratic predictions result in fewer iterations than the constant predictor case, they are outperformed by the use of data-driven-based initial guesses. In Figure \ref{fig:someIterationsCase1} and for two time steps, we show how the data-driven initial prompts a faster rate of convergence, as we can see that the data-driven predictor results in a first iteration with a much lower residual than with a quadratic extrapolation.

\subsection{Example 2: Hyperelastic flaps in a channel behind a cylinder wake}\label{casTest2}
\begin{figure}
\begin{center}
\includegraphics[width=1\textwidth]{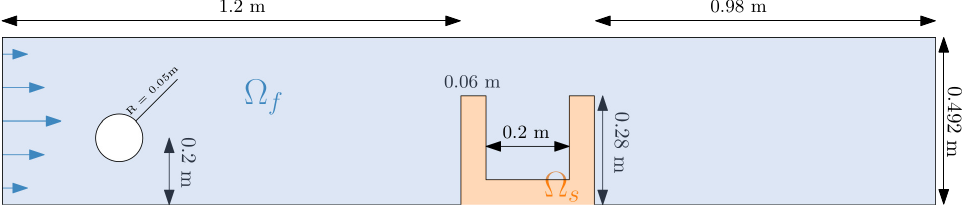}
\end{center}
\caption{Example 2 - test case schematic explanation and dimensions.}
\label{testcase2}
\end{figure}

%
In this section, we consider the problem first introduced in \cite{tiba2023nonintrusive}, and illustrated in figure \ref{testcase2}, where an incompressible flow in a $2D$ channel faces a massless elastic body with two mounted flaps behind a rigid cylinder.

For the fluid, the Navier-Stokes equations read
    \begin{equation}\label{fluid_eq}
        \begin{cases}
            \rho_f \dfrac{\partial \pmb{v}}{\partial t}_{|\Tilde{\mathcal{A}}} + \rho_f [(\pmb{v} - \pmb{w}).\nabla]\pmb{v} + \nabla p - 2 \,\pmb{\text{div}}(\mu_f \pmb{D}(\pmb{v})) = 0 \quad \text{in}\ \Omega_f(t)\\[1.3ex]
           
            \nabla\cdot\pmb{v}= 0 \quad \text{in }\ \Omega_f(t)\\[1.3ex]
        \end{cases}    
    \end{equation}
with $p$ the fluid pressure, $\rho_f$ the fluid density, $\mu_f$ the fluid dynamic viscosity and $\pmb{D}(\pmb{v})$ is the fluid strain rate tensor. The fluid equations are described on a moving domain (using the ALE moving frame) $\Omega_f(t)$. The notation $\Tilde{\mathcal{A}}$ represents the ALE mapping from the reference domain (the $t=0$ configuration) to the computational domain and $\pmb{w}$ is the ALE velocity. 

In this case, we have $\rho_f = 1000\ kg/m^3$, $\mu_f = 0.001\ m^2/s$ and a fully developed Poiseuille inlet flow is applied, with a maximum velocity of $v_{max} = 2.5\ m/s$ starting from $v = 0\ m/s$ at $t = 0s$ and increased by a sinusoidal ramp until reaching $v_{max}$ at $t = 1 s$. This corresponds to a Reynolds number of $Re = 250$, based on $v_{max}$ and the cylinder diameter. The boundary condition at the top and bottom walls is a no slip condition, and a zero pressure on the right boundary.

For the solid subproblem, the equilibrium and constitutive equations for a static hyperelastic solid are:

        \begin{equation}\label{solid_eq}
        \begin{cases}
             \nabla_{\pmb{X}} \pmb{P} = \pmb{0} \quad \text{in}\; \Omega_{s} \\[1.3ex]
            \pmb{P} = \dfrac{\partial W}{\partial \pmb{F}_s} \\[1.3ex]
            \pmb{u} = \pmb{0} \; \quad \text{in}\; \Gamma_{D, s}\\[1.3ex]
        \end{cases}  .
        \end{equation}

The equations are written in the Lagrangian frame with $\nabla_{\pmb{X}}$ the gradient operator in the original configuration, $\pmb{P}$ is the first Piola-Kirchoff stress tensor (PK1) and $\Gamma_{D, s}$ is the Dirichlet boundary. The material model is described in the stored energy density function~$W$, here using the hyperelastic Neo-Hookean model:
\begin{equation}
    W(\pmb{F}_s) =  \frac{\lambda_s}{2}\,(ln(J))^2 - \mu_s ln(J) + \frac{\mu_s}{2}(trace(\pmb{C_s}) - 3)
\end{equation}
where $J = det(\pmb{F}_s)$ is the determinant of the deformation gradient tensor $\pmb{F}_s$, $\pmb{C}_s = \pmb{F}_s^T \pmb{F}_s$ is the right Cauchy Green deformation tensor, $\mu_s = \dfrac{E}{2 (1 + \nu_s)}$ and $\lambda_s =\frac{E \nu_s}{(1 + \nu_s)(1 - 2  \nu_s)}$. In this example, $E = 10 \times 10^6$~Pa and $\nu_s = 0.3$. The coupling conditions (\ref{fsiCondition1}) and (\ref{fsiCondition2}) are imposed on the FSI interface.

\begin{figure}
     \centering
     \includegraphics[width=1\textwidth]{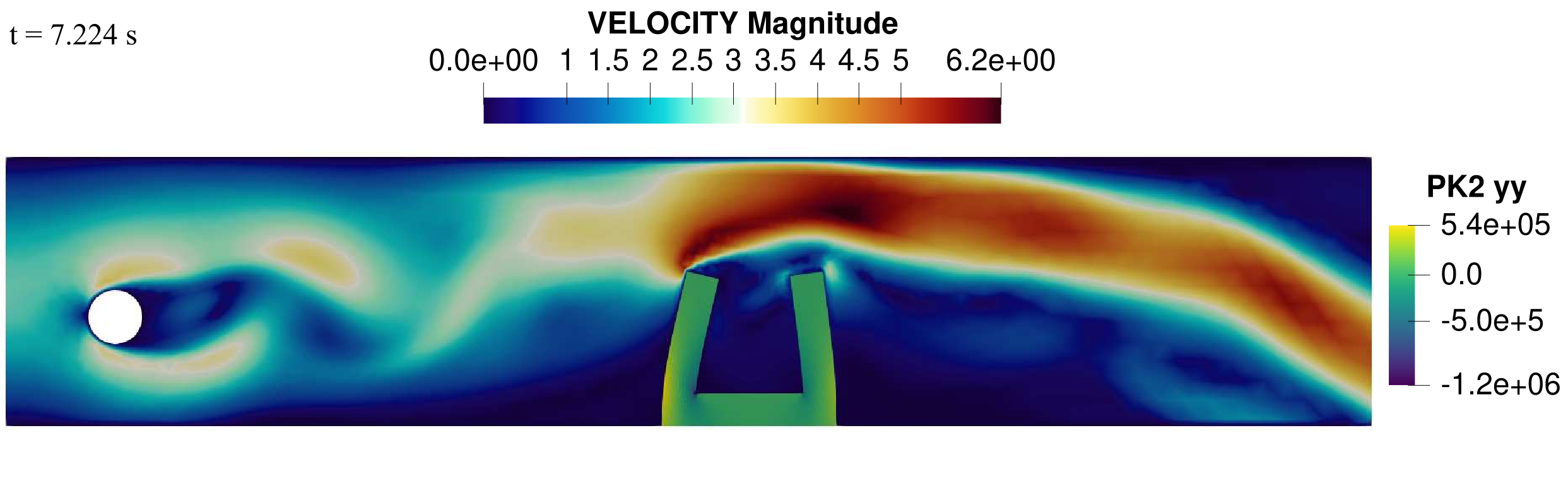}
     \caption{ROM-FOM solution at $t = 7.224 s$: The velocity magnitude on the fluid domain and the PK2 stress on the solid domain.}
     \label{fig:contoursCase2}
 \end{figure}

The fluid problem is discretized using $5440$ variational multiscale (VMS) finite elements \cite{codina_stabilized_2001} and $1640$ quadrilateral plane strain finite elements are used for the structural problem, with 8 ($X$ and $Y$) displacement degrees of freedom at element nodes, making a total of $N_S = 3610$ solid dofs. \texttt{KratosMutiphysics} \cite{dadvand_object-oriented_2010} was used as the finite elements code for both problems, using the modules \texttt{FluidDynamicsApplication} and \texttt{StructuralMechanicsApplication} as separate solvers in a partitioned coupling. The fluid time step is 
$\Delta t = 8\times10^{-3}$~s and the second-order "Bossak" time integration scheme is used\cite{bossak}. The interface grid has matching nodes from the solid and fluid sides and consists of $265$ nodes at the interface, giving $N = 530$. The relative convergence tolerance used is $\delta = 0.005$ and IQN-ILS with reuse $q = 3$ is used as the QN acceleration scheme. Figure \ref{fig:contoursCase2} shows the ROM-FOM solution at $t=7.224~s$.

We define the parameter as the Reynolds number $\pmb{\mu} = Re$ and we use the results from simulations of three points : $\mu \in \{180, 205, 250\}$ solved for $t \in [0, 3.6 s]$, generating a total of $m = 7410$ snapshots. The evaluation of the new ROM predictor will be then done on an unseen parameter $\mu = 192$ and on a larger simulation time $t \in [0, 8 s]$.

For the dimensionality reduction, $r_u=9$ displacement modes and $r_f=45$ force modes are used. For the \textit{SROM} regressor $\mathcal{I}_S(\cdot)$, a $2^{nd}$ order polynomial regression is used with an $L_1$ regularization term as described in (\ref{lasso}). The \textit{FROM} regressor $\mathcal{I}_F(\cdot)$ is chosen here as an RBF function with a cubic kernel
\begin{equation}
    \phi( \pmb{x} ) = \pmb{x}^3 .
\end{equation}
For the local iterations convergence, the tolerance is $\delta_r = 2 \%$, the \textit{FROM} is retrained every $Z=200$ iterations, and the reduced data batch size $p$ is $6900$.

\begin{figure}
  \makebox[\textwidth][c]{\includegraphics[width = 1.2\textwidth]{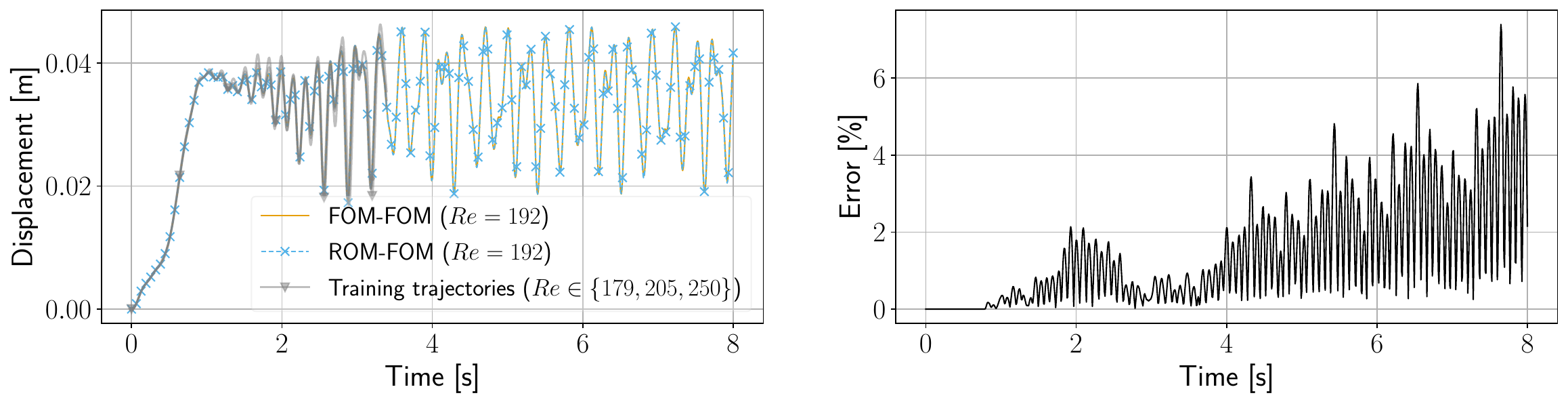}}%
    \caption{Example 2 - (\textbf{Left}) The left tip displacement evolution of the solid comparing the ROM-FOM and the FOM-FOM solution in time, while the training trajectories (in the different Reynolds numbers) of the same quantity are shown in grey. (\textbf{Right}) The relative displacement error of the SROM at $Re = 192$ is computed using (\ref{case2RelErr}) and shown as a function of the simulation time.}
    \label{fig:dispResultsCase2}
\end{figure}

\begin{figure*}
    \centering
    \begin{subfigure}[b]{0.475\textwidth}
        \centering
        \includegraphics[width=.7\textwidth]{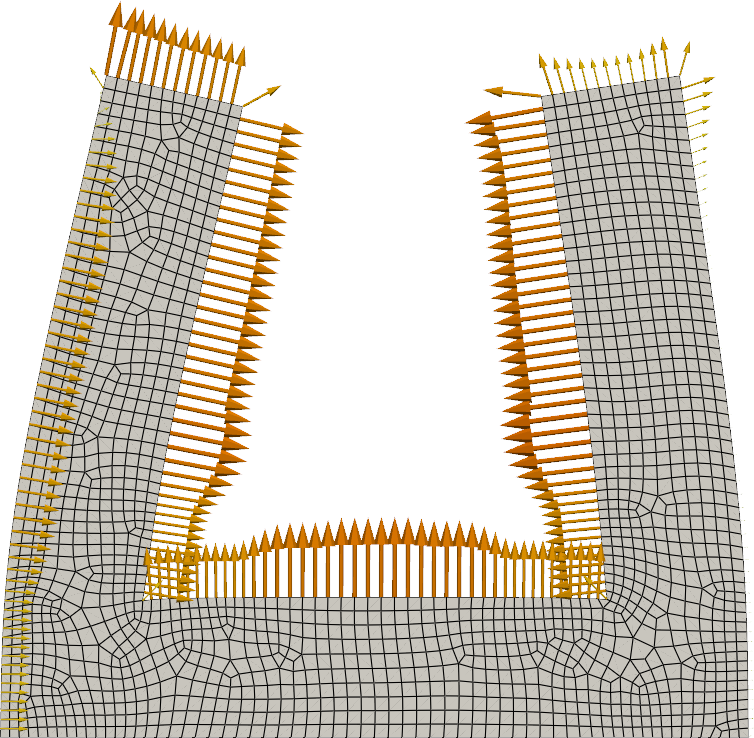}
        \caption[Network2]%
        {{\small }}    
        \label{fig:loadsOnSolid}
    \end{subfigure}
    \hfill
    \begin{subfigure}[b]{0.475\textwidth}  
        \centering 
        \includegraphics[width=\textwidth]{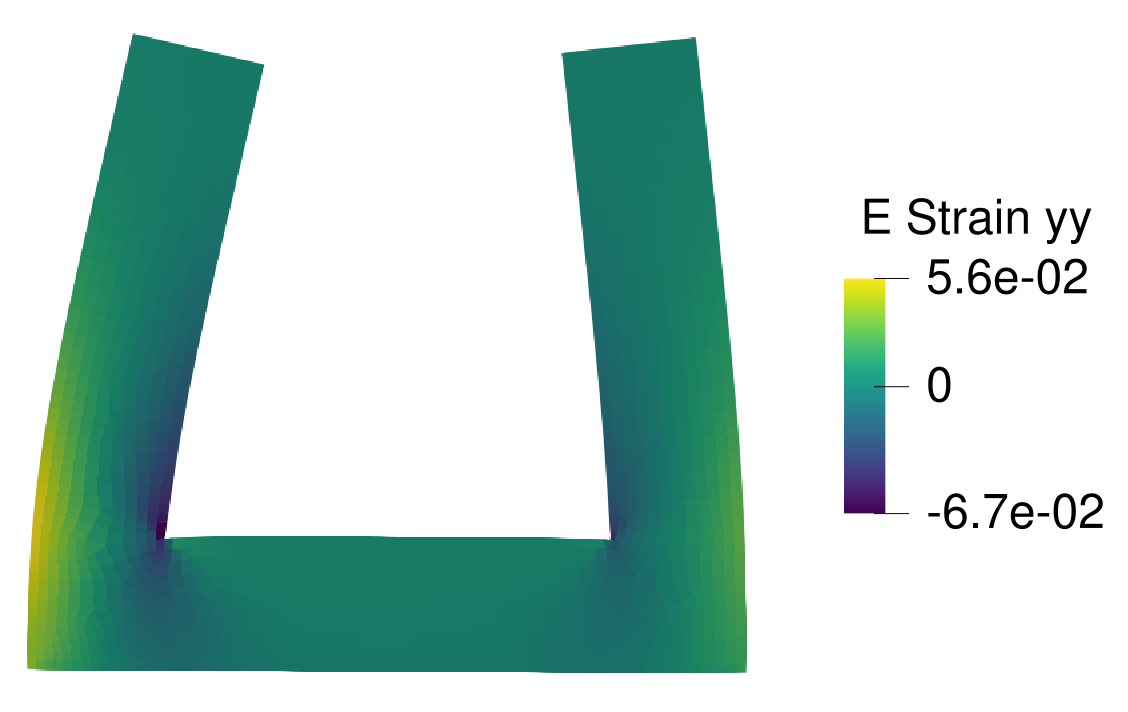}
        \caption[]%
        {{\small FOM-FOM solution.}}    
        \label{fig:FOMFOMGreenLagrange}
    \end{subfigure}
    \vskip\baselineskip
    \begin{subfigure}[b]{0.475\textwidth}   
        \centering 
        \includegraphics[width=0.9\textwidth]{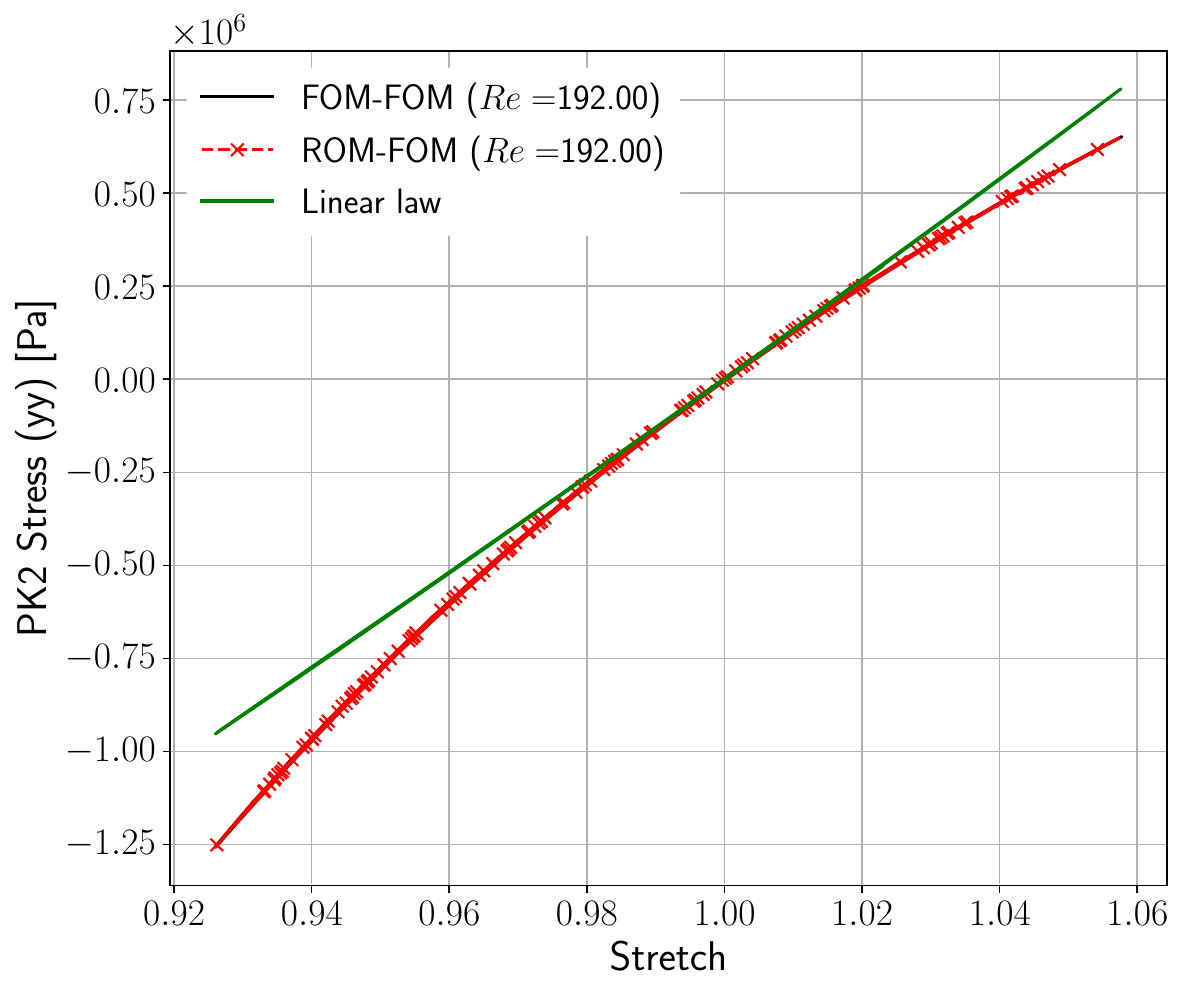}
        \caption[]%
        {{\small Plot of the relationship between the $yy$ component of the second Piola Kirchoff stress (PK2) and the stretch $\lambda$ at the same direction. The linear law is plotted for comparison.}}    
        \label{fig:nonlinMatcase2}
    \end{subfigure}
    \hfill
    \begin{subfigure}[b]{0.475\textwidth}   
        \centering 
        \includegraphics[width=\textwidth]{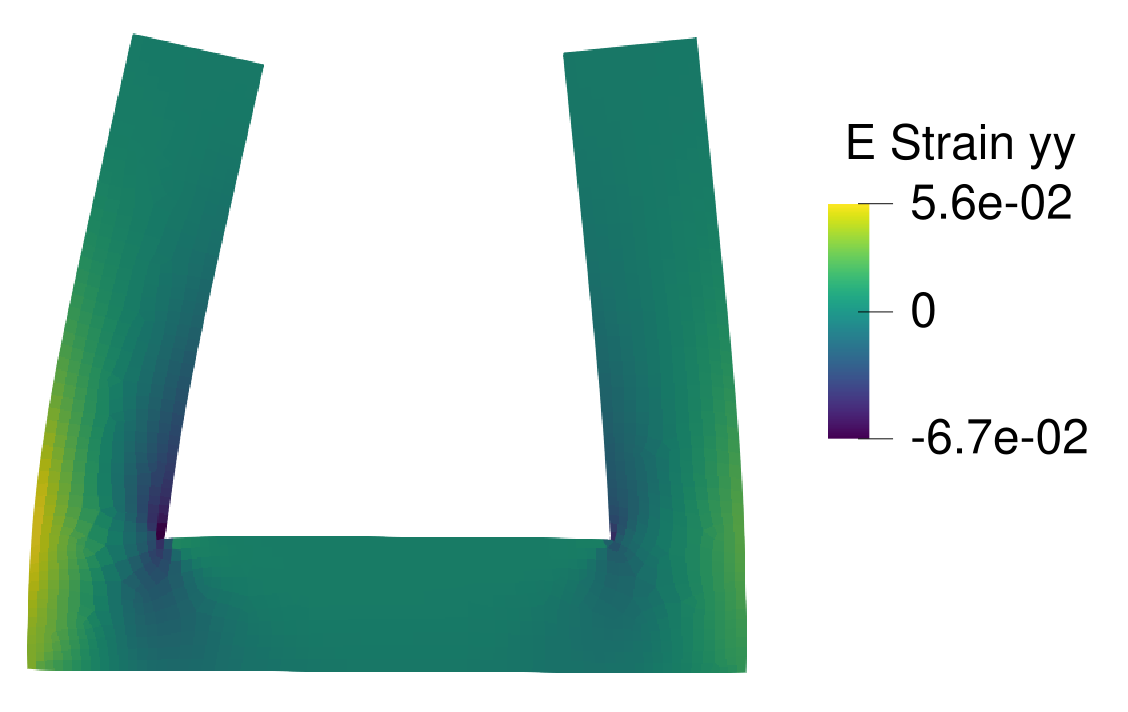}
        \caption[]%
        {{\small ROM-FOM solution.}}    
        \label{fig:ROMFOMGreenLagrange}
    \end{subfigure}
    \caption
    {\small The SROM prediction evaluated on the solid nonlinear behaviour in example 2. In (a) we see the loads applied on the deformed solid at $t = 4.7 s$, and (b) and (d) show the $yy$ component of the Green Lagrange strain at the same time step, comparing the ROM-FOM and FOM-FOM solutions. In (c) the nonlinear behaviour of the Neo-Hook solid is represented using the results of the FOM-FOM and ROM-FOM.} 
    \label{fig:nonlinearStressCase2}
\end{figure*}

The solid ROM gives an accurate model prediction when coupled with the fluid FOM as shown in Figure \ref{fig:dispResultsCase2} where the left tip x-displacement is plotted and compared using the ROM-FOM and FOM-FOM coupling. The displacement values used in the offline training at different $Re$ numbers are also shown in the same figure. The accuracy of the displacement field is reported as the relative error 
\begin{equation}\label{case2RelErr}
e(t) = \frac{||\pmb{d}(t) - {\pmb{d}}(t)_{FOM}||_2}{ <{||{\pmb{d}}(t)_{FOM}||_2}>}    
\end{equation}
with $<\cdot>$ represnting the time-average.
We can see in the figure it remains under $7\%$ even for such a long simulation time. In figure \ref{fig:nonlinearStressCase2}, we show the strain prediction through the Green-Lagrange strain tensor $\pmb{E} = 1/2 (\pmb{C}_s - \pmb{I})$ and the overall nonlinear material behavior using the response of one element as an example, showing the significant accuracy of the SROM prediction. This reaffirms that the SROM is indeed able to capture the nonlinearity.

\begin{figure}
    \centering
    \includegraphics[width = 1\textwidth]{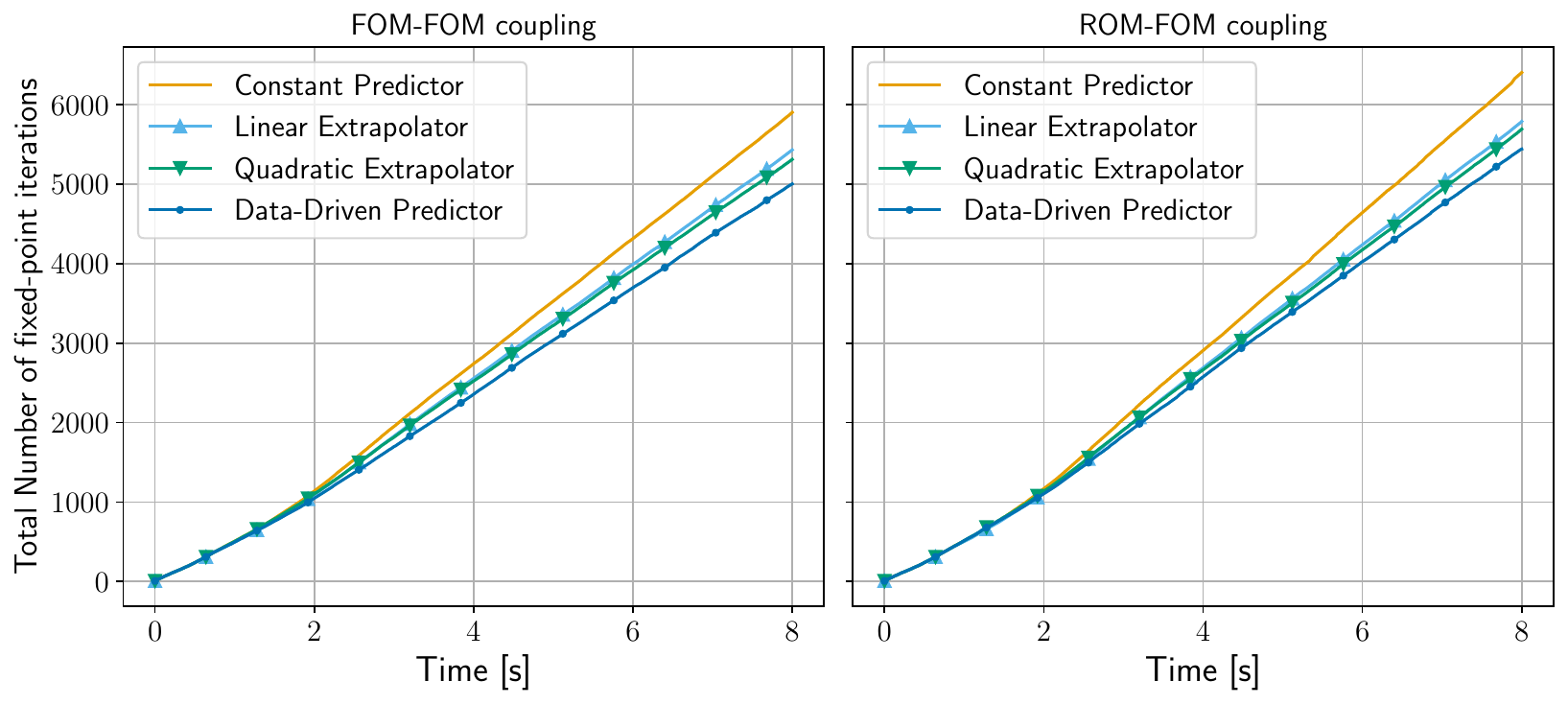}
    \caption{Example 2 - Comparison of total fixed-point iterations needed for the simulation at an unseen $Re = 192$, using different predictors.}
    \label{fig:case2-iterations}
\end{figure}

The accumulated number of fixed-point iterations over the simulation time is reported in Figure \ref{fig:case2-iterations} for the different predictors. Once again, the novel data-driven predictor produces the least number of FSI iterations. More details about the average number of iterations are given in Table \ref{tab:case2-speed}. In addition, in the second column of Table \ref{tab:case2-speed}, we also demonstrate the efficiency in terms of the CPU time spent when using the different predictors. As the evaluation of the \textit{FROM} and the local \textit{SROM} are inexpensive at the start of each time step, and keeping in mind the negligible cost of the \textit{FROM} update, the data-driven predictor does indeed result in a smaller overall CPU time. Using the solid ROM to replace the FOM allows even greater speedup, especially since, in this problem, the solid FOM solver takes nearly twice as CPU time as the fluid FOM solver. We note that, taking advantage of the fluid FOM, the presented approach performs very well, in terms of the SROM accuracy, and the faster fixed-point convergence in spite of the complex dynamics of this FSI problem. The on-the fly update of the FROM also prevents inaccurate predictions when these unseen complex dynamics take place.

\begin{figure}
     \centering
     \includegraphics[width = 1\textwidth]{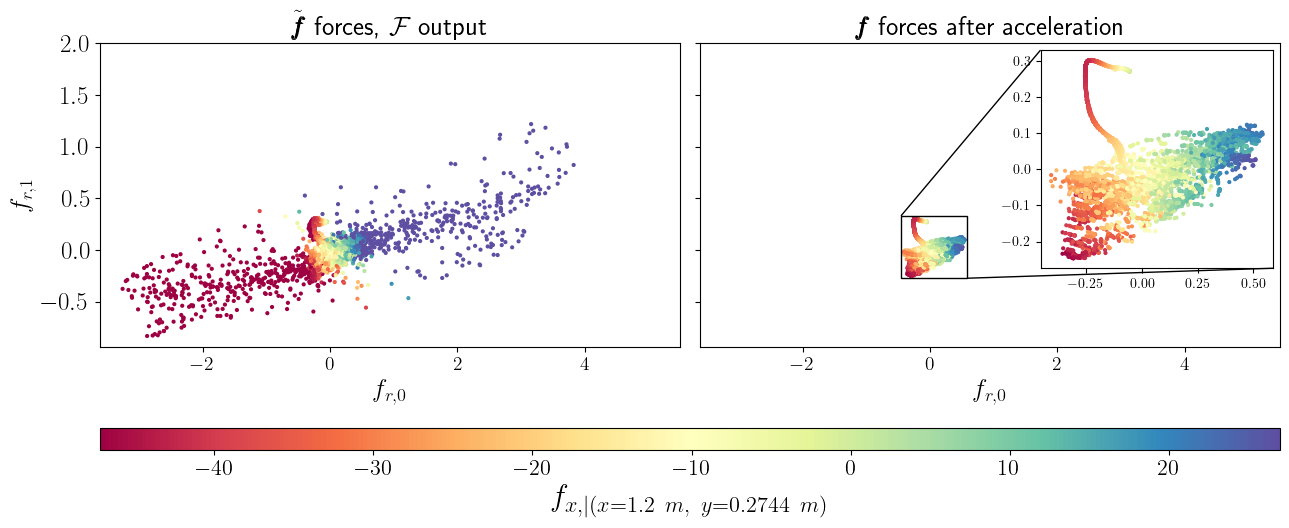}
     \caption{Comparison between the fluid solver output $\Tilde{\pmb{f}}$ (Left) and the accelerated forces $\pmb{f}$ (Right) using the first two components $f_{r, 0}$ and $f_{r, 1}$ of their reduced representations $\pmb{\Phi}_f^T \Tilde{\pmb{f}}$ and $\pmb{\Phi}_f^T {\pmb{f}}$. The colormap corresponds to the amplitude of the X-force at a node close to the left tip.}
     \label{fig:Comparison_f_tilde}
 \end{figure}

In order to highlight the importance of using the accelerated forces $\pmb{f}$ and not the fluid output $\Tilde{\pmb{f}}$ as the SROM input (See the section \ref{CouplingSchemeParagraph} above), we show in Figure \ref{fig:Comparison_f_tilde} the phase space composed of the first 2 components of their latent representations $\Tilde{\pmb{f}}_r$ and $\pmb{f}_r$. The values shown correspond to the forces obtained during the training simulations. We can see that the variance is much bigger with $\Tilde{\pmb{f}}_r$ and this poses difficulties on the training of SROM. Bigger data variance also means that a bigger range of the solid nonlinearity is included in the data, since bigger force amplitudes are reached with $\Tilde{\pmb{f}}_r$ than with $\pmb{f}_r$.

\begin{figure}
    \centering
    \includegraphics[width = 1\textwidth]{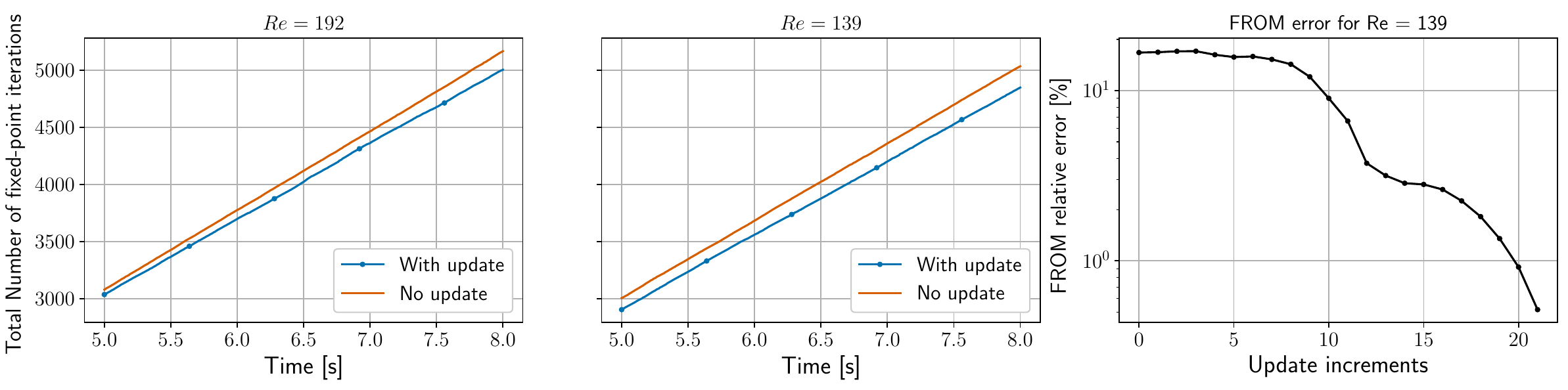}
    \caption{Effect of the FROM update on the accuracy for the $2^{nd}$ example. The total number of global iterations achieved when using an FROM with and without update, evaluated on unseen Reynolds numbers : (\textbf{Left}) $Re = 192$. (\textbf{Center}) $Re = 139$. In (\textbf{Right}), the test error of the FROM is plotted along the model update increments for $Re = 139$.}
    \label{fig:UpdateEffectCase2}
\end{figure}

As a last indication, the effect of the FROM update is first assessed by looking at the difference of total number of fixed-point iterations achieved with and without the update. For the unseen parameters $Re=192$, the benefits of the update is clearly shown in Figure  \ref{fig:UpdateEffectCase2} left where the number of fixed-point iterations is always diminished. More importantly, when considering a parameter $Re = 139$ outside of the Reynolds number training interval $[180, 250]$. Figure \ref{fig:UpdateEffectCase2} center shows that hundreds less convergence iterations are needed overall, \textit{i.e} for all the time steps. Second, we wish to demonstrate the evolution of the model error along the update iterations. Offline data are stored from the last $1000$ iterations' results of the FOM-FOM simulation on the unseen parameter $Re=139$. The FROM is then evaluated on this test data and incrementally updated on each $Z$-sized batches of offline results from the beginning of the simulations onward. The error is computed after each FROM update as the mean relative maximum error and reported on Figure \ref{fig:UpdateEffectCase2} right. We can clearly see the decrease of the FROM error along the update increments, highlighting the ability of the model to leverage its updates for extrapolation.

\begin{table}[]
    \centering
    \begin{tabular}{c||c|c|c|c}
         & \multicolumn{2}{c}{FOM-FOM} & \multicolumn{2}{c}{ROM-FOM} \\
         & Avg. iterations & CPU Time & Avg. iterations & CPU Time \\
    \midrule
        Data-Driven & \textbf{5.11} & 2.7 & 5.6 & \textbf{1.0} \\
        Constant & 6.12 & $3.43$ & 6.67 & 1.13 \\
        Linear & 5.59 & \textbf{3.16} & 5.99 & 1.24 \\
        Quadratic & \textbf{5.46} & 3.35 & 5.88 & 1.01 \\
    \end{tabular}
    \caption{Example 2 - Average number of iterations and the total CPU time of the simulation at the unseen $Re$ number using the FOM-FOM and ROM-FOM coupling.}
    \label{tab:case2-speed}
\end{table}

\subsection{Example 3 : 3D hyperelastic incompressible flaps:}
\begin{figure}
\begin{center}
\includegraphics[width=1\textwidth]{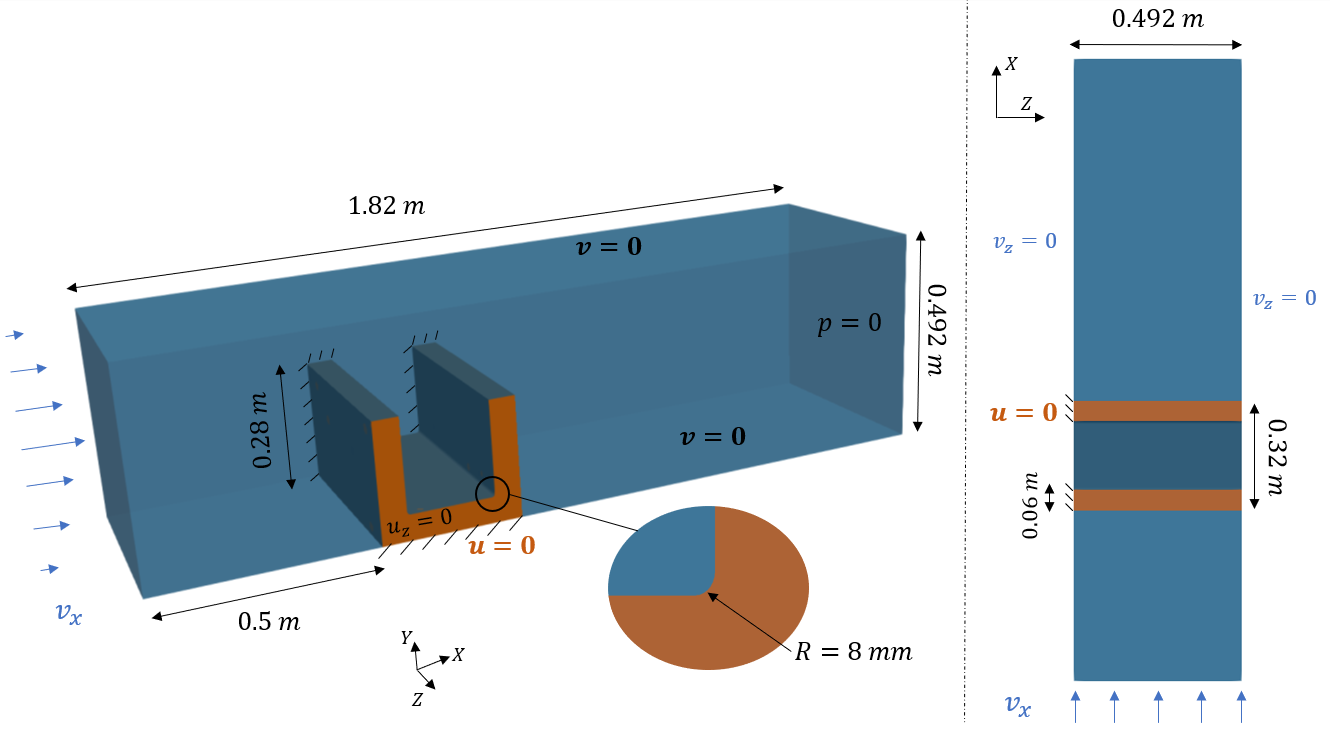}
\end{center}
\caption{Example 3 - Test case schematic explanation and dimensions.}
\label{testcase3}
\end{figure}

\begin{figure}
\begin{center}
\includegraphics[width=0.6\textwidth]{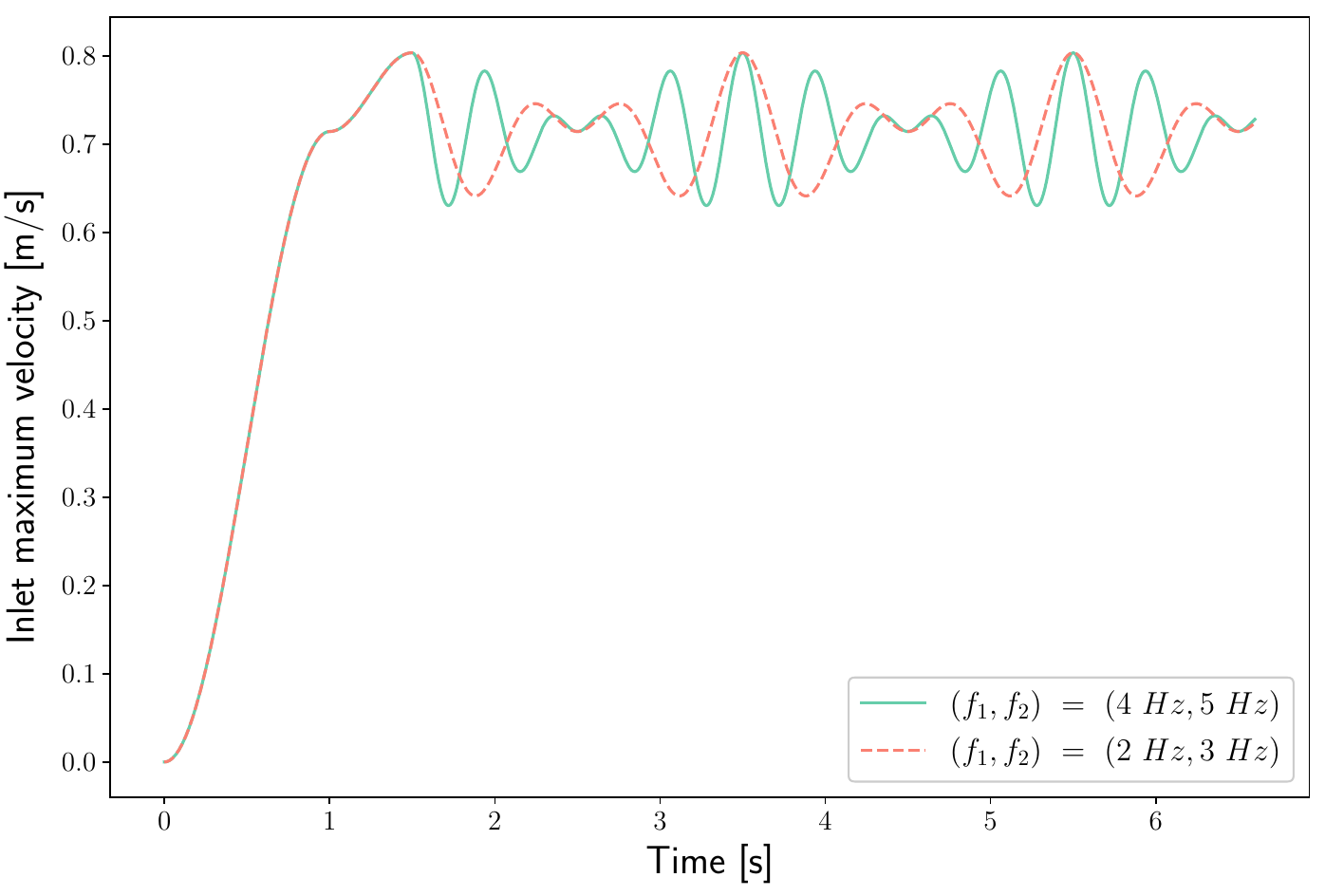}
\end{center}
\caption{Inlet velocity at the center of the inlet section.}
\label{fig:case3-inletV}
\end{figure}

To show the performance of our methodology in larger scale problems, we consider a $3D$ extension of the previous problem. The figure \ref{testcase3} gives a brief description. The boundary faces corresponding to the top, bottom, back and front walls of the geometry in the left of Figure \ref{testcase3} will be called henceforward $y+$, $y-$, $z-$ and $z+$ respectively. To simplify the configuration, no cylinder is present in this problem but a fully developed pulsatile inlet flow is applied with a signal composed of two frequencies $(f_1, f_2)$ as illustrated in figure \ref{fig:case3-inletV} and expressed as:
\begin{equation}
    \begin{cases}
        v_{|x=0}(y, t) = \frac{1}{2}(1-cos(\pi t)) g(y)\,\,\,\,\,\,\,\,\,\,\,\,\,\,\,\,\,\,\,\,\,\,\,\,\,\,\,\,\,\,\,\,\,\,\,\,\,\,\,\,\,\,\,\,\,\,\,\,\,\,\,\,\,\,\,\,\,\,\,\,\,\,\,\,\,\,\,\,\,\,\,\,\,\,\,\,\,\,\,\,\,\,\,\, \forall t \in [0, 1]\\
        v_{|x=0}(y, t) = \frac{1}{16}(16+(1-cos(2\pi(t-1)))) g(y)\,\,\,\,\,\,\,\,\,\,\,\,\,\,\,\,\,\,\,\,\,\,\,\,\,\,\,\,\,\,\,\,\,\,\,\,\,\,\,\,\,\,\,\,\,\, \forall t \in [1, 1.5]\\
        v_{|x=0}(y, t) = \frac{1}{16}(16+(cos(f_1\pi(t-1.5)))+(cos(f_2\pi(t-1.5)))) \,\,\,\, \forall t \in [1.5, 6.6]\\
        g(y) = 11.8\,y\,(0.492-y)
    \end{cases}
\end{equation}

with $\pmb{\mu}_1 = (f_1, f_2) = (4~Hz, 5~Hz)$ the parameter values used for training and $\pmb{\mu}_2 = (f_1, f_2) = 2~Hz, 3~Hz)$ for testing.
Slip conditions are used on the $z-$ and $z+$ faces, no-slip conditions are imposed on the $y-$ and $y+$ faces and a zero pressure is imposed on the outlet. The discretization uses 437039 elements with 84988 nodes. This flow setting corresponds to a Reynolds number $Re = 225$ based on the maximum inlet velocity and the length of the solid flap.

\begin{figure}
\begin{minipage}{.5\linewidth}
\centering
\subfloat[]{\label{StressCase3Fom}\includegraphics[width=.8\textwidth]{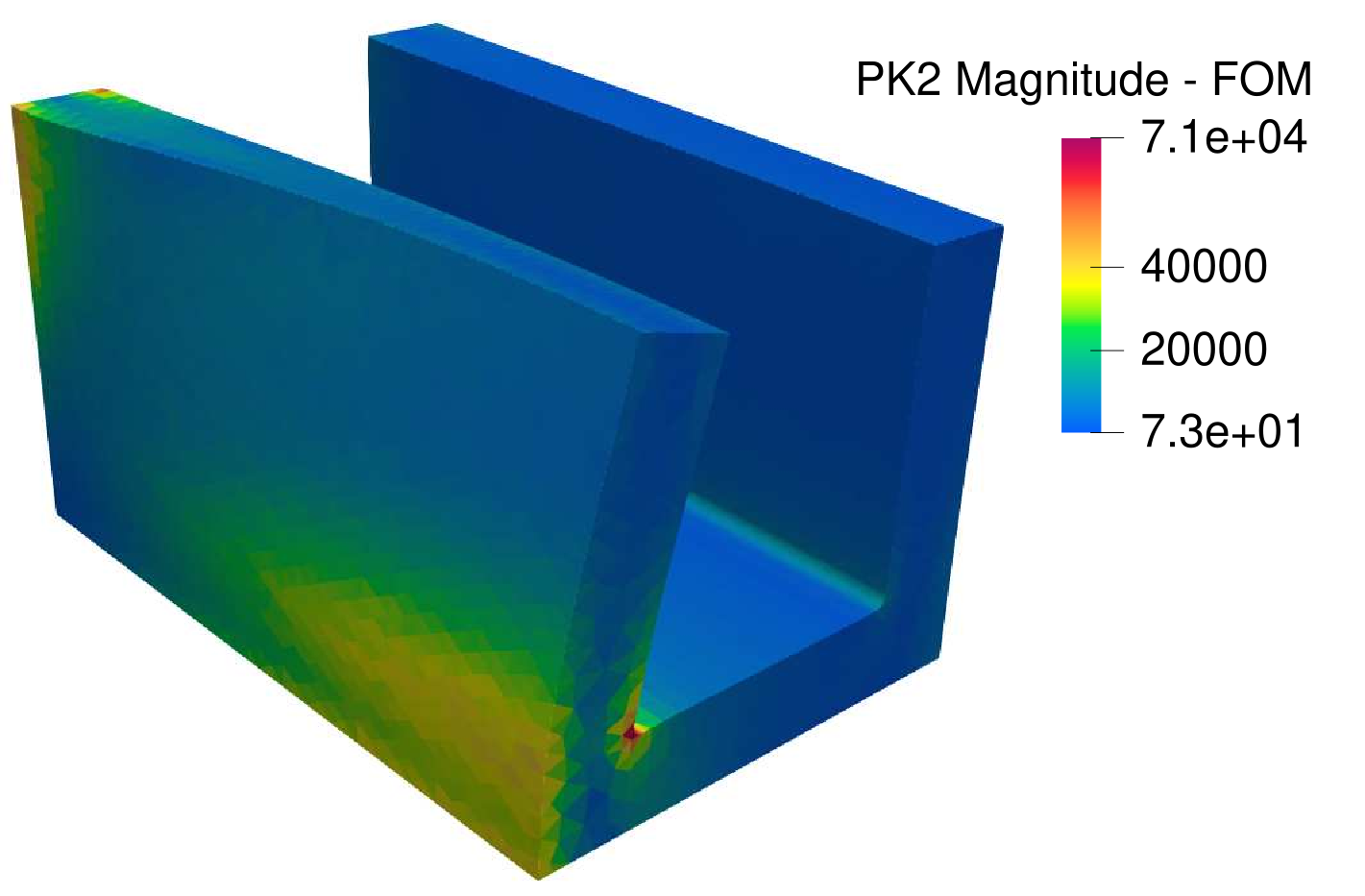}}
\end{minipage}%
\begin{minipage}{.5\linewidth}
\centering
\subfloat[]{\label{StressCase3Rom}\includegraphics[width=.8\textwidth]{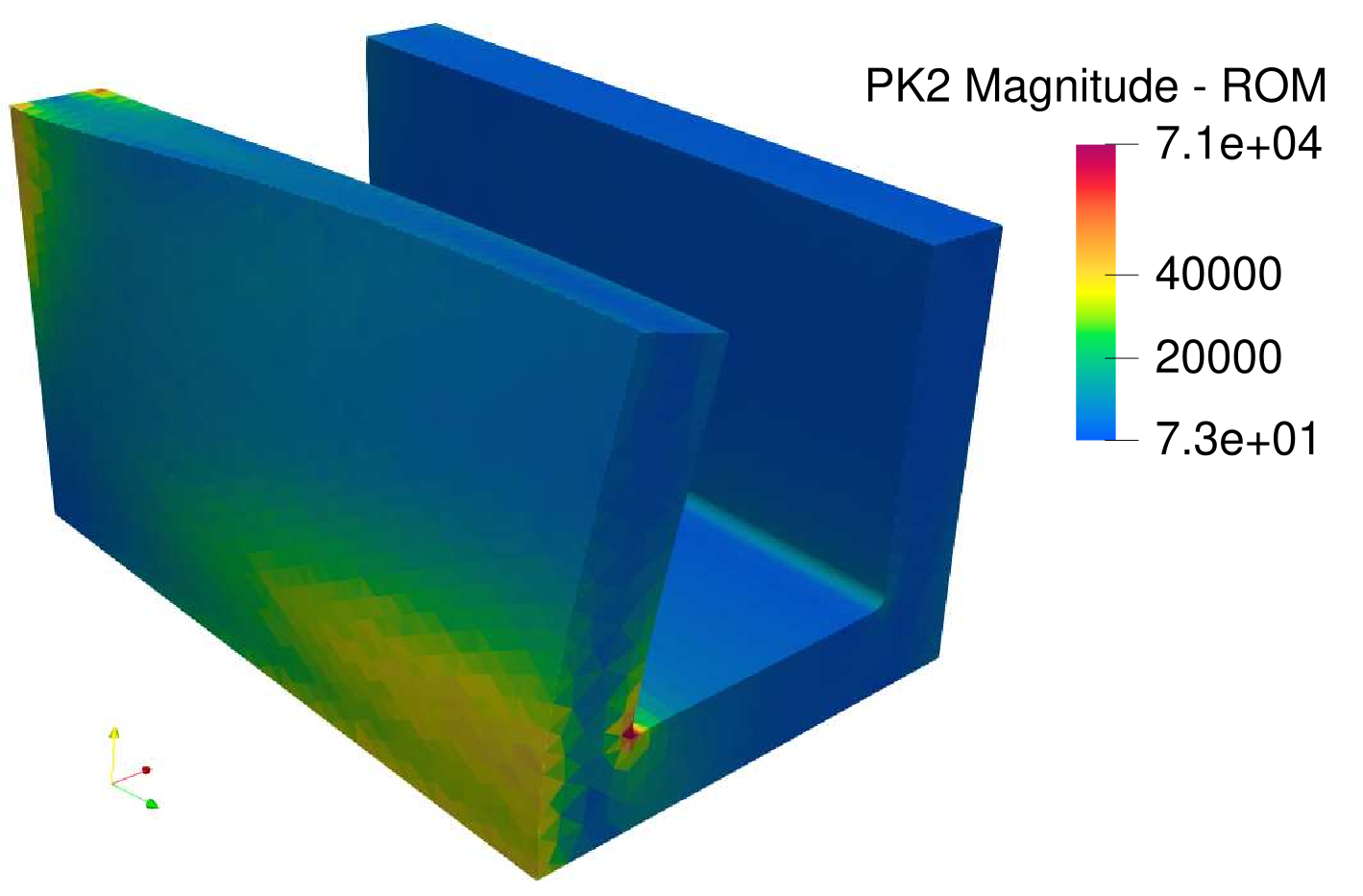}}
\end{minipage}\par\medskip
\centering
\subfloat[]{\label{deformedSolidStreams}\includegraphics[width=.9\textwidth]{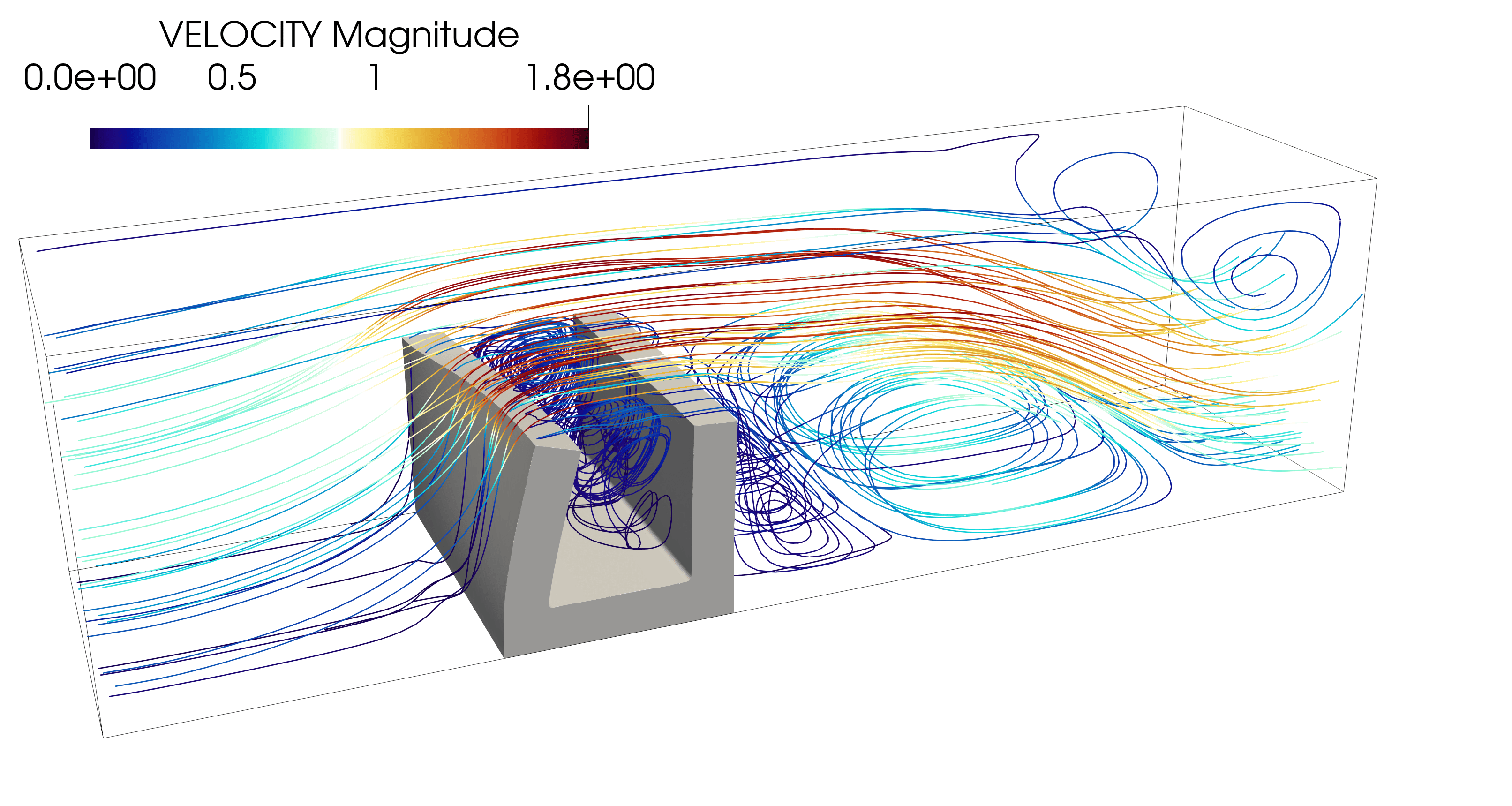}}\par\medskip
\caption{Evaluation of SROM prediction at $t=3.37 s$ of the PK2 stress magnitude comparing  (\textbf{a}) the FOM-FOM solution and (\textbf{b}) the ROM-FOM solution. In (\textbf{c}) the deformed solid and the velocity streamlines are shown at the same time step. The deformation is $\times10$ amplified}
\label{fig:StressCase3}
\end{figure}

For the solid material, a nearly-incompressible Neo-Hookean material is used with $\nu_s = 0.485$, and $\pmb{u}-\theta$ mixed tetrahedral elements are used where $\theta$ represents the Jacobian determinant $\theta = J \approx 1$. The mixed formulation is stabilized based on the VMS approach (See \cite{SCOVAZZI2023116076} for more details on the element used here). As a consequence, two solid ROMs are constructed: an SROM in a similar manner to the previous cases, and a second $\theta$-ROM for the $\theta$ field. In fact, in order to compute the strain and stress \textit{a posteriori}, an accurately computed $\theta$ field must be available, and in our ROM-FOM approach this is done only after convergence of each time step, since only the displacement at the interface is needed at the other fixed-point iterations. The solid mesh consists of 65400 tetrahedral elements with 15338 nodes.
In order to ensure a valid discretization, a mesh independence study was done on the fluid and the solid domains separately. Details on this study are reported on the Appendix~\ref{appx:mesh}. The second-order Bossak time integration is used by the fluid solver with a time step of $\Delta t = 0.01~s$. Similarly to the previous test case, matching interface grids are used, consisting of $7575$ nodes at the interface, giving $N = 22725$, and IQN-ILS with reuse $q = 3$ is used with a fixed-point tolerance of $\delta = 5\times10^{-4}$. In Figure~\ref{deformedSolidStreams}, we show the solution of the problem at $t = 3.37~s$.

The solid regressor $\mathcal{I}_S(\cdot)$ used here is a thin plate spline RBF interpolator, and $\mathcal{I}_F(\cdot)$ is a ridge regression with a regularization parameter $\lambda = 1\times10^{-5}$. For dimensionality reduction, $r_u=12$ displacement modes and $r_f=280$ forces modes are kept. For the data-driven predictor, the model update is computed after each $Z=240$ fixed-point iterations and the batch of the reduced snapshots is of size $p = 6900$. The tolerance used here is $\delta_r = 1\times10^{-4}$. The FOM-FOM simulation using $\pmb{\mu}_1$ and for all the time steps until $t = 6.6~s$ generates $m=2962$ snapshots for the training of SROM and FROM.

\begin{figure}
    \centering
    \includegraphics[width = 1\textwidth]{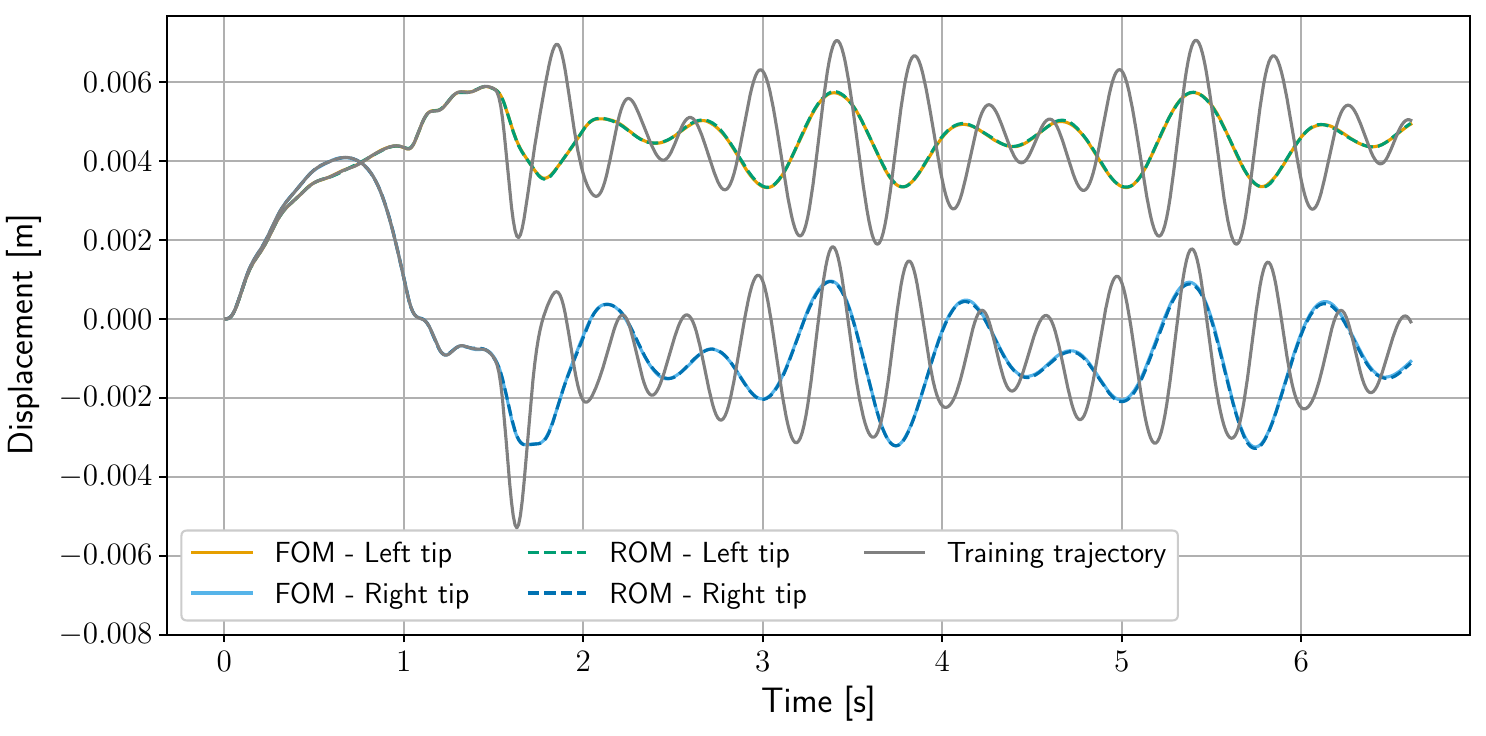}
    \caption{Example 3 - The left and right tip displacement evolution of the solid, comparing the ROM-FOM solution to the FOM-FOM solution.}
    \label{fig:dispResultsCase3}
\end{figure}

The SROM shows a significant accuracy on the displacement prediction, as seen in Figure \ref{fig:dispResultsCase3}, where the displacement evolution of the left tip and right tip of the $z+$ face are shown while comparing the FOM-FOM and ROM-FOM solutions. The prediction of the stress field, namely the PK2 stress is also obtained with high accuracy as seen in Figures \ref{StressCase3Fom} and \ref{StressCase3Rom} where the SROM and $\theta$ROM accurate output clearly lead to accurate stress predictions as well.

\begin{figure}
    \centering
    \includegraphics[width = 1\textwidth]{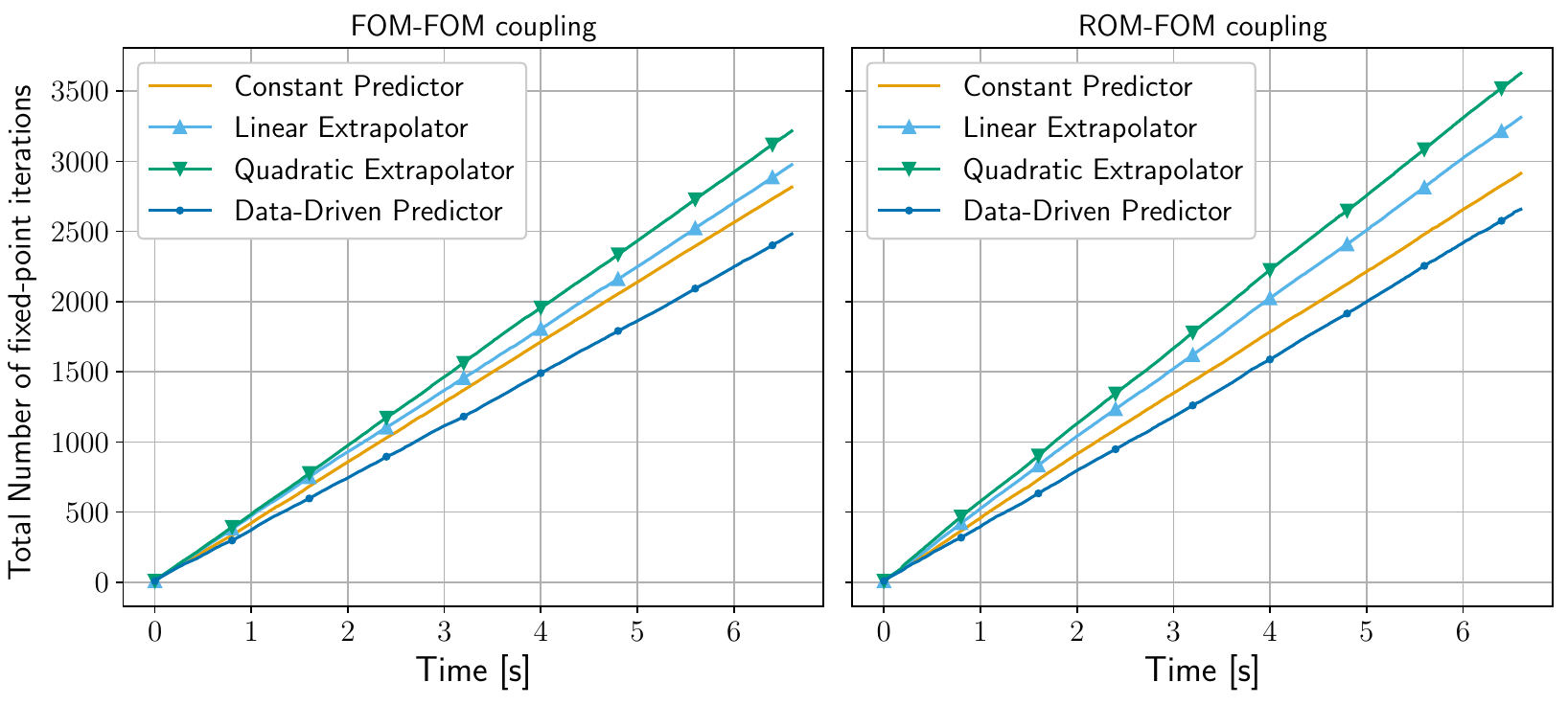}
    \caption{Example 3 - Comparison of total fixed-point iterations needed for the simulation at an unseen parameter $\pmb{\mu}_2$, using different predictors.}
    \label{fig:case3-iterations}
\end{figure}

The addition of the FROM and the local iterations for the data-driven initial guess results in much less total fixed-point iterations. This is clearly seen in Figure \ref{fig:case3-iterations}, where the data-driven predictor outperforms the classical approaches. The average number of fixed-point iterations and the total CPU time are also reported in Table \ref{tab:case3-speed}. Remarkably, the linear and quadratic extrapolators result in slower convergence than the constant extrapolation, while the data-driven predictor ensures a faster convergence of the fixed-point FSI problem, showing the improved robustness of the data-driven approach for predicting the initial guess.

\begin{table}[]
    \centering
    \begin{tabular}{c||c|c|c|c}
         & \multicolumn{2}{c}{FOM-FOM} & \multicolumn{2}{c}{ROM-FOM} \\
         & Avg. iterations & CPU Time & Avg. iterations & CPU Time \\
    \midrule
        Data-Driven & \textbf{3.75} & 1.06 & 4.0 & \textbf{1.0} \\
        Constant & \textbf{4.24} & \textbf{1.20} & 4.3 & 1.06 \\
        Linear & 4.5 & 1.28 & 4.97 & 1.25 \\
        Quadratic & 4.86 & 1.36 & 5.49 & 1.34 \\
    \end{tabular}
    \caption{Example 3 - Average number of iterations and the total CPU time of the simulation at the unseen $\pmb{\mu}$ parameter using the FOM-FOM and ROM-FOM coupling.}
    \label{tab:case3-speed}
\end{table}

\section{Conclusions}\label{sect:concl}

In this work, a novel data-driven predictor for the acceleration of convergence of unsteady partitioned fluid-structure interaction coupling has been proposed. This predictor provides an enhanced initial guess for the FSI fixed-point problem at each time step. It is obtained by resolving a reduced fixed-point problem that can be solved at the beginning of the time step for a small fraction of the computational time of the regular FSI problem. This is achieved using two reduced order models for the solid and the fluid problems by approximating the force-to-displacement and the displacement-to-force relationships respectively. The two models are then strongly coupled to predict the initial guess when this reduced fixed-point converges. Each reduced order model is constructed from three components: An encoder, a regressor and a decoder. The encoder-decoder uses the POD and quadratic manifolds for the dimensionality reduction and the regressors use either RBF functions or polynomial regression. The data-driven nature of this predictor makes it more robust and efficient than the classical approach, since it uses the information from the results of the high-fidelity solver, instead of using finite-differences from the last few time steps. Moreover, the regression model of the fluid ROM is updated online using the high-fidelity forces from the fluid FOMs, enriching further the ROM from the latest available HF data. Overall, the proposed methodology leverages physics-based insights from the high fidelity fluid solver, thus establishing a physics-aware machine learning predictor. This enables the use of the predictor in extrapolating regions of the time-parameter space. This predictor can also be combined with the solid ROM presented in \cite{tiba2023nonintrusive} for replacing the solid solver as a whole to predict the solution at an even cheaper computational cost. In the paper, and through three examples with strong FSI coupling and neglected solid inertia, we have demonstrated the performance of this novel predictor in achieving faster convergence of the fixed-point problem compared to classical extrapolators. In particular, we showed the significant computational gain that can be achieved with this predictor, even when applying it for unseen time and parameters, even in an extrapolation setting, and even for fairly complex dynamics of the FSI problem, all while keeping a very high accuracy of the ROM when replacing the solid FOM. 

We showed that designing an FSI predictor with such a data-driven strategy makes it more robust for easing convergence than the classical extrapolators, since the data-driven ROM benefits from recent HF data more judiciously. The ROM update strategy presented in this work could be further enhanced in order to obtain more accurate and adapted ROMs: for example, instead of a straightforward retraining of the regressor component of the ROM only (as done in this work), an update strategy of the dimensionality reduction part (namely the POD bases update), could also be done using the online HF data. We believe that such a strategy will eventually lead to faster fixed-point convergence, since new force values at prediction time can lie outside the available POD subspace obtained from offline data. This adaptive encoder/decoder approach will be pursued in future researches.

\section{Acknowledgements}
This work has been funded by the ANR (Agence Nationale de la Recherche, France), Altair Engineering and Michelin under the project AHEAD.

\newpage

\begin{appendices}
\section{IQN-ILS Algorithm}

\begin{algorithm}[H]
\caption{Interface Quasi-Newton Inverse Least-Squares \cite{degroote_performance_2009}}\label{alg:iqnils}
\KwData{Iteration $k$

Time step $n$

Force $\Tilde{\pmb{f}}^k$

Force residual $\pmb{r}^k$

Reuse $q$

Relaxation $w_0$}
\KwResult{$\Delta \pmb{f}^k = Acc(\Tilde{\pmb{f}}^k, {\pmb{r}}^k)$}
\vline

\uIf{$n == 1$ and $k==1$}{

\nl $\Delta \pmb{f}^k = w_0 \pmb{r}^k$
}\Else{
\nl $\pmb{V} = [ \Delta\pmb{r}^k ~ \cdots ~ \Delta\pmb{r}^{k-d}]$  from the previous $d$ fixed-point iterations (including past $q$ time steps)

\nl $\pmb{W} = [ \Delta\Tilde{\pmb{f}}^k ~ \cdots ~ \Delta\Tilde{\pmb{f}}^{k-d}]$  

\nl QR decomposition $\pmb{V} = \pmb{Q} \pmb{R}$

\nl QR filtering using e.g the approach in  \cite{haeltermannFilter}

\nl $\pmb{b} = -\pmb{Q}^T \pmb{r}^k$

\nl Solve $\pmb{R} \pmb{c} = \pmb{b}$ for $\pmb{c}$ using back-substitution

\nl $\Delta \pmb{f}^k = \pmb{W} \pmb{c} + \pmb{r}^k$
}
\end{algorithm}

\section{Mesh convergence study}\label{appx:mesh}
To conduct the mesh convergence study efficiently, the solid and fluid problems are treated independently, i.e in a decoupled way, as the goal is merely to check the validity of the chosen mesh sizes even if the actual FSI problem is different, assuming the other conditions are close enough

A static surface load is applied on the left solid face, with a load amplitude comparable of that occurring during the FSI problem. The fluid problem for this study differs from the coupled problem in that only a constant inlet velocity is applied instead of a pulsatile inlet.

\subsection{Solid mesh:}
\begin{figure*}
  \begin{subfigure}{0.33\textwidth}
    \includegraphics[width=\linewidth]{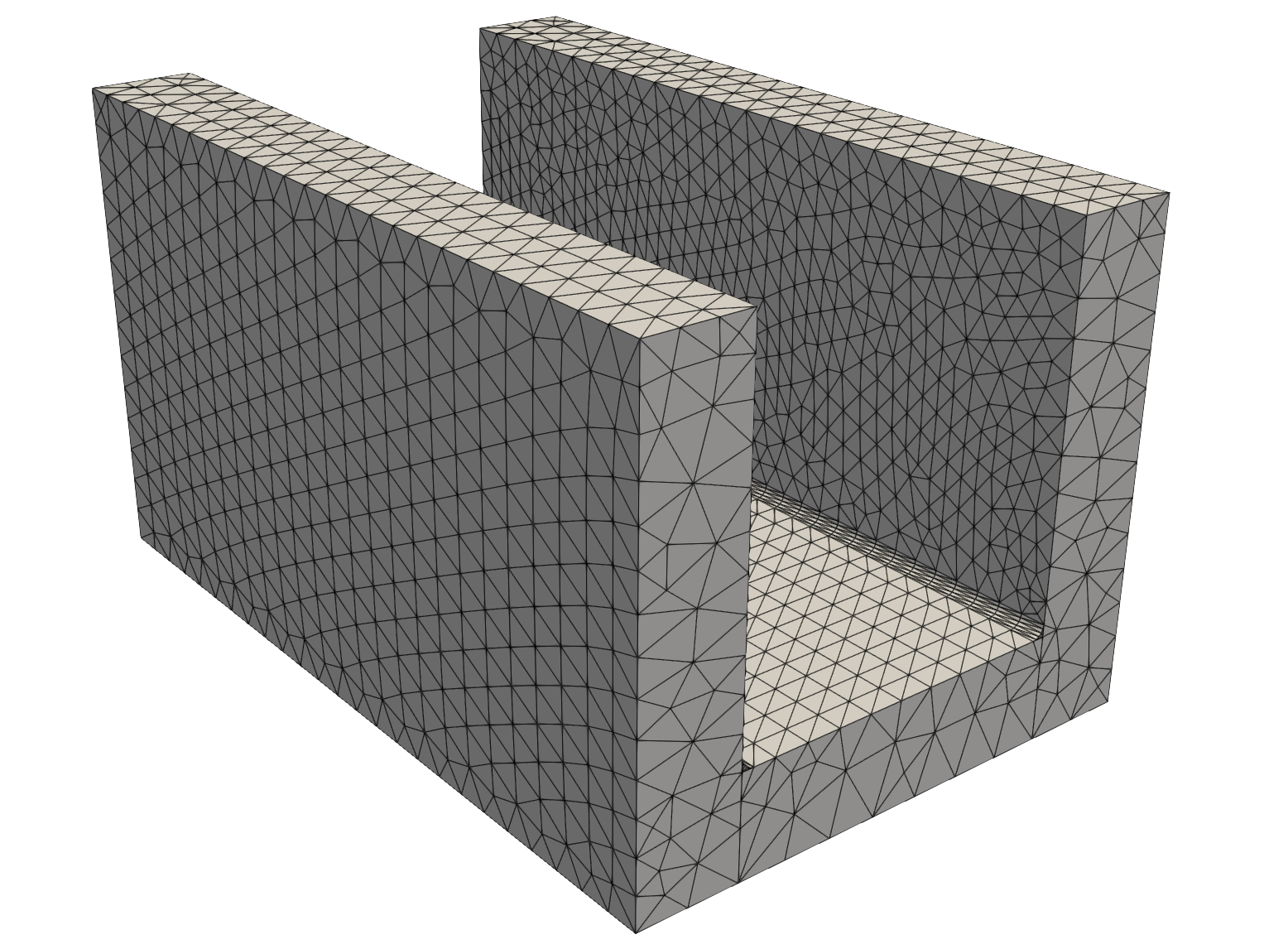}
    \caption{$h_1= 20~mm$}
    \label{fig:figure1}
  \end{subfigure}%
  \hfill
  \begin{subfigure}{0.33\textwidth}
    \includegraphics[width=\linewidth]{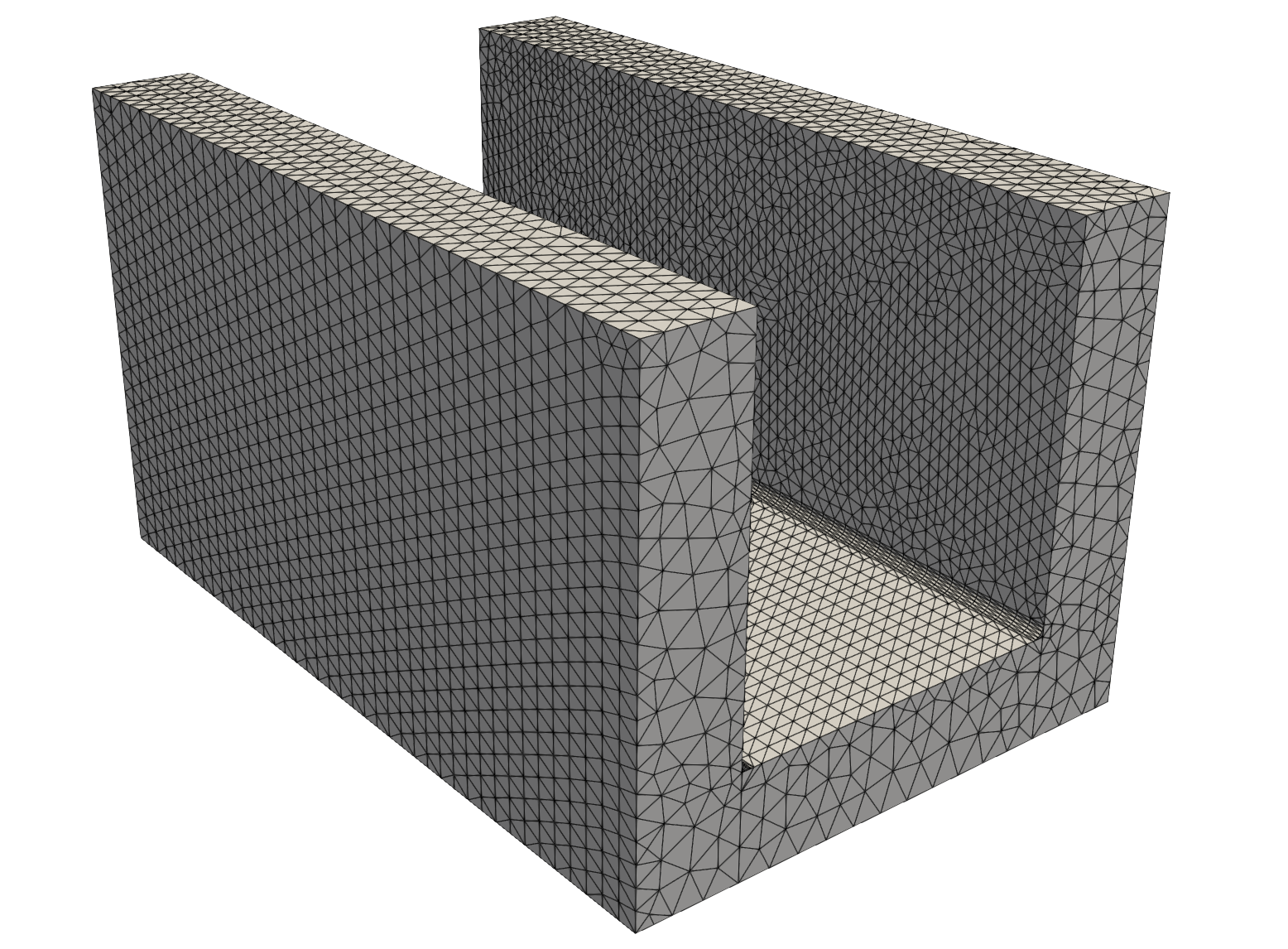}
    \caption{$h_2= 11.75~mm$}
    \label{fig:figure2}
  \end{subfigure}%
  \hfill
  \begin{subfigure}{0.33\textwidth}
    \includegraphics[width=\linewidth]{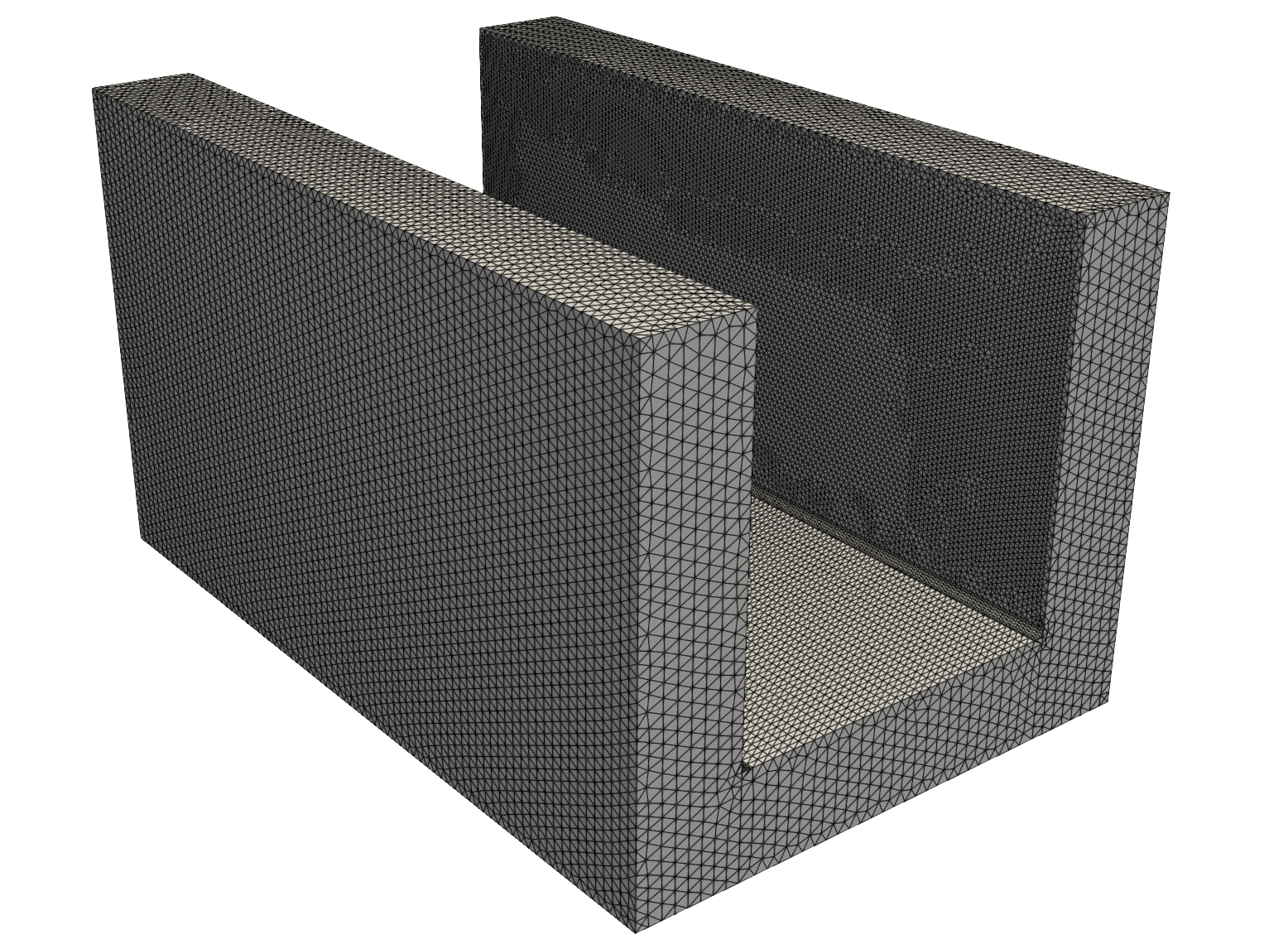}
    \caption{$h_3= 6.25~mm$}
    \label{fig:figure3}
  \end{subfigure}
  \caption{Solid meshes used in the mesh convergence study.}\label{fig:SolidMeshes}
\end{figure*}

A constant pressure of $p = 1800~Pa$ is applied on the left face and three different meshes with three different (average) mesh sizes on the boundaries are used: $h_1 = 20~mm$, $h_3=6.25~mm$ and $h_2=11.75~mm$. 
The three meshes are illustrated in Figure \ref{fig:SolidMeshes}. Note that the mesh size changes locally as it is reduced near the corners.

\begin{figure}
    \centering
    \includegraphics[width=.6\textwidth]{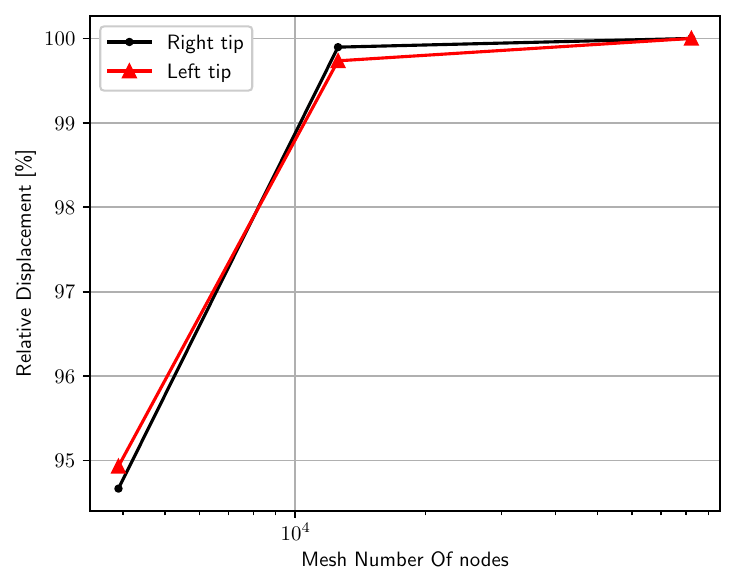}
    \caption{The displacement of the two tips of the solid plotted for each considered mesh. The displacement quantity is relative to that associated to the finer (reference) mesh.}
    \label{fig:solidConvDisp}
\end{figure}

The most interesting quantity in the problem - the displacement - is reported on the right tip and left tip at the $z+$ face for the three grids and plotted on Figure \ref{fig:solidConvDisp}, from which we see that the $h_2$ displacement falls under $1\%$ from that of the reference value, assumed to be the one associated to the fine $h_3$ mesh. From there, we conclude that the $h_2$ grid is sufficient for the use of our FSI problem.

\subsection{Fluid mesh:}
Similar to the FSI problem, a Poiseuille inlet flow velocity is applied until $t=6.6~s$ for this study, with the difference here is that the inlet flow is constant in time whether a pulsatile flow is applied on the FSI example. The value of the inlet velocity corresponds to the maximum attained on the FSI problem inlet.

\begin{figure}
\centering
\begin{subfigure}[b]{0.7\textwidth}
   \includegraphics[width=1\linewidth]{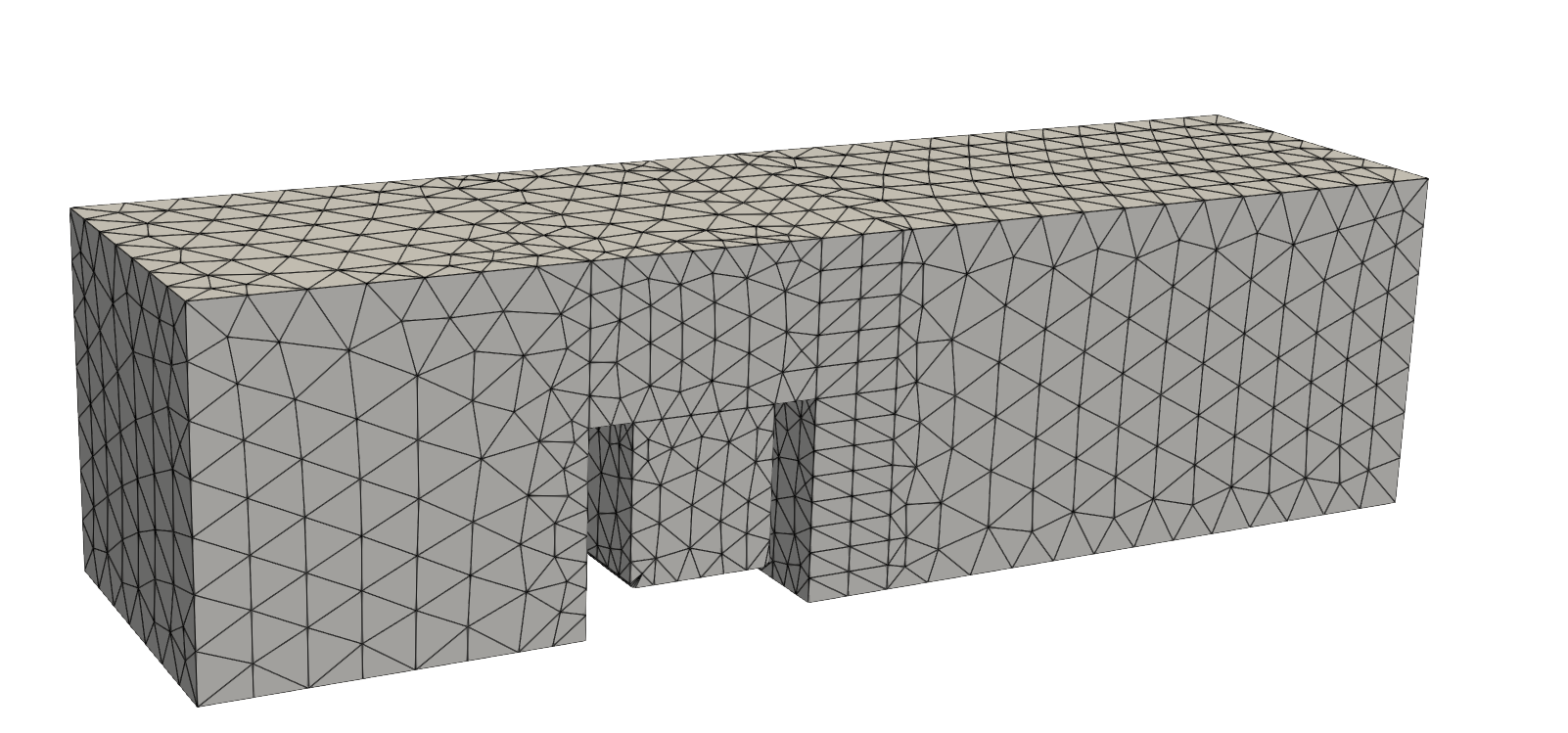}
   \caption{$h_1=40~mm$}
   \label{fig:crsFld} 
\end{subfigure}

\begin{subfigure}[b]{0.7\textwidth}
   \includegraphics[width=1\linewidth]{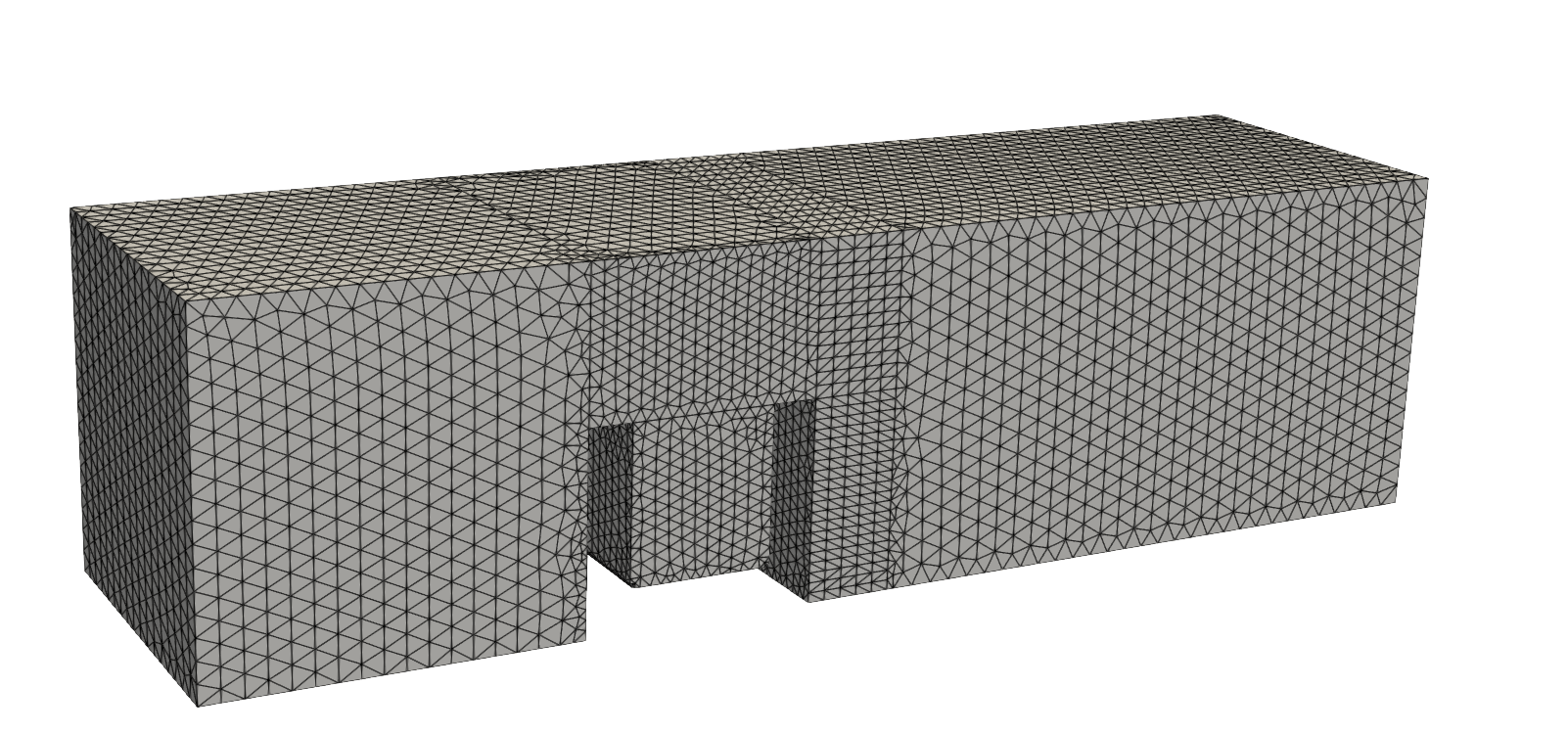}
   \caption{$h_2=18~mm$}
   \label{fig:midFld}
\end{subfigure}

\begin{subfigure}[b]{0.7\textwidth}
   \includegraphics[width=1\linewidth]{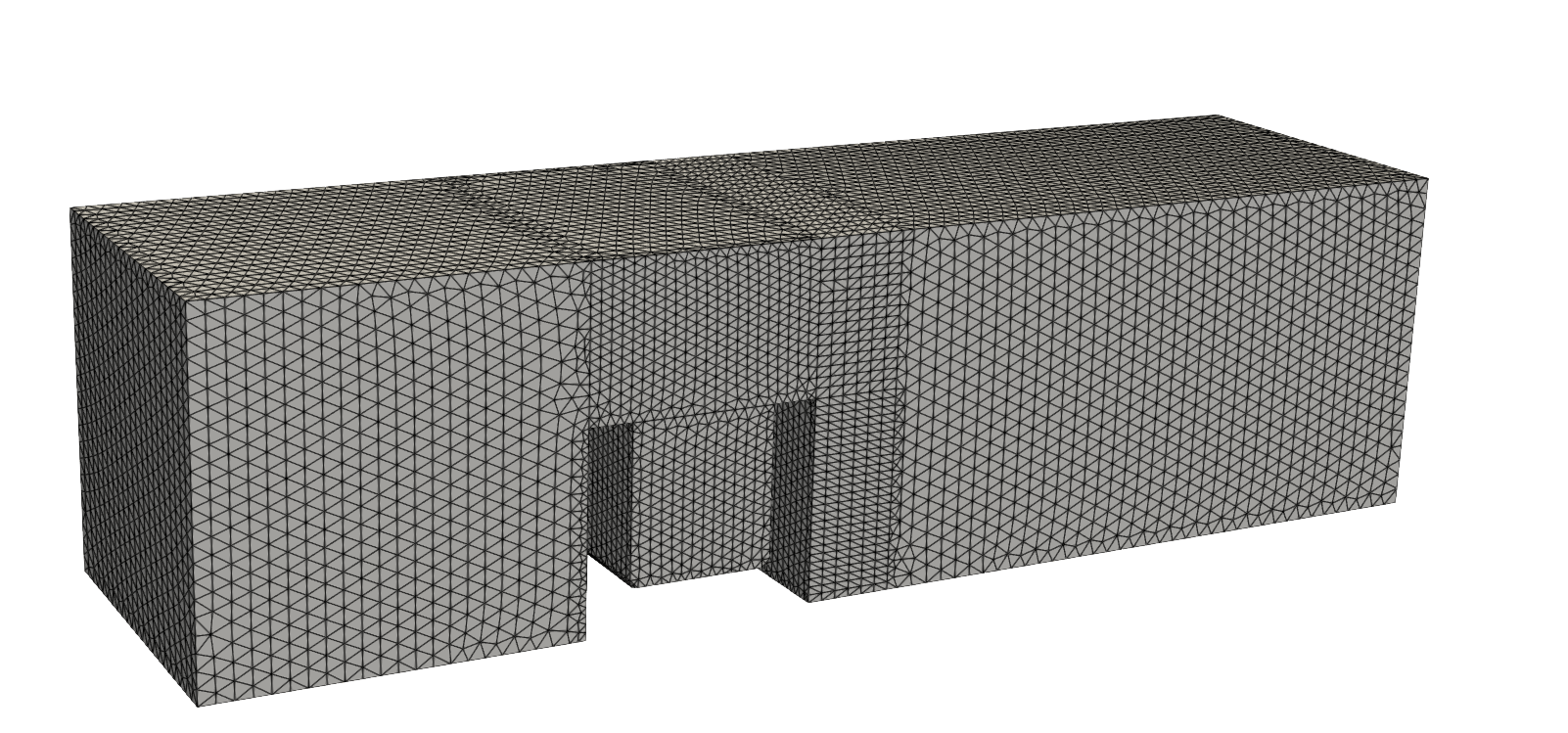}
   \caption{$h_3=12~mm$}
   \label{fig:finFld}
\end{subfigure}

\caption{Outer surfaces of the fluid meshes used in the mesh convergence study.}\label{fig:FldMeshes}
\end{figure}

Three grids with varying average mesh sizes on the boundaries $h_1=40~mm$, $h_2=18~mm$ and $h_3 = 12~mm$. Note again, that  locally, the mesh size changes as it is refined near the FSI interface and in the region between the flaps. The three grids are shown in Figure \ref{fig:FldMeshes}.

\begin{figure}
\centering
\begin{subfigure}[b]{0.7\textwidth}
   \includegraphics[width=1\linewidth]{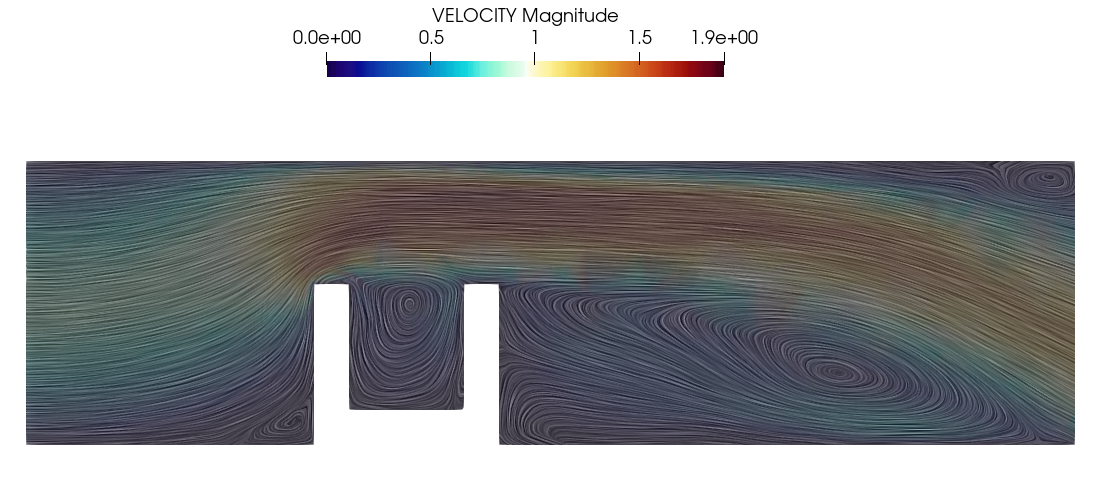}
   \caption{$h_1=40~mm$}
   \label{fig:crsFldStrm} 
\end{subfigure}

\begin{subfigure}[b]{0.7\textwidth}
   \includegraphics[width=1\linewidth]{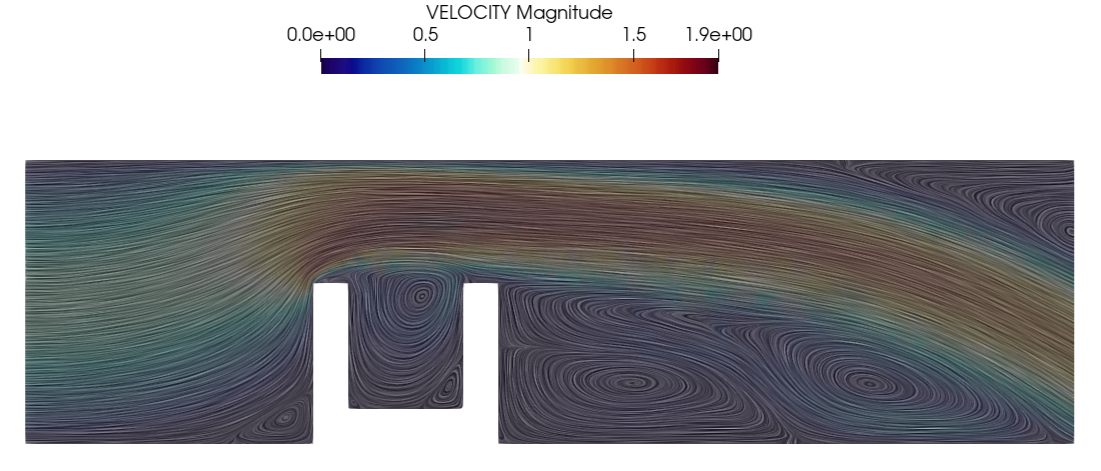}
   \caption{$h_2=18~mm$}
   \label{fig:midFldStrm}
\end{subfigure}

\begin{subfigure}[b]{0.7\textwidth}
   \includegraphics[width=1\linewidth]{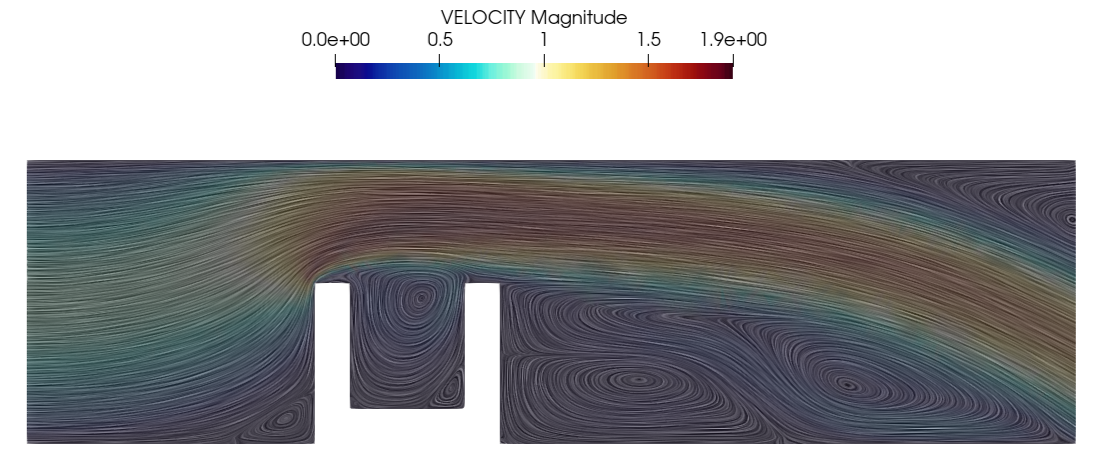}
   \caption{$h_3=12~mm$}
   \label{fig:finFldStrm}
\end{subfigure}

\caption{Streamlines of the velocity field on the mid section for the three considered grids at $t=6.6~s$.}\label{fig:FldMeshesStreams}
\end{figure}

\begin{figure}
    \centering
    \includegraphics[width=.6\textwidth]{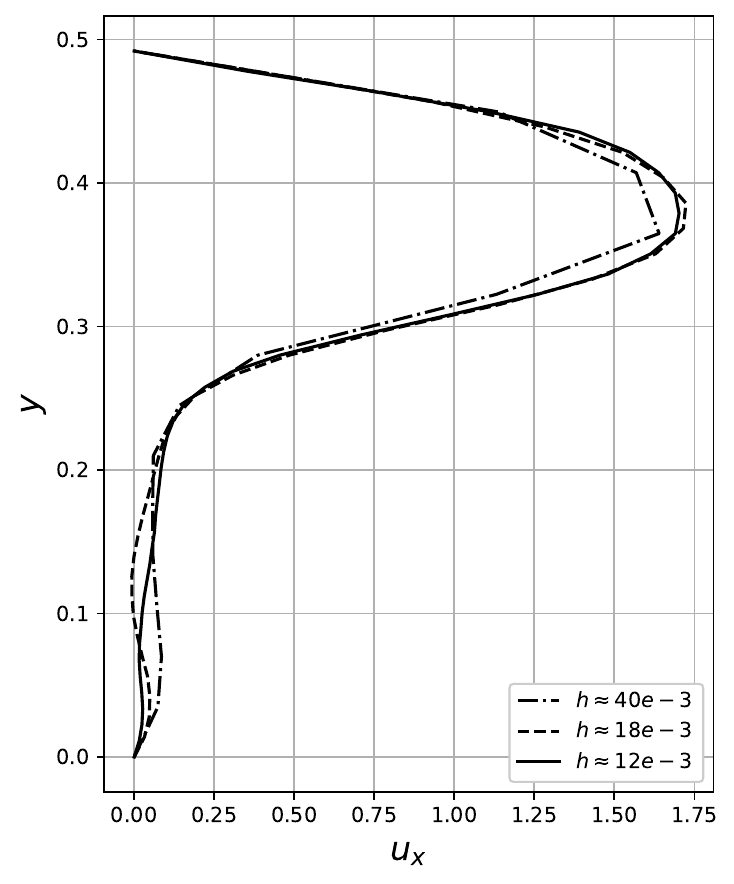}
    \caption{The axial velocity profile for the three considered grids.}
    \label{fig:fluidConvDisp}
\end{figure}

The axial velocity profile $120~mm$ on the right FSI interface is shown in Figure \ref{fig:fluidConvDisp} for the three different grids, using the mean of the time steps corresponding to the last $4.6~s$. The streamlines of the velocity field at $t=6.6~s$ are also shown on the mid section of the channel in Figure \ref{fig:FldMeshesStreams}. In addition, we used the Grid Convergence Index (GCI) method from the ASME Journal of Fluids Engineering policy for mesh convergence \cite{asme}. The grid refinement factor between each two meshes is indeed greater than $1.3$ and using the maximum velocity of the profiles reported in Figure \ref{fig:fluidConvDisp} as the main quantity, the GCI index obtained is $4.9\%$ which is we considered acceptable.

From all these results, we concluded the choice of using the grid with $h_2$ for our FSI problem.

\end{appendices}

\clearpage

\bibliographystyle{elsarticle-num-names} 
\bibliography{elsarticle-template-num-names}

\end{document}